\documentclass{jfm}
\usepackage{graphicx}
\usepackage{epstopdf, epsfig}

\usepackage{amsmath, bm}

\shorttitle{Viscous Flow in a 2-D Shock Tube}
\shortauthor{G. Zhou, K. Xu and F. Liu}

\title{Grid-converged Solution and Analysis of the Unsteady Viscous Flow in a Two-dimensional Shock Tube}

\author{Guangzhao Zhou\aff{1}
  \corresp{\email{zgz@pku.edu.cn}},
  Kun Xu\aff{2}
 \and Feng Liu\aff{3}}

\affiliation{\aff{1}College of Engineering, Peking University, Beijing 100871, China
\aff{2}Department of Mathematics, Hong Kong University of Science and Technology, Kowloon, Hong Kong, China
\aff{3}Department of Mechanical and Aerospace Engineering, University of California, Irvine, CA 92697-3975, United States}

\begin{document}

\maketitle

\begin{abstract}
The flow in a shock tube is extremely complex with dynamic multi-scale structures of sharp fronts, flow separation, and vortices due to the interaction of the shock wave, the contact surface, and the boundary layer over the side wall of the tube.  Prediction and understanding of the complex fluid dynamics is of theoretical and practical importance. It is also an extremely challenging problem for numerical simulation, especially at relatively high Reynolds numbers. Daru \& Tenaud (Daru, V. \& Tenaud, C. 2001 Evaluation of TVD high resolution schemes for unsteady viscous shocked flows. Computers \& Fluids 30, 89--113) proposed a two-dimensional model problem as a numerical test case for high-resolution schemes to simulate the flow field in a square closed shock tube. Though many researchers have tried this problem using a variety of computational methods, there is not yet an agreed-upon grid-converged solution of the problem at the Reynolds number of 1000. This paper presents a rigorous grid-convergence study and the resulting grid-converged solutions for this problem by using a newly-developed, efficient, and high-order gas-kinetic scheme. Critical data extracted from the converged solutions are documented as benchmark data. The complex fluid dynamics of the flow at $\Rey=1000$ are discussed and analysed in detail. Major phenomena revealed by the numerical computations include the downward concentration of the fluid through the curved shock, the formation of the vortices, the mechanism of the shock wave bifurcation, the structure of the jet along the bottom wall, and the Kelvin-Helmholtz instability near the contact surface.

\end{abstract}

\begin{keywords}
\end{keywords}

\section{Introduction}
\begin{figure}
  \centerline{\includegraphics[width=0.5\textwidth]{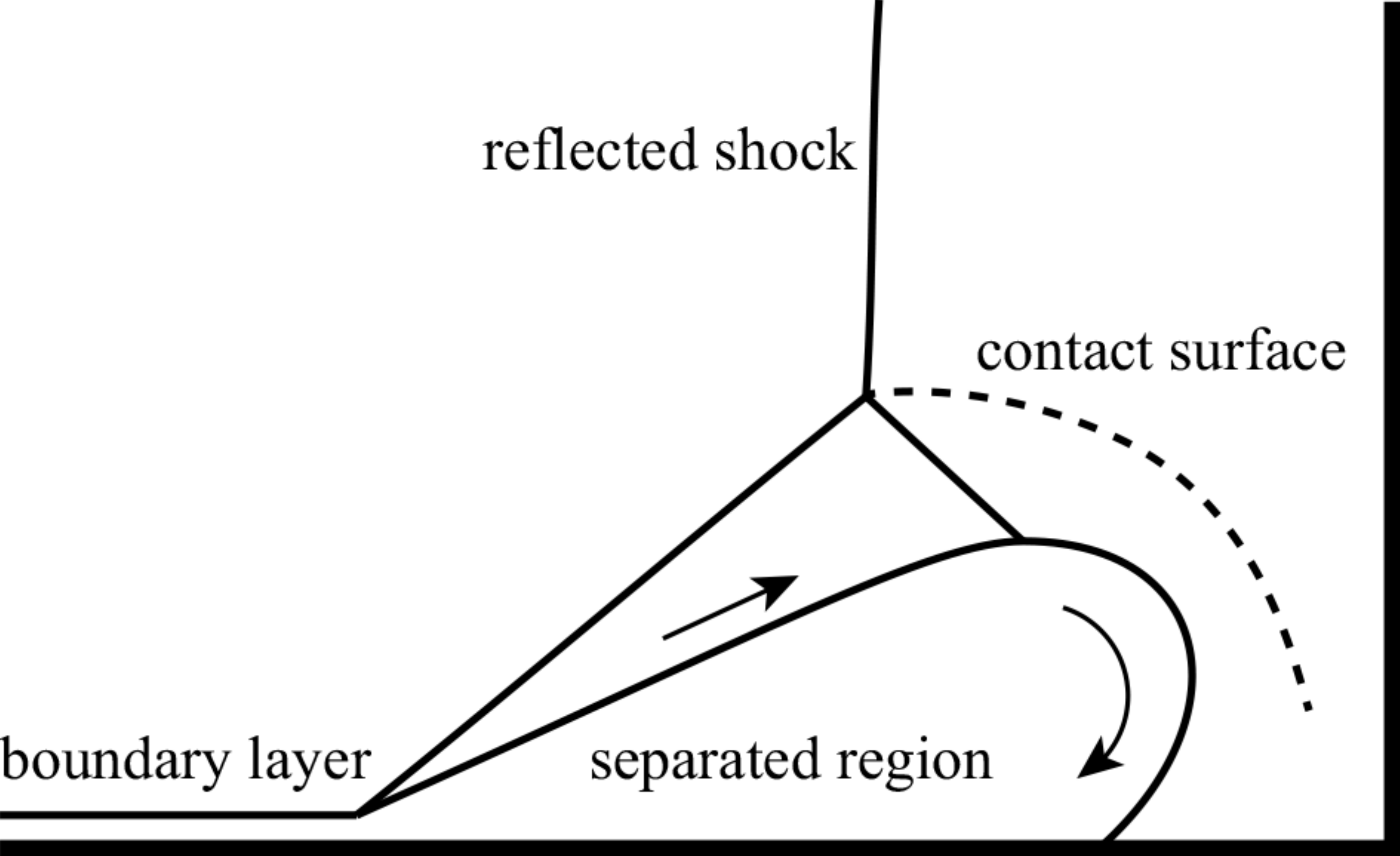}}
  \caption{Main flow structure of the interaction between the boundary layer and the reflected shock.}
\label{vst}
\end{figure}

The shock tube is used as an experimental apparatus for studies of hypersonic flow and chemical reactions. The shock wave reflected from the end wall interacts with the boundary layer on the side wall induced by the incident shock as shown schematically in figure~\ref{vst}. Compression by the main high-energy flow from the left causes the fluid at the end wall to `leak' backwards near the bottom wall where the fluid dynamic pressure is low because of the wall boundary layer. In time, the forward and backward flow in the boundary layer separates from the bottom wall resulting in a complex system of vortices, shock wave bifurcation, and other various flow structures. The homogeneity of the flow conditions in that region, however, is important for experimental tests using the shock tube \citep{Bull1968}. \citet{Mark1958} was the first to study this type of shock-wave/boundary-layer interaction. He developed a model based on the experimental results for analysis and prediction of the flow configuration and primary geometric parameters.  \citet{Byron1961} used a more realistic model, which is applicable for higher Mach numbers compared to Mark's model. Subsequent theoretical analyses can be found in \citet{Davies1969} and \citet{Stalker1978}.

In recent decades, experiments and numerical simulations of this problem have been reported by other authors \citep{Kleine1992, Wilson1995, Weber1995}. As the viscosity plays an important role in the development of the flow field, the Reynolds number is a key parameter determining the features of the interaction. Differences of the Reynolds numbers used in the above papers make it difficult for comparison and analysis between their reported results.

\citet{Daru2001} proposed a two-dimensional model problem for numerical simulation of the flow field in a viscous shock tube, which is designed for evaluating different numerical methods. This is a time-dependent unsteady problem. At moderate Reynolds numbers, a number of vortices appear in the computational domain due to high shearing effect, with length scales varying in a wide range. The multi-scale nature and the complicated flow field make it a good test case for high-order high-resolution schemes. As very fine grids are needed to resolve small structures, a practical problem is whether the computation could be completed within acceptable computational time. Therefore this case is a challenge for the robustness, accuracy, resolution as well as efficiency of a numerical method. 

Since presented by \citet{Daru2001}, the viscous shock tube problem has been tested in many articles \citep{Sjogreen2003, Daru2004, Kim2005a, Kim2005b,  Daru2009, Li2011, Houim2011, Wan2012, Sun2014, Kotov2014, Tenaud2015, Wang2015, Pan2016a, Pan2016b}. The cases with Reynolds numbers of 200 and 1000 are most frequently used. The results for the $\Rey = 200$ case by different schemes are generally similar. But for the $\Rey = 1000$ case, a range of solutions that are noticeably different have been reported in different papers. It seems that a grid-converged solution has not been shown at this Reynolds number. In this paper, grid-converged solutions are successfully obtained at both Reynolds numbers. The results for $\Rey = 1000$ are in good agreement with the solution by \citet{Daru2009}. We recommend the current results be a reference solution.

The gas-kinetic scheme has been developed in the past years and shown great success in various categories of flows. The method employs the BGK equation \citep{BGK} instead of the Navier--Stokes equations. A gas distribution function is modelled to represent the flow status. Since all macroscopic variables are simply the moments of the distribution function, the inviscid and viscous fluxes are treated simultaneously \citep{Xu1998, Xu2001}. Based on the high-order gas-kinetic scheme proposed by \citet{Luo2013a} which employs the WENO-JS reconstruction technique \citep{Liu1994, Jiang1996} and a high-order gas evolution model, several simplifications are made by the authors and the resultant scheme enhances the efficiency by about $400\%$ for two-dimensional flows \citep{Zhou2017}. With this efficient high-order gas-kinetic scheme, we are able to simulate the viscous shock tube problem with finer grids and achieve grid-converged solutions at both $\Rey = 200$ and $\Rey = 1000$ in an acceptable CPU time.

In the following section, we will first outline the numerical method.  \S\ref{section problem} spells out the specification and computational conditions of the shock tube problem. The solutions at $\Rey=200$ and $\Rey=1000$ are presented in \S\ref{section 200} and \S\ref{section 1000}. \S\ref{section 1000} focuses on the difficult case at $\Rey=1000$. A procedure making use of the Grid-Convergence Index (GCI) is presented and used to prove grid convergence of our computations on a sequence of successively refined grids.  The grid-converged solution provides fine details of the complex flow structure for the $\Rey=1000$ case. In \S\ref{section process} we discuss and analyse the detailed evolution of the fluid dynamics revealed by the numerical solution starting from the initiation of the incident shock wave and contact surface through a sequence of phenomena including the downward concentration of the fluid through the curved shock, the formation of the vortices, the bifurcation of the shock wave, creation of a jet-like flow towards the bottom wall, and vortex structures created by Kelvin-Helmholtz instability near the contact surface. Finally, we draw the conclusions in \S\ref{section conclusion}.

\section{Numerical Procedure} \label{section method}
In this section, we give a brief introduction to the numerical method. More details can be found in \citet{Luo2013a} and \citet{Zhou2017}. 

We start from the BGK equation \citep{BGK}:
\begin{equation} \label{bgk}
f_t + \bm{u \cdot\nabla} f = \frac{g-f}{\tau},
\end{equation}
where $f$ is the gas distribution function, $g$ is the equilibrium state that $f$ approaches, $\bm{u} = (u,v)^T$ is the particle velocity, and $\tau$ is the collision time. For two-dimensional flow, the equilibrium (Maxwellian) distribution is
\begin{equation} \label{equilibrium}
g = \rho \left( \frac{\lambda}{\pi} \right)^{\frac{K+2}{2}} \!\! e^{-\lambda \left[ (u-U)^2 + (v-V)^2 + \xi^2 \right]},
\end{equation} 
where $\rho$ is the density, $U$, $V$ are macroscopic velocities in the $x$ and $y$ directions. $\lambda = m/2kT$, where $m$ is the molecular mass, $k$ is the Boltzmann constant and $T$ is the temperature. $K$ is the number of internal degrees of freedom which equals to $3$ for diatomic molecules. $\xi$ is the internal variable with $\xi^2 = \xi_1^2 + \xi_2^2 + \cdots + \xi_K^2$.

\eqref{bgk} has an analytical integral solution:
\begin{equation}\label{solution}
f(\bm{x},t,\bm{u},\xi) = \frac{1}{\tau} \int_0^t g\left( \bm{x}^\prime, t^\prime,  \bm{u}, \xi \right) e^{-(t-t^\prime) / \tau} dt^\prime + e^{-t/\tau} f_0 \left( \bm{x} - \bm{u} t, \bm{u}, \xi \right),
\end{equation}
where $\bm{x}^\prime = \bm{x} - \bm{u} (t-t^\prime)$ is the particle trajectory. Therefore $f$ depends on the equilibrium distribution function $g(\bm{x},t,\bm{u},\xi) $  and the initial distribution function $f_0(\bm{x},\bm{u},\xi) $.

Let $g = g(\bm{0},0,\bm{u},\xi) $ denote the Maxwellian distribution at the point $(x,y,t) = (0,0,0)$. Then $\tilde{g}$, the equilibrium distribution in the neighbourhood, can be expressed via the Taylor expansion to the second order:
\begin{equation}
\tilde{g}(\bm{x},t,\bm{u},\xi) = g + g_x x + g_y y + g_t t + \frac{1}{2}g_{xx} x^2 + \frac{1}{2}g_{yy} y^2 + \frac{1}{2}g_{tt} t^2 + g_{xy} xy + g_{xt} xt + g_{yt} yt.
\end{equation}

According to the Chapman-Enskog expansion, to the order of the Navier-Stokes equations, the non-equilibrium distribution $f$ has the following relation with the equilibrium distribution $g$  \citep{Ohwada2004}:
\begin{equation}
f = g - \tau Dg =  g - \tau \left( g_t + ug_x + vg_y \right).
\end{equation}
Expand each term of $f$ at the point $(x,y,t) = (0,0,0)$, and neglect high-order derivatives of $g$, we have
\begin{equation}
\begin{aligned}
\tilde{f} (\bm{x},t,\bm{u},\xi) =   & g + g_x x + g_y y + g_t t + \frac{1}{2} g_{xx} x^2 + \frac{1}{2} g_{yy} y^2 + \frac{1}{2} g_{tt}t^2 \\
& + g_{xy} xy + g_{xt} xt + g_{yt} yt  - \tau ( g_t + g_{xt} x  + g_{yt} y + g_{tt}t ) \\
& -  \tau u ( g_x + g_{xx} x + g_{xy} y + g_{xt} t ) - \tau v ( g_y + g_{xy} x + g_{yy} y + g_{yt} t  ).
\end{aligned}
\end{equation}

Note that for an arbitrarily given equilibrium state $g$, there exist $\tilde{g}$ and $\tilde{f}$ corresponding to $g$. Then we have the form $\tilde{g} = \tilde{g}(g,\bm{x},t,\bm{u})$, $\tilde{f} = \tilde{f}(g,\bm{x},t,\bm{u})$. The initial state at the cell interface should be discontinuous:
\begin{equation}
f_0 \left( \bm{x}, \bm{u},\xi \right) = \left\{
\begin{aligned}
& f_0^l  \left( \bm{x}, \bm{u},\xi \right) = \tilde{f}^l  \left( g_0^l, \bm{x}, 0, \bm{u}\right) , \quad x \leq 0, \\
& f_0^r  \left( \bm{x}, \bm{u},\xi \right) = \tilde{f}^r  \left( g_0^r, \bm{x}, 0, \bm{u} \right) , \quad x > 0,
\end{aligned}\right.
\end{equation}
where $g_0^l$ and $g_0^r$ correspond to the reconstructed conservative variables at the left and right sides of the cell interface, respectively, i.e.,
\begin{equation}
\bm{W}^l = \int g_0^l \bm{\psi} d\Xi, \quad \bm{W}^r = \int g_0^r \bm{\psi} d\Xi,
\end{equation} 
where $d\Xi = du dv d\xi$, $d\xi = d\xi_1 d\xi_2 \cdots d\xi_K$, and $\bm{\psi}$ is the vector of moments:
\begin{equation}
\bm{\psi} = \left( \psi_1, \psi_2, \psi_3, \psi_4 \right)^T = \left( 1, u, v, \textstyle\frac{u^2 + v^2 + \xi^2}{2} \right)^T.
\end{equation}

On the other hand, the equilibrium distribution function in the integral of the solution is replaced by
\begin{equation}
g \left( \bm{x}, t, \bm{u},\xi \right)  = \tilde{g} \left( g^e, \bm{x}, t, \bm{u}\right),
\end{equation}
where the equilibrium distribution $g^e$ is obtained from the statuses of both sides:
\begin{equation}
\int g^e \bm{\psi} d\Xi =  \bm{W}^e = \int_{u\geq 0} g_0^l \bm{\psi} d\Xi  + \int_{u < 0} g_0^r \bm{\psi} d\Xi.
\end{equation}

Substitute the expressions of $f_0$ and $g$ into the solution \eqref{solution}, and neglect some unimportant terms \citep{Zhou2017},  the final form of the distribution function reads:
\begin{equation} \label{final}
\begin{aligned}
 f(0,y,t,\bm{u},\xi)  =& g^e + \frac{1}{2} g^e_{yy} y^2 + g^e_t t + \frac{1}{2} g^e_{tt} t^2 - \tau \left[ \left( g^e_t + u g^e_x + v g^e_y \right) + \left( g^e_{tt} + u g^e_{xt} + v g^e_{yt} \right) t \right] \\
& - e^{-t/\tau_n} \left[ g^e  -  \left( u g^e_x + v g^e_y \right) t  \right]  + e^{-t/\tau_n} \left\{ \begin{aligned}
g^l & -  \left( u g^l_x + v g^l_y \right) t ,\quad u \geq 0\\
 g^r & -  \left( u g^r_x + v g^r_y \right) t , \quad u<0
\end{aligned} \right\}.
\end{aligned}
\end{equation}

The collision time is determined by
\begin{equation}
\tau = \frac{\mu}{p^e}, \quad \tau_n = \tau + \alpha \Delta t e^{1 - \eta^{-10}}, \quad \eta = \left| \frac{p^l - p^r}{p^l + p^r} \right|,
\end{equation}
where $\mu$ is the dynamic viscosity and $p^e$ is the pressure corresponding to $g^e$. $\tau_n$ is the numerical collision time which contains artificial dissipation \citep{Luo2013a}. Note that an adaptive function $e^{1- \eta^{-10}}$ is designed for the numerical collision time. This function ensures that $\tau_n$ differs from $\tau$ only when the normalized pressure difference $\eta$ is large enough. By doing this we aim to provide a necessary but minimum artificial dissipation.  $\alpha$ is a constant and is taken to be $0.3$ for all computations in this paper.

Once the distribution function $f$ is obtained, the flux at a vertically placed cell interface can be expressed as
\begin{equation}
\bm{F} = \int uf \bm{\psi} d\Xi.
\end{equation}

For a rectangular cell $[x_{i-1/2},x_{i+1/2}] \times [y_{j-1/2}, y_{j+1/2}]$ with dimensions $\Delta x_i = x_{i+1/2} - x_{i-1/2}$ and $\Delta y_j = y_{j+1/2} - y_{j-1/2}$, the cell-averaged conservative variable $\bm{W}_{ij}$ is updated from the time $t_n$ to $t_{n+1}$ as follows:
\begin{equation} \label{update}
\begin{aligned}
\bm{W}_{ij}^{n+1} = \bm{W}_{ij}^{n} & - \frac{1}{\Delta x_i \Delta y_j} \int_{t_n}^{t_{n+1}} \int_{-\frac{1}{2} \Delta y_j}^{\frac{1}{2} \Delta y_j} \left[ \bm{F}_{i+1/2	}(t,y)  - \bm{F}_{i-1/2} (t,y) \right] dy dt  \\
&- \frac{1}{\Delta x_i \Delta y_j} \int_{t_n}^{t_{n+1}} \int_{-\frac{1}{2} \Delta x_i}^{\frac{1}{2} \Delta x_i} \left[ \bm{F}_{j+1/2}(t,x) - \bm{F}_{j-1/2} (t,x) \right] dx dt. \\
\end{aligned}
\end{equation}
Since $\bm{F}$ is an explicit function of $t$ and $x,y$, the integrations in \eqref{update} can be easily obtained. 

Finally, we give the coefficients for representing the derivatives of $g$ in \eqref{final}:
\begin{equation}
\begin{aligned}
& a_{x} = g_x/g, \quad a_{y} = g_y/g, \quad a_{t} = g_t/g,  \\
& a_{xx} = g_{xx}/g, \quad a_{yy} = g_{yy}/g,  \quad a_{xy} = g_{xy}/g,  \\
& a_{xt} = g_{xt}/g, \quad a_{yt} = g_{yt}/g, \quad a_{tt} = g_{tt}/g.
\end{aligned}
\end{equation}
Each coefficient can be written as $\Lambda = \Lambda_1 \psi_1 +  \Lambda_2 \psi_2 + \Lambda_3\psi_3 + \Lambda_4\psi_4$. Define the moment of a variable as:
\begin{equation}
\left\langle \cdots \right\rangle = \int g(\cdots) \bm{\psi} d\Xi,
\end{equation}
then the coefficients are derived  as follows:
\begin{equation}
\begin{aligned}
& \left\langle a_{x} \right\rangle = \bm{W}_x \rightarrow a_{x}, \quad \left\langle a_y \right\rangle = \bm{W}_y \rightarrow a_y, \quad \left\langle a_x u + a_y v + a_t \right\rangle = \bm{0} \rightarrow a_t, \\
& \left\langle a_{xx} \right\rangle = \bm{W}_{xx} \rightarrow a_{xx}, \quad  \left\langle a_{yy} \right\rangle = \bm{W}_{yy} \rightarrow a_{yy}, \quad \left\langle a_{xy} \right\rangle = \bm{W}_{xy} \rightarrow a_{xy}, \\
& \left\langle a_{xx} u +  a_{xy} v + a_{xt} \right\rangle = \bm{0} \rightarrow a_{xt}, \quad \left\langle a_{xy} u +  a_{yy} v + a_{yt} \right\rangle = \bm{0} \rightarrow a_{yt}, \\
& \left\langle a_{xt} u +  a_{yt} v + a_{tt} \right\rangle = \bm{0} \rightarrow a_{tt}.
\end{aligned}
\end{equation}
All moments can be obtained explicitly. See \citet{Xu2001} for details.

To provide the initial values for the evolution process, the macroscopic variables and their derivatives need to be constructed before each computational step.
In the perpendicular direction of the cell interface, a standard 5th-order WENO-JS method  \citep{Jiang1996} is used to determine the value of the variables on both sides of the interface. Following the suggestion in \citet{Shu1997}, the characteristic variables are used instead of conservative variables. For a scalar variable $Q$, assume $\bar{Q}_i$ is the averaged value in the $i$-th cell, $Q_i^l$ and $Q_i^r$ are the values to be reconstructed at the left and right boundaries of the $i$-th cell, then the process is as below:
\begin{equation}
Q^r_i = \sum_{s=0}^2 w_s q_s, \quad Q^l_i = \sum_{s=0}^2 \tilde{w}_s \tilde{q}_s,
\end{equation}
where
\begin{equation}
\begin{aligned}
& q_0 = \frac{1}{3} \bar{Q}_i +  \frac{5}{6} \bar{Q}_{i+1} -  \frac{1}{6} \bar{Q}_{i+2},\quad \tilde{q}_0 = \frac{11}{6} \bar{Q}_i -  \frac{7}{6} \bar{Q}_{i+1} + \frac{1}{3} \bar{Q}_{i+2}, \\
& q_1 = -\frac{1}{6} \bar{Q}_{i-1} +  \frac{5}{6} \bar{Q}_{i} +  \frac{1}{3} \bar{Q}_{i+1}, \quad \tilde{q}_1 = \frac{1}{3} \bar{Q}_{i-1} +  \frac{5}{6} \bar{Q}_{i} -  \frac{1}{6} \bar{Q}_{i+1}, \\
& q_2 = \frac{1}{3} \bar{Q}_{i-2} -  \frac{7}{6} \bar{Q}_{i-1} +  \frac{11}{6} \bar{Q}_{i}, \quad \tilde{q}_2 = -\frac{1}{6} \bar{Q}_{i-2} +  \frac{5}{6} \bar{Q}_{i-1} +  \frac{1}{3} \bar{Q}_{i},  \\
& w_s = \frac{\alpha_s}{\sum_{p=0}^2 \alpha_p}, \quad \tilde{w}_s = \frac{\tilde{\alpha}_s}{\sum_{p=0}^2 \tilde{\alpha}_p}, \quad  s = 0,1,2,\\
& \alpha_s = \frac{d_s}{\left( \epsilon + \beta_s \right)^2}, \quad  \tilde{\alpha}_s = \frac{\tilde{d}_s}{\left( \epsilon + \beta_s \right)^2}, \quad  s = 0,1,2, \\
& \beta_0 = \frac{13}{12}\left( \bar{Q}_{i} - 2\bar{Q}_{i+1} + \bar{Q}_{i+2} \right)^2 + \frac{1}{4}\left( 3\bar{Q}_{i} - 4\bar{Q}_{i+1} + \bar{Q}_{i+2} \right)^2, \\
& \beta_1 = \frac{13}{12}\left( \bar{Q}_{i-1} - 2\bar{Q}_{i} + \bar{Q}_{i+1} \right)^2 + \frac{1}{4}\left( \bar{Q}_{i-1} - \bar{Q}_{i+1} \right)^2, \\
& \beta_2 = \frac{13}{12}\left( \bar{Q}_{i-2} - 2\bar{Q}_{i-1} + \bar{Q}_{i} \right)^2 + \frac{1}{4}\left( \bar{Q}_{i-2} - 4\bar{Q}_{i-1} + 3\bar{Q}_{i} \right)^2, \\
& d_0 = \tilde{d}_2 = \frac{3}{10}, \quad d_1 = \tilde{d}_1 = \frac{6}{10}, \quad d_2 = \tilde{d}_0 = \frac{1}{10}.
\end{aligned}
\end{equation}

We set $\epsilon = 10^{-6}$ in our computations. The results of the one-dimensional WENO scheme are line-averaged values. A third-order interpolation is then used to obtain the value at the midpoint of the interface. After that, the first- and second-order derivatives in both  $x$ and $y$ directions can be calculated from the reconstructed variables.

\section{Description of the Viscous Shock Tube Problem} \label{section problem}
The viscous shock tube problem was proposed by \citet{Daru2001}. A diaphragm is vertically located in the middle of a square 2-D shock tube with unit side length, separating the space into the left and right parts. The initial state in non-dimensional form is given by 
\begin{equation}
\left( \rho, u, v, p \right) = \left\{ \begin{aligned}
& (120,0,0,120/\gamma), \quad x \leq 0.5, \\
& (1.2,0,0,1.2/\gamma), \quad x > 0.5,
\end{aligned} \right.
\end{equation} 
where $\gamma = 1.4$ is the specific heat ratio of air.  The Prandtl number is taken to be $\Pran = 0.73$. No-slip adiabatic conditions are applied at all boundaries of the tube.

\begin{figure}
\centering
\begin{minipage}[t]{0.33\textwidth}
\centering
\includegraphics[width=\textwidth]{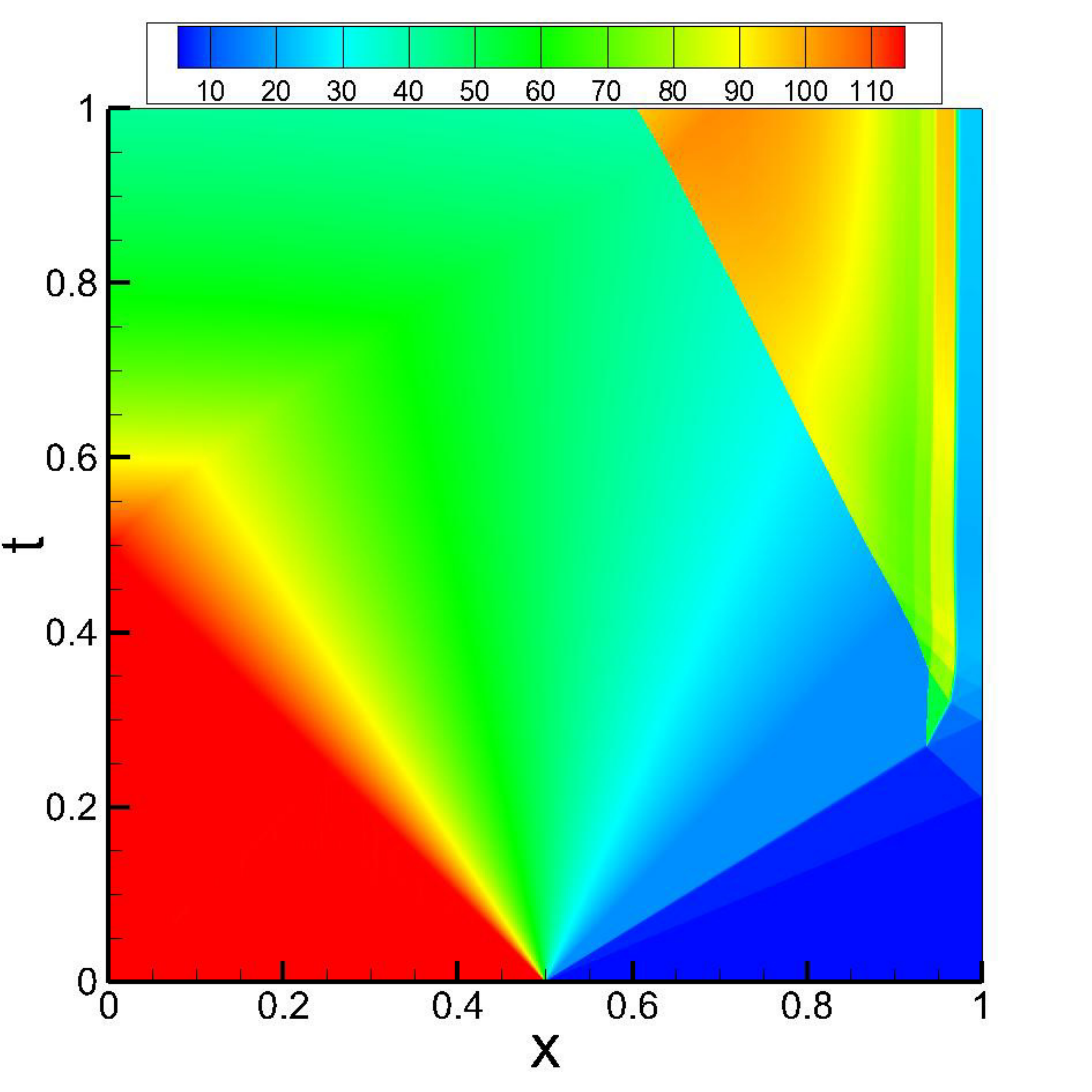}
\centerline{\footnotesize (a)}
\end{minipage}%
\begin{minipage}[t]{0.33\textwidth}
\centering
\includegraphics[width=\textwidth]{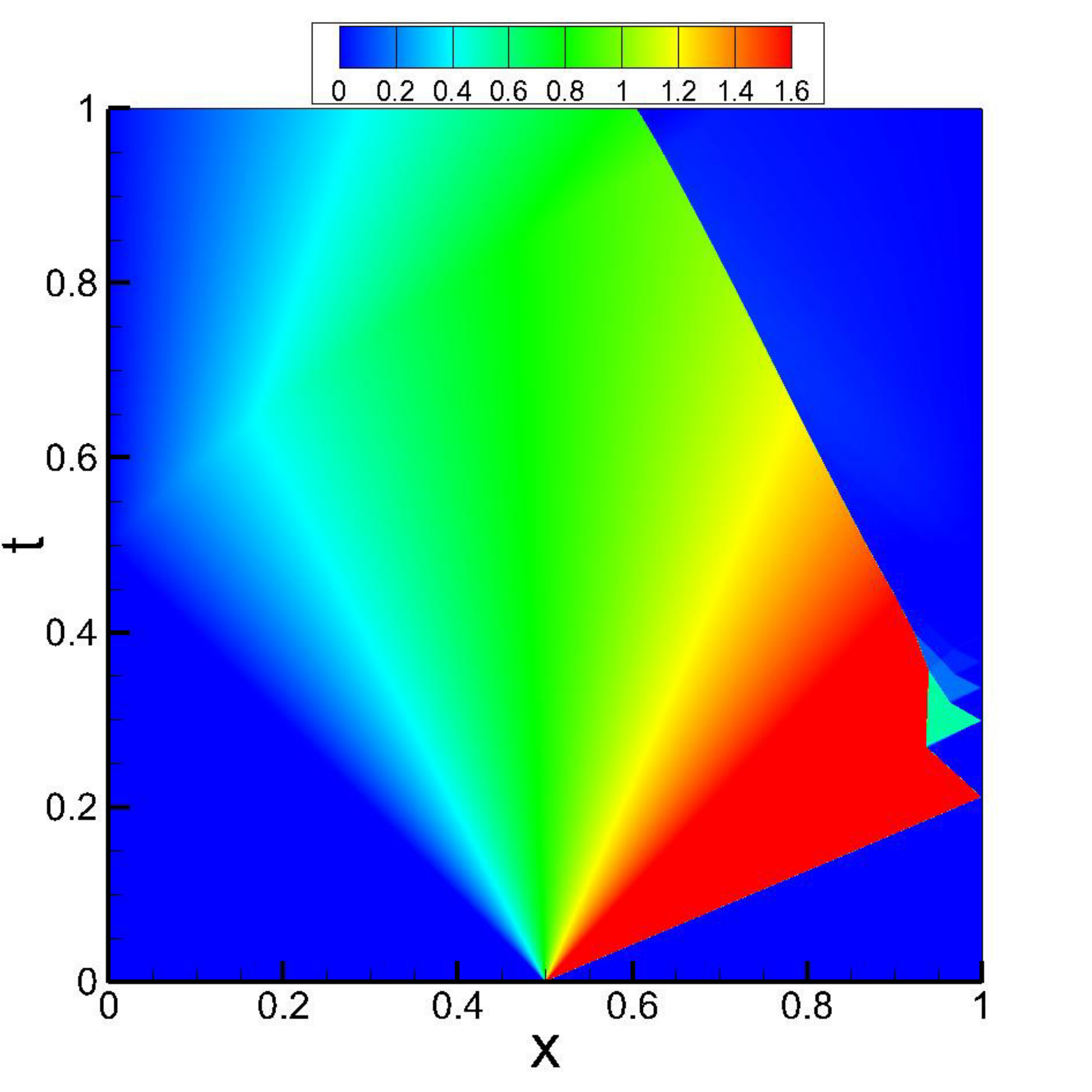}
\centerline{\footnotesize (b)}
\end{minipage}%
\begin{minipage}[t]{0.33\textwidth}
\centering
\includegraphics[width=\textwidth]{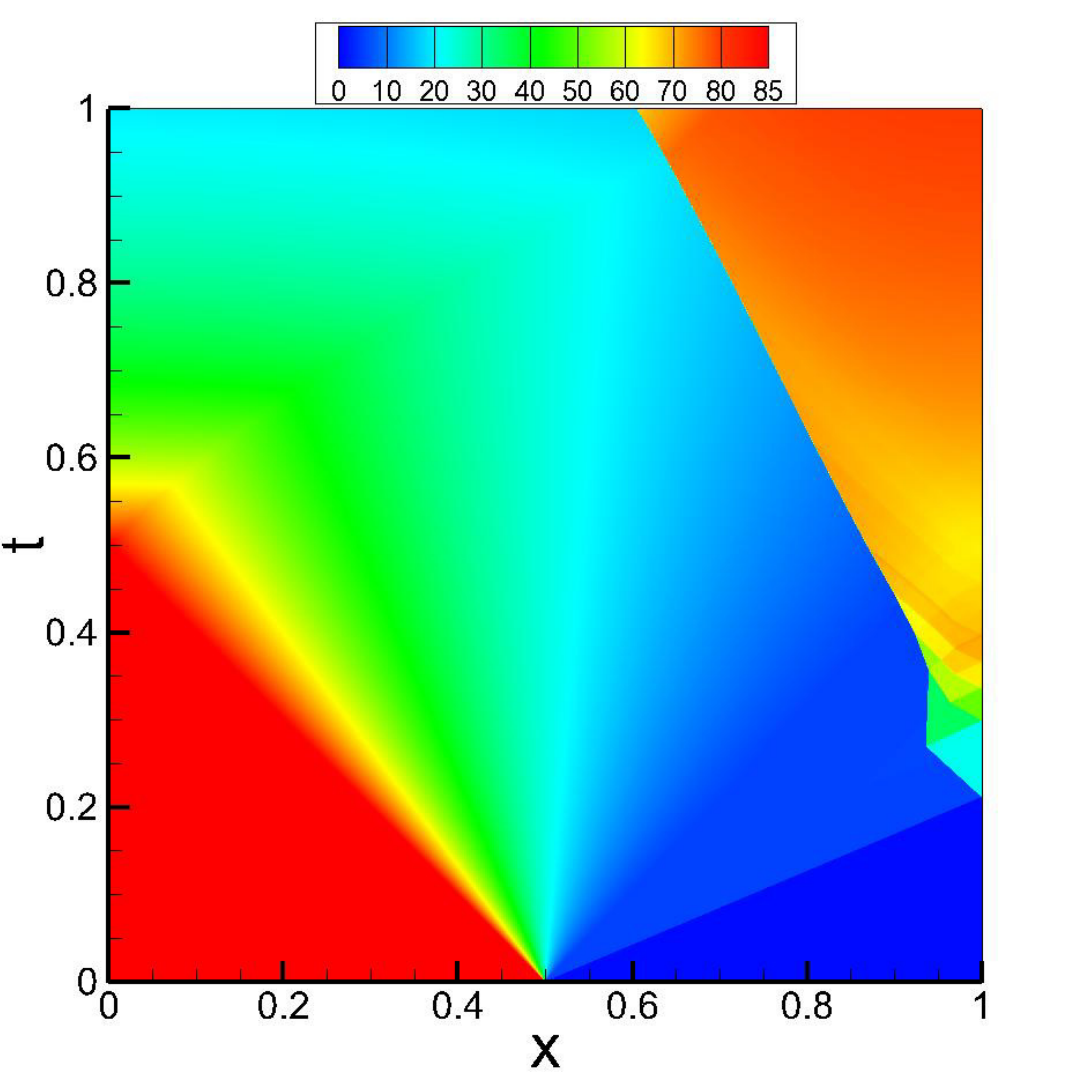}
\centerline{\footnotesize (c)}
\end{minipage}
\centering
\caption{$x-t$ diagrams of (\textit{a}) density, (\textit{b}) velocity and (\textit{c}) pressure for the inviscid case.}\label{xt}
\end{figure}

The diaphragm is broken instantly at $t = 0$. A shock wave with the Mach number $M\!a=2.37$ forms and moves towards the right, followed by a contact discontinuity. Simultaneously, a rarefaction wave expands in both directions. Figure~\ref{xt} shows the evolution of density, velocity and pressure from $t=0$ to $t=1$ in the inviscid case (hence the flow is one-dimensional). It is seen from the figures that the incident shock reaches the right wall at about $t=0.21$. Then it is reflected back to the left, later interacting with the contact discontinuity. 

With presence of viscosity, the incident shock wave induces boundary layers along the horizontal walls of the tube. They will then interact with the incident and reflected shock, as well as other structures appearing later. In figure~\ref{xt}, we can observe a number of wave reflections and interactions in the region close to the right wall for the one-dimensional inviscid case. For the two-dimensional viscous case, the flow field will surely be more complicated.

Since the configuration is symmetric about the line $y=0.5$, only half of the tube $[0,1] \times [0,0.5]$ is computed. And we focus on the evolution of the flow field from $t=0$ to $t=1$ at both Reynolds numbers of 200 and 1000. The viscosity is assumed to be constant (so that $\mu=1/\Rey$). All grids used are uniform with $\Delta x = \Delta y$. The \textit{CFL} number is 1.0 for all computations.

\section{The $\Rey=200$ Case} \label{section 200}
The $\Rey = 200$ case has been simulated by many authors \citep{Daru2001, Sjogreen2003, Daru2004, Kim2005a, Kim2005b, Daru2009, Houim2011, Wan2012, Sun2014, Kotov2014, Tenaud2015, Wang2015, Pan2016a, Pan2016b}. At this relatively low Reynolds number, the results presented in different papers are quite consistent when the grid is fine enough. As reported in \citet{Daru2009}, the sufficient grid resolution is $1000 \times 500$ for the high-order scheme presented therein. Other computations \citep{Daru2001, Sjogreen2003} indicate the behaviour of high-order methods is obviously better than that of the second-order ones.

An important problem might be the lack of criteria for the judgement of convergence and for the comparison between results. \citet{Daru2001, Daru2009} used the plot of density distribution along the bottom wall to demonstrate convergence. This method was also adopted by some other authors \citep{Kim2005a, Kim2005b, Pan2016b}. Another commonly used criterion is to compare the height of the primary vortex \citep{Kim2005a, Kim2005b, Wang2015, Pan2016a, Pan2016b}. On the same uniform $500 \times 250$ grid, the reported vortex height varies from 0.163 to 0.171 by different schemes. However, it is found that the flow structures are not necessarily the same even when the vortex heights are very close. 

\begin{figure}
\centering
\begin{minipage}[t]{0.5\textwidth}
\flushright
\includegraphics[width=\textwidth]{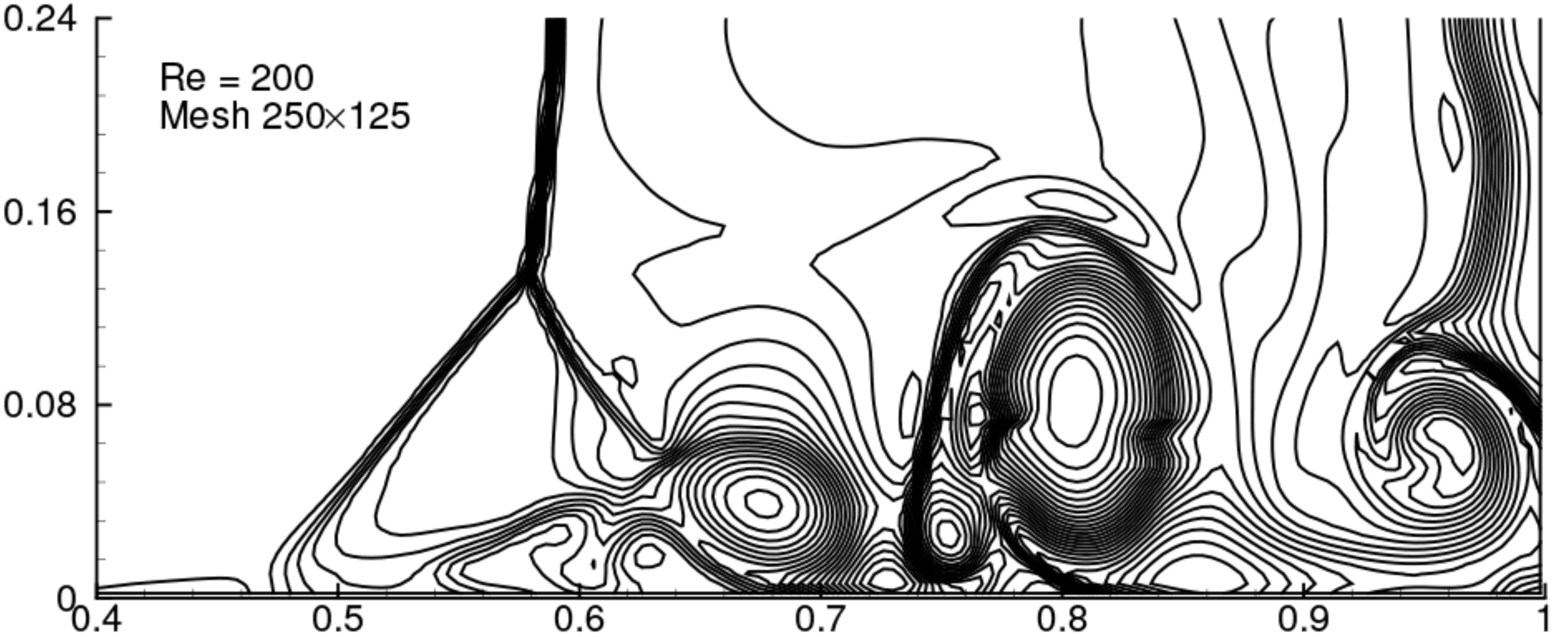}
\centerline{\footnotesize (a)}
\end{minipage}%
\begin{minipage}[t]{0.5\textwidth}
\centering
\includegraphics[width=\textwidth]{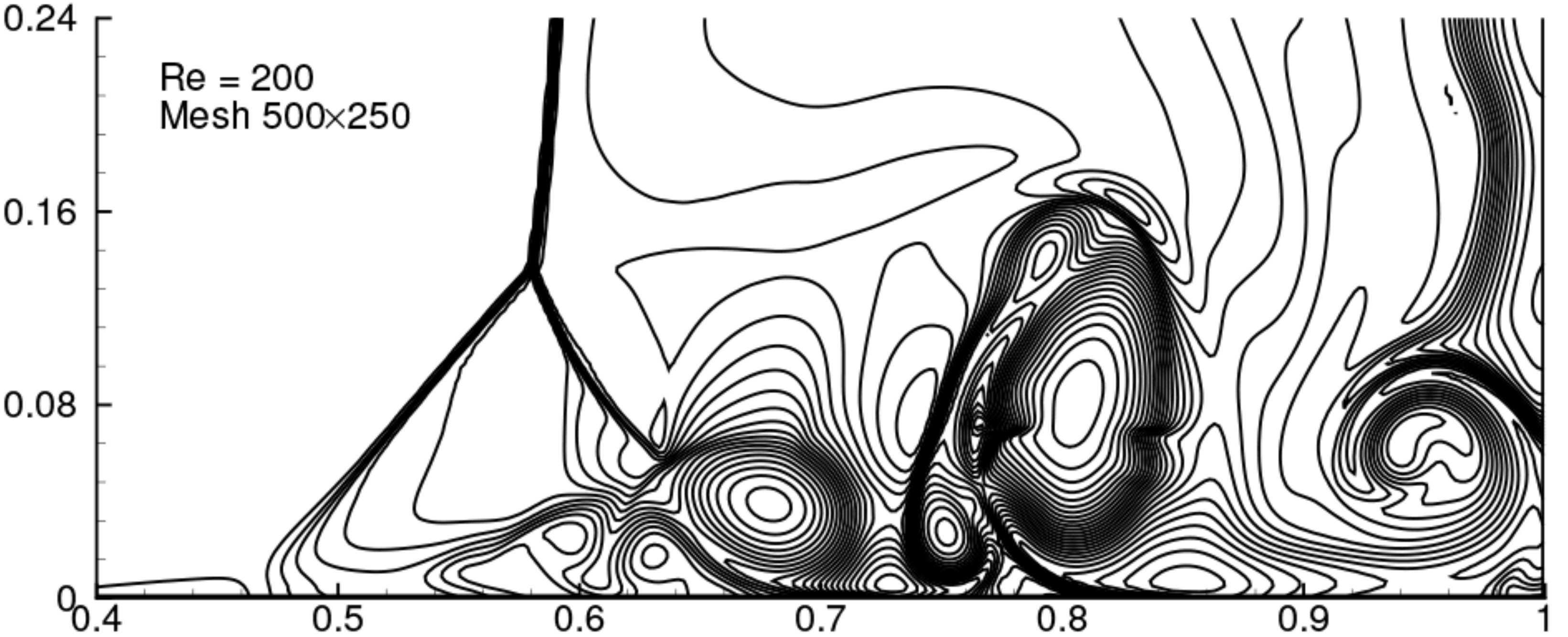}
\centerline{\footnotesize (b)}
\end{minipage}
\begin{minipage}[t]{0.5\textwidth}
\centering
\includegraphics[width=\textwidth]{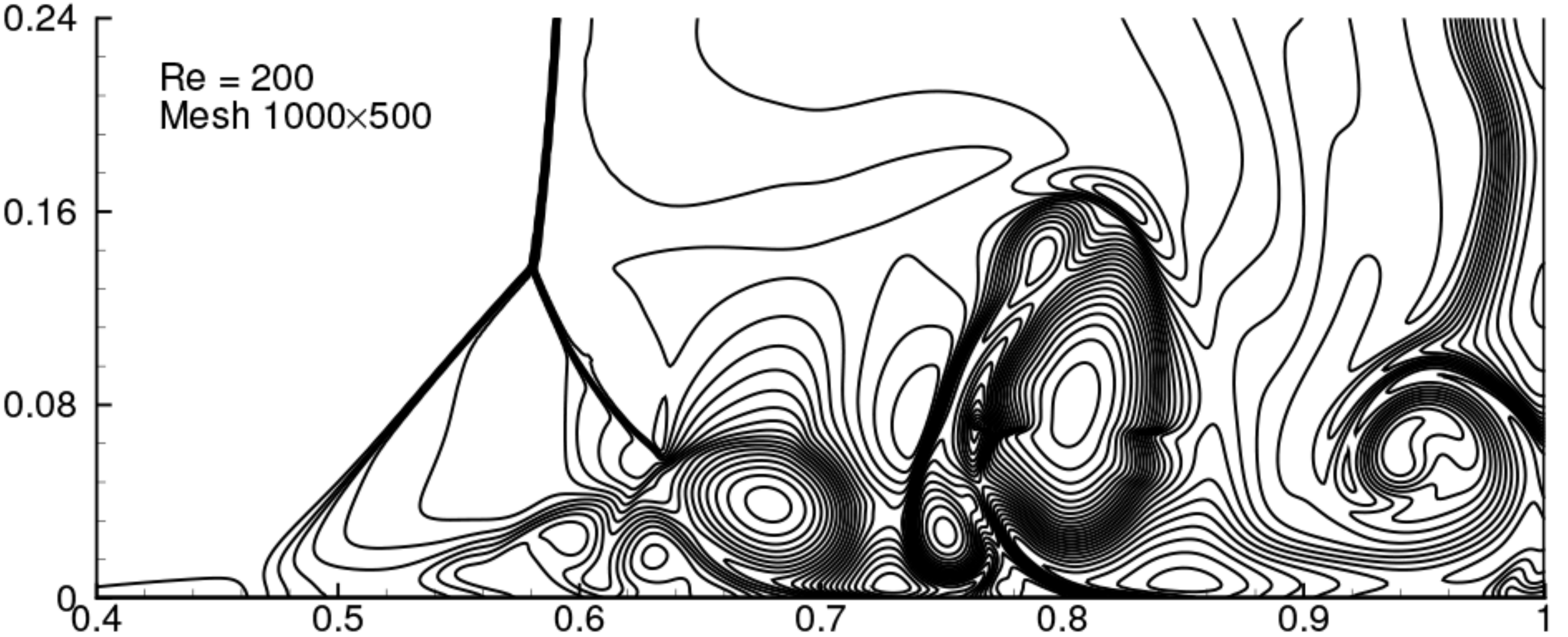}
\centerline{\footnotesize (c)}
\end{minipage}%
\begin{minipage}[t]{0.5\textwidth}
\centering
\includegraphics[width=\textwidth]{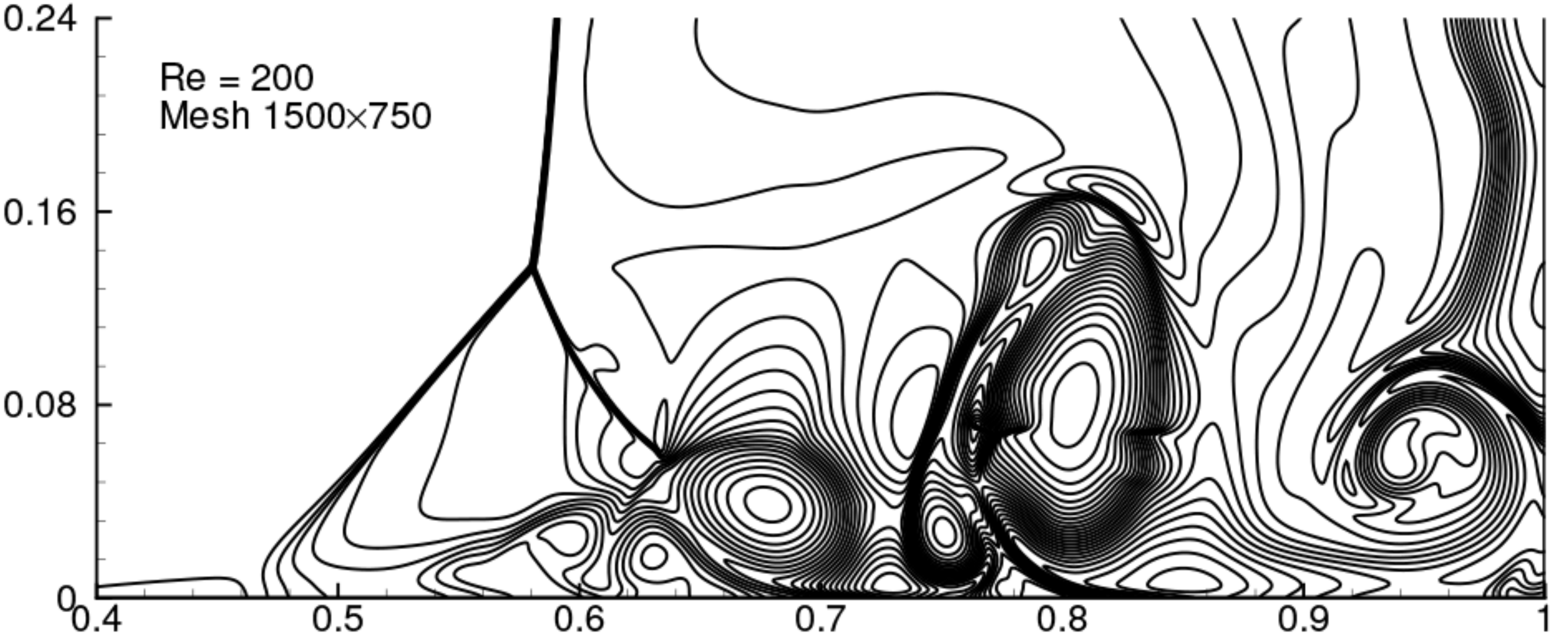}
\centerline{\footnotesize (d)}
\end{minipage}
\centering
\caption{Density distribution at $t=1$ for the $\Rey = 200$ case. $25$ contours are equally spaced from $22$ to $121$ with the grids of (\textit{a}) $250 \times 125$, (\textit{b}) $500 \times 250$, (\textit{c}) $1000 \times 500$ and (\textit{d}) $1500 \times 750$.}\label{vst_200_density}
\end{figure}

\begin{figure}
\centering
\begin{minipage}[t]{0.5\textwidth}
\flushright
\includegraphics[width=\textwidth]{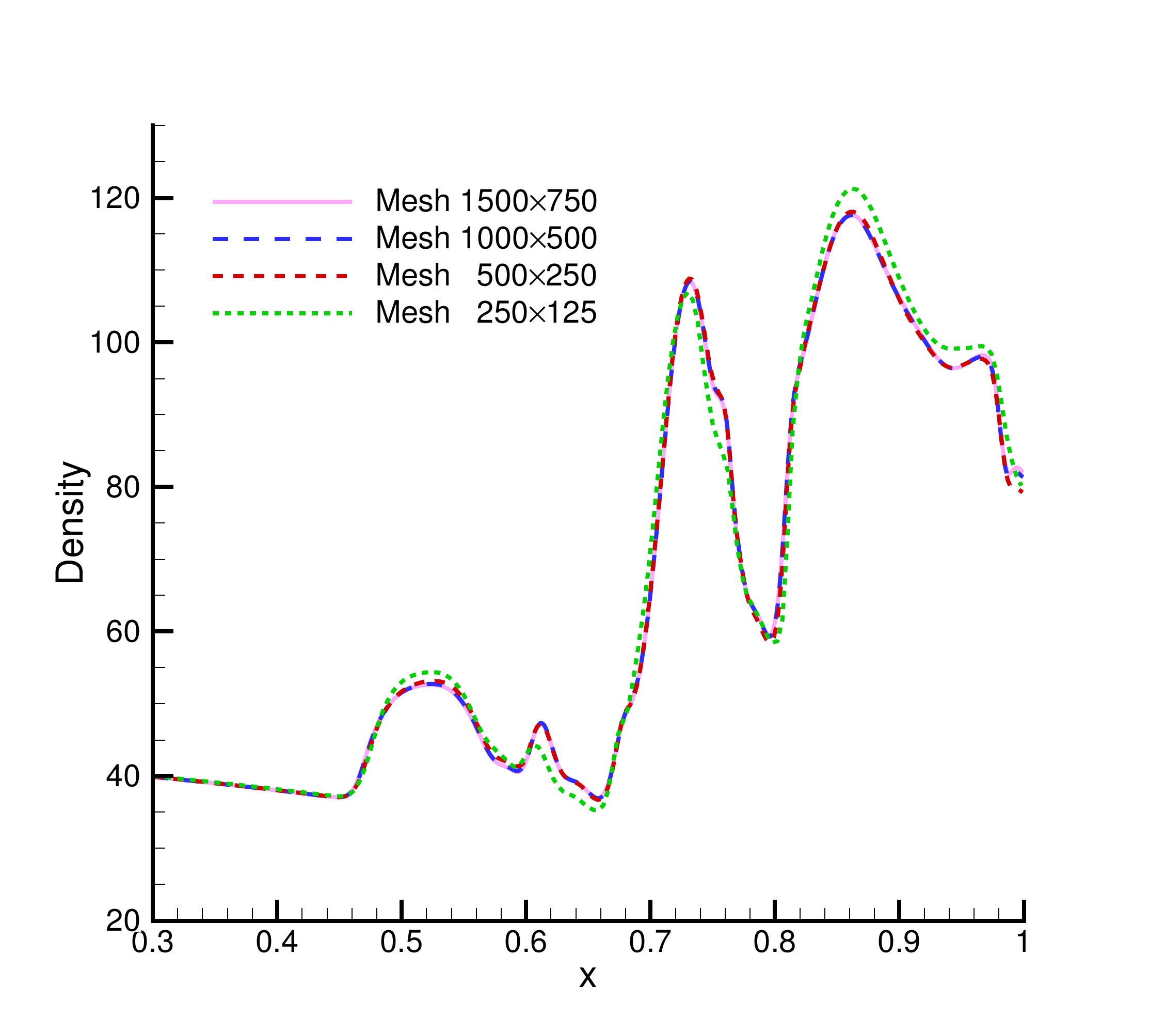}
\centerline{\footnotesize (a)}
\end{minipage}%
\begin{minipage}[t]{0.5\textwidth}
\centering
\includegraphics[width=\textwidth]{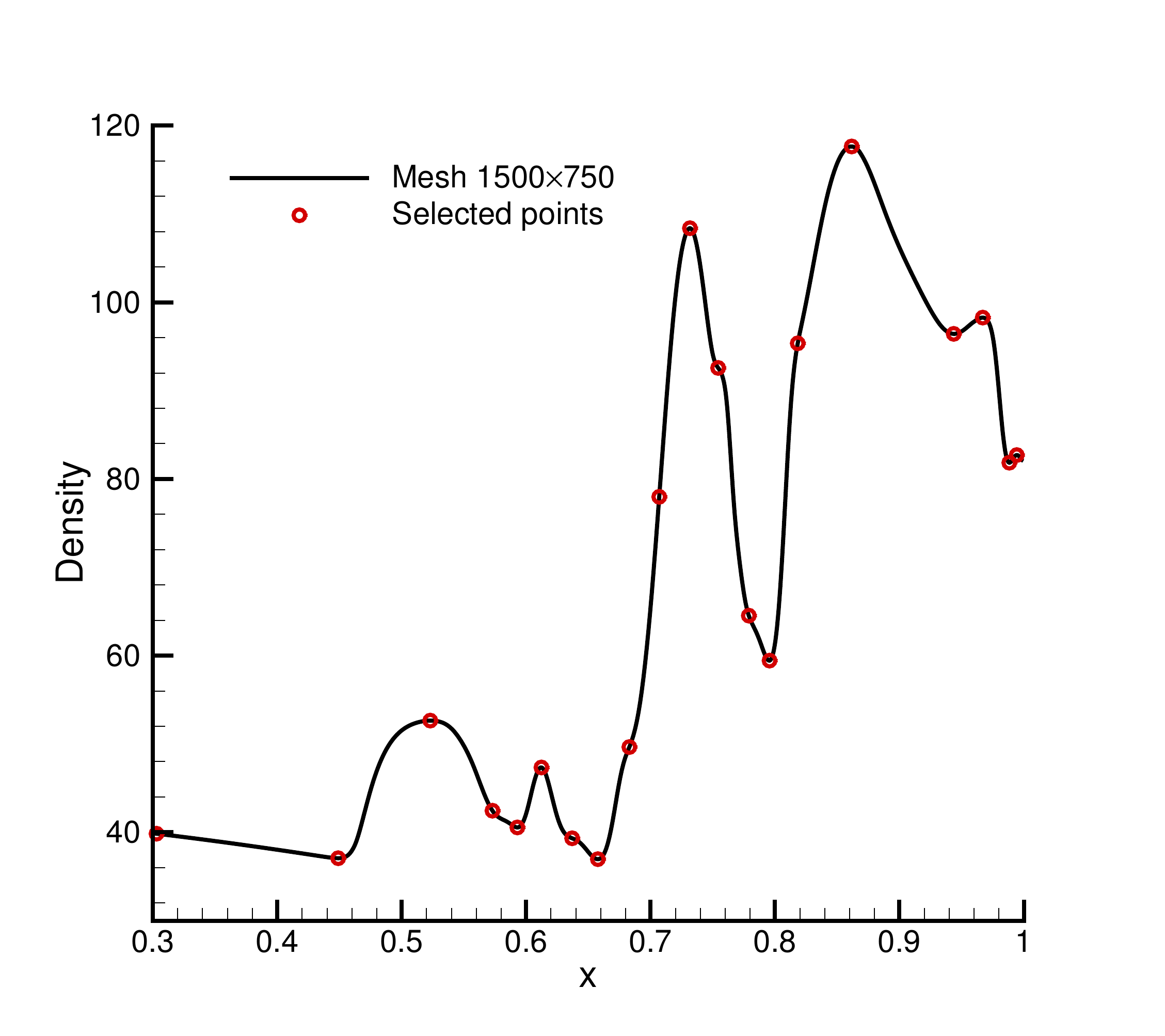}
\centerline{\footnotesize (b)}
\end{minipage}
\centering
\caption{Density distribution along the bottom wall at $t=1$ for the $\Rey = 200$ case. \protect\\(\textit{a}) Comparison of different grids; (\textit{b}) Positions of the selected points in table~\ref{vst_200_extract}.}\label{vst_200_bottom}
\end{figure}

The grid convergence for the present scheme is illustrated in figure~\ref{vst_200_density}, where the density contours at $t=1$ are presented. The results by the $500 \times 250$ grid and the $1000 \times 500$ grid are almost indistinguishable. Figure~\ref{vst_200_bottom}(a) shows the density distribution along the bottom wall. The curves from the $500 \times 250$ grid to the $1500 \times 750$ grid are nearly identical. Even with a coarser $250 \times 125$ grid, a very good result is obtained.

\begin{figure}
\centering
\includegraphics[width=0.6\textwidth]{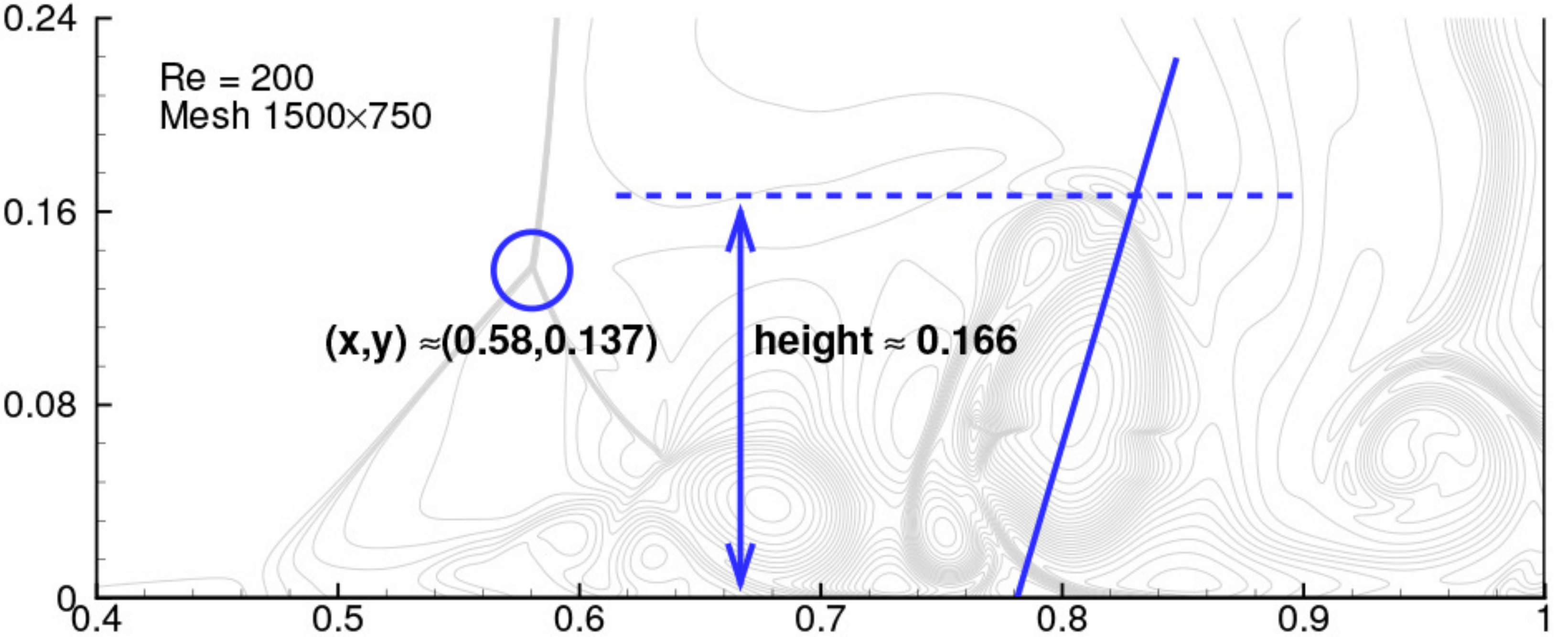}
\centering
\caption{Accuracy evaluation criteria for the $\Rey = 200$ case.}\label{vst_200_criteria}
\end{figure}

We think that the density distribution along the bottom wall is a good criterion for convergence study. Some critical points on the curve of the finest grid are extracted and listed in table~\ref{vst_200_extract} as a reference for comparison.  The positions of the selected points are given in figure~\ref{vst_200_bottom}(b). For macroscopic evaluations of the computed results, we recommend the following three criteria which are easily measured in the density contour plot, see figure~\ref{vst_200_criteria}:
\begin{enumerate}[(1)]
\item The position of the triple point, which is approximately $(x,y) = (0.58, 0.137)$.
\item The height of the primary vortex, which is approximately 0.166.
\item The orientation of the long axis of the primary vortex. This is an obvious criterion for qualitative evaluation.
\end{enumerate}

\begin{table}
  \begin{center}
\def~{\hphantom{0}}
  \begin{tabular}{cc|cc|cc|cc}
   $x$  &  $\rho$ & $x$ & $\rho$ & $x$ & $\rho$ & $x$ & $\rho$  \\
   0.3030 & 39.8418    & 0.6123   & 47.3367   & 0.7317 &  108.3916  & 0.8617   & 117.6452 \\
   0.4490 & 37.0662    & 0.6370   & 39.3203   & 0.7543 &  92.5760    & 0.9437  &  96.4287 \\
   0.5230 & 52.6465    & 0.6577   & 36.9558   & 0.7790 &  64.5319   &  0.9670   &  98.2689 \\
   0.5730 & 42.4400    & 0.6830   & 49.6513   & 0.7957 &  59.4386   &  0.9883   &  81.8465 \\
   0.5930 & 40.5506    & 0.7070   & 77.9810   & 0.8183 &   95.3607  &  0.9943   &  82.7077 \\
  \end{tabular}
  \caption{Extracted data of the density along the bottom wall. $\Rey = 200$.} 
  \label{vst_200_extract}
  \end{center}
\end{table}

\section{The $\Rey=1000$ Case: Numerical Simulation} \label{section 1000}

The above case at $\Rey = 200$ serves as verification for the present computational code. When the Reynolds number is increased to 1000, many fine flow structures appear hence the flow field becomes more complex. This case has also been simulated in several papers \citep{Daru2001, Sjogreen2003, Daru2004, Daru2009, Li2011, Wan2012, Kotov2014, Pan2016b}. The results from different papers or even from different methods in the same paper are very different. One reason is the sensitivity of the problem to the computational conditions, another reason is that the grids used in previous studies are not fine enough to achieve grid convergence due to practical limit on computational time. Grid-convergence studies were performed in \citet{Daru2001}, \citet{Sjogreen2003}, and \citet{Daru2009} with different numerical methods including classical TVD schemes and various high-order schemes. The most successful result is obtained by \citet{Daru2009}, where two high-order schemes (RK3-WENO5 and OSMP7) showed the same trend of convergence, and the results on the two finest grids ($3000 \times 1500$ and $4000 \times 2000$) are very similar. However, some small visible differences still exist on the two sets of grids, as noted in \citet{Daru2009}. Armed with the new accurate and efficient gas-kinetic scheme, we perform in this section a rigorous systematic grid-convergence study of the viscous shock tube problem at $\Rey = 1000$. 

\subsection{Numerical results}
\begin{figure}
\centering
\begin{minipage}[t]{0.5\textwidth}
\flushright
\includegraphics[width=\textwidth]{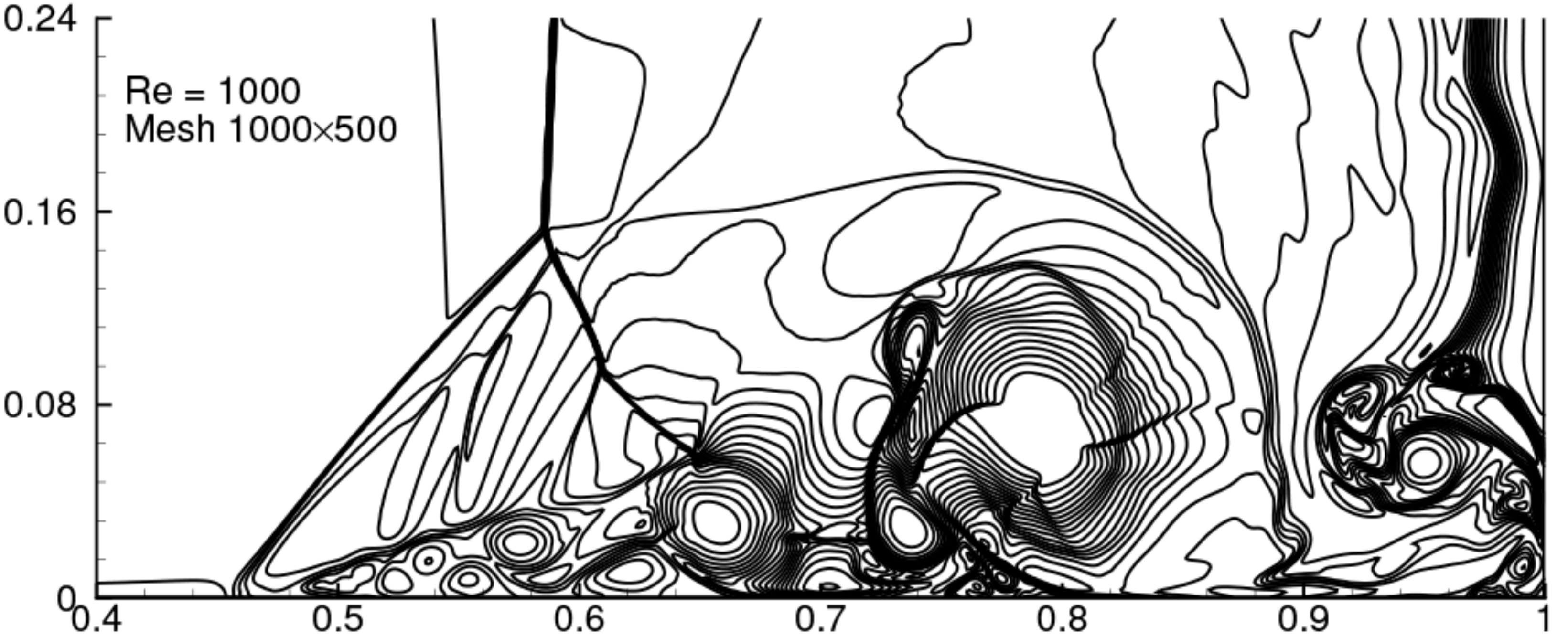}
\centerline{\footnotesize (a)}
\end{minipage}%
\begin{minipage}[t]{0.5\textwidth}
\centering
\includegraphics[width=\textwidth]{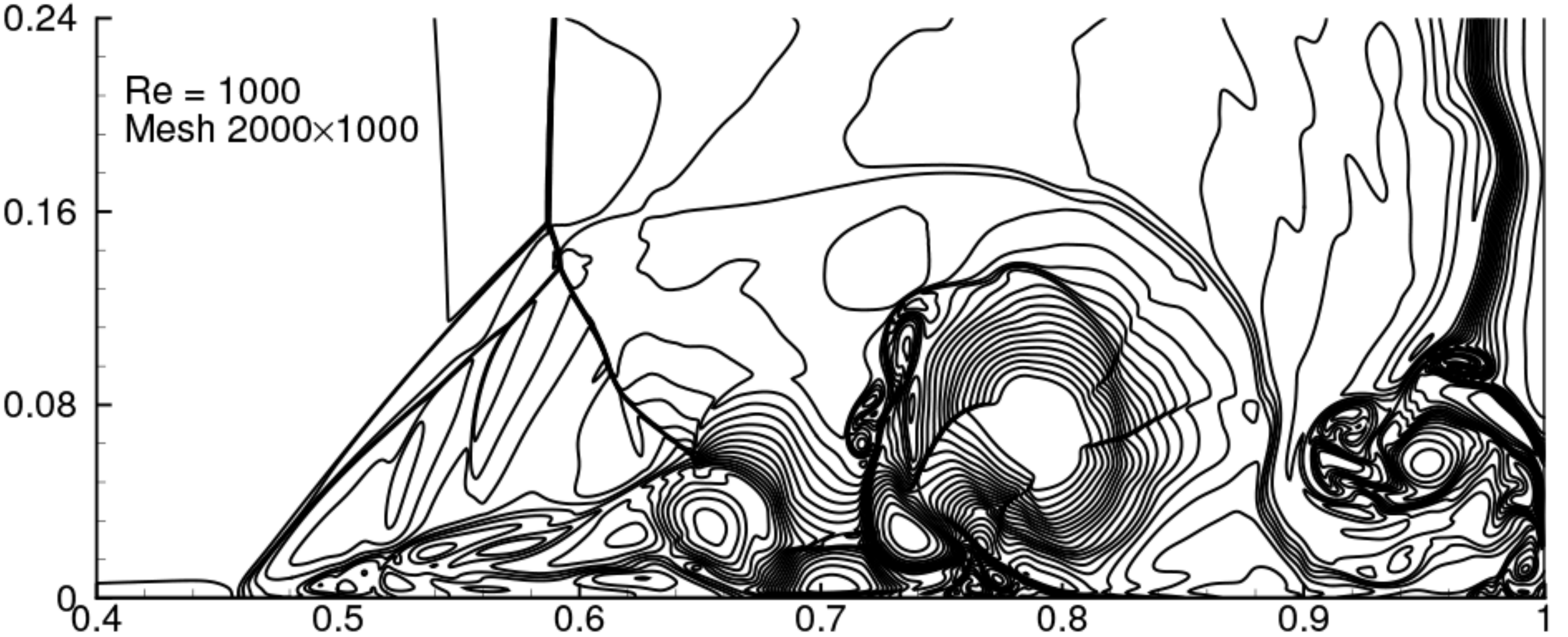}
\centerline{\footnotesize (b)}
\end{minipage}
\begin{minipage}[t]{0.5\textwidth}
\centering
\includegraphics[width=\textwidth]{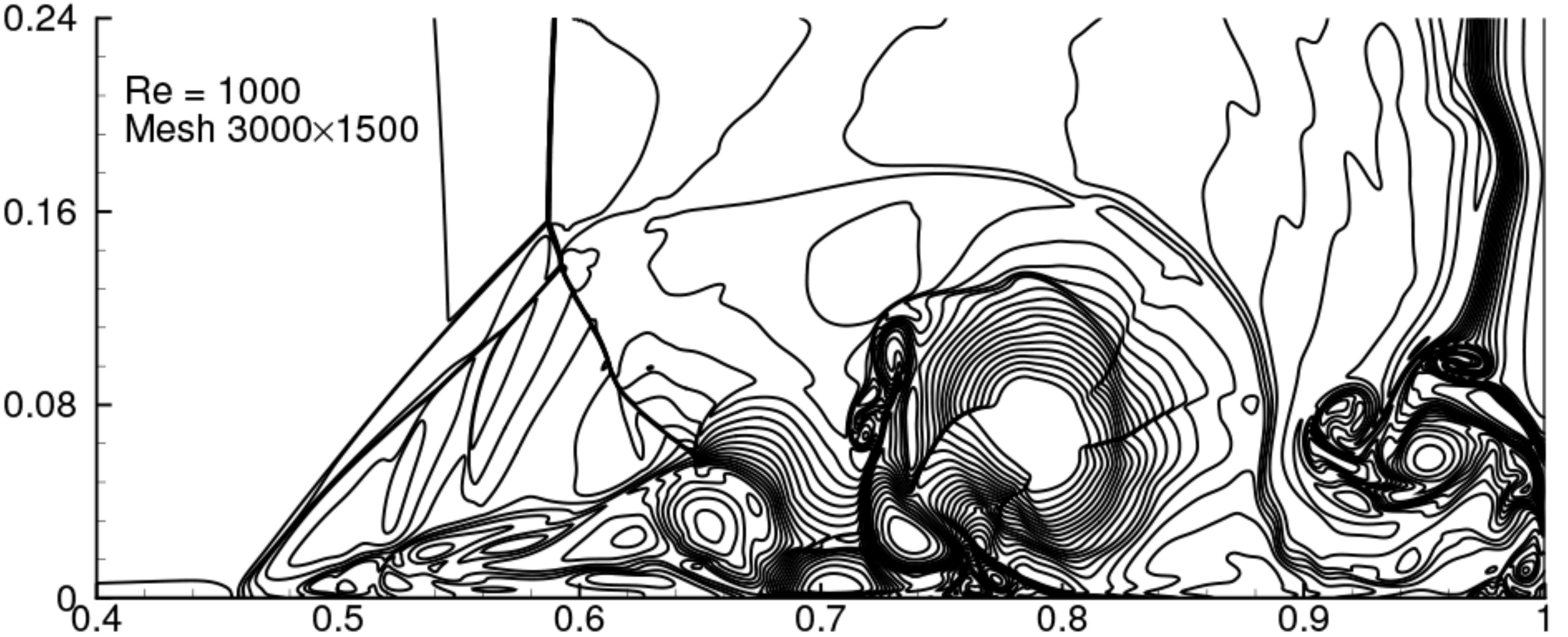}
\centerline{\footnotesize (c)}
\end{minipage}%
\begin{minipage}[t]{0.5\textwidth}
\centering
\includegraphics[width=\textwidth]{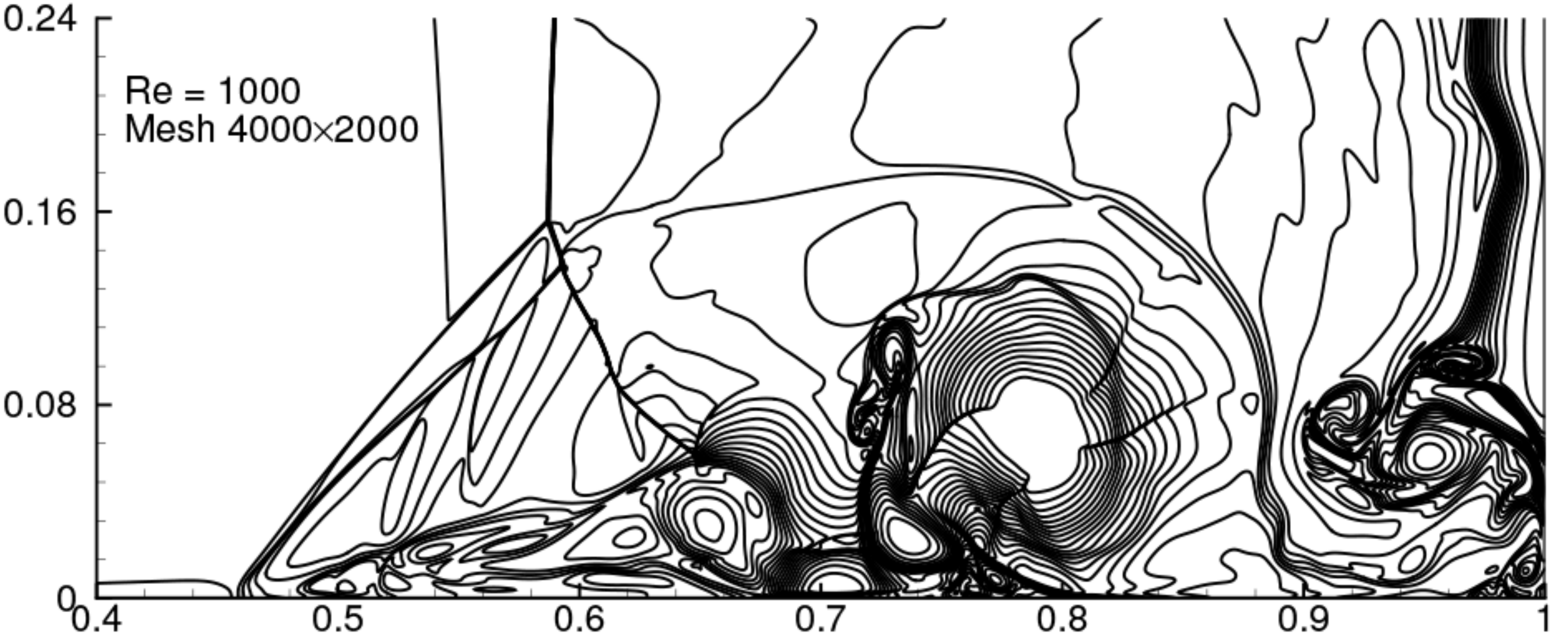}
\centerline{\footnotesize (d)}
\end{minipage}
\begin{minipage}[t]{0.5\textwidth}
\centering
\includegraphics[width=\textwidth]{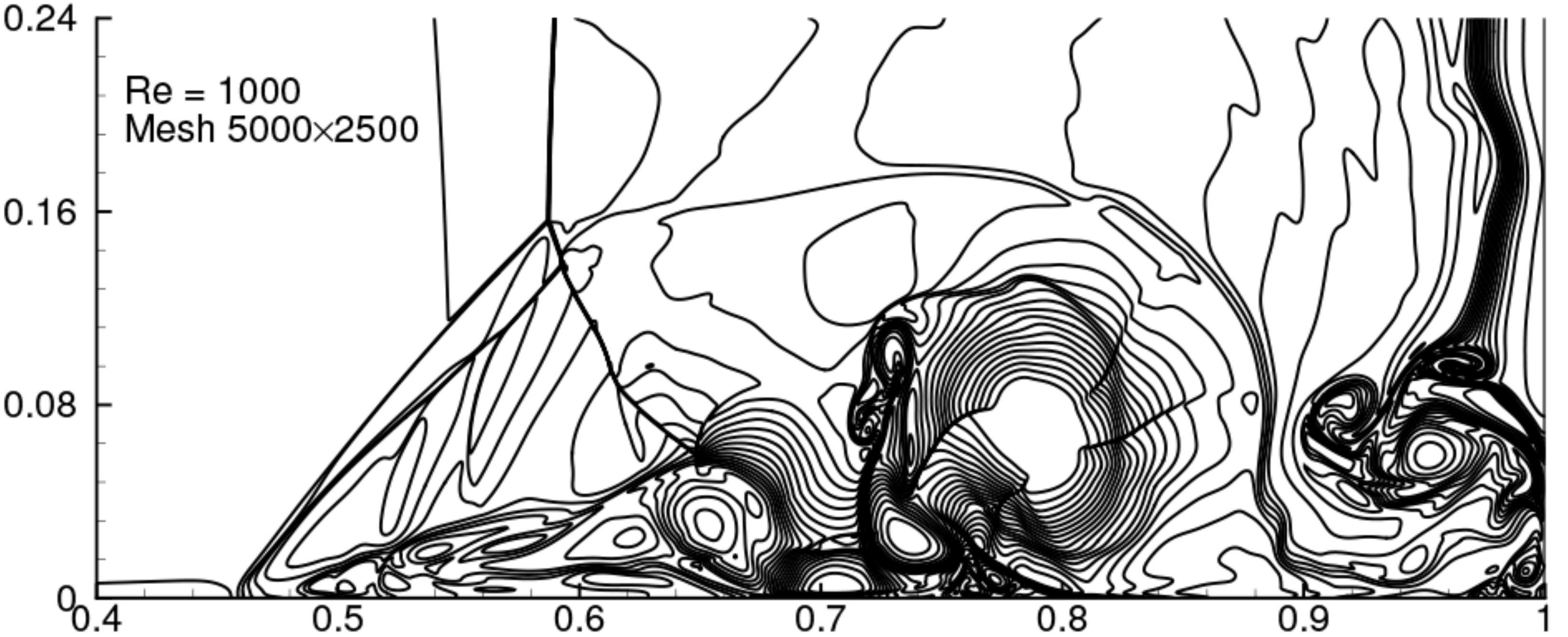}
\centerline{\footnotesize (e)}
\end{minipage}
\begin{minipage}[t]{\textwidth}
\centering
\caption{Density distribution at $t=1$ for the $\Rey = 1000$ case. 20 contours are equally spaced from 20 to 115 with the grids of (\textit{a}) $1000 \times 500$, (\textit{b}) $2000 \times 1000$, (\textit{c}) $3000 \times 1500$, (\textit{d}) $4000 \times 2000$ and (\textit{e}) $5000 \times 2500$.}\label{vst_1000_density}
\end{minipage}
\end{figure}

Five successively refined grids are used for investigation, which are $1000 \times 500$, $2000 \times 1000$, $3000 \times 1500$, $4000 \times 2000$, and $5000 \times 2500$, respectively. Figure~\ref{vst_1000_density} shows the density distribution at $t=1$ on different grids. It is clear that a converged solution in terms of the density field is obtained on the $3000 \times 1500$ grid. And the main features of the vortex structures are able to be predicted on the $2000 \times 1000$ grid. 

The converged computational density distribution agrees well with the result on the finest $4000 \times 2000$ grid in \citet{Daru2009}, providing evidence that the results obtained by \citet{Daru2009} and by our current scheme are both accurate and reliable, thus can be regarded as a reference solution.

\begin{figure}
\centering
\begin{minipage}[t]{0.5\textwidth}
\flushright
\includegraphics[width=\textwidth]{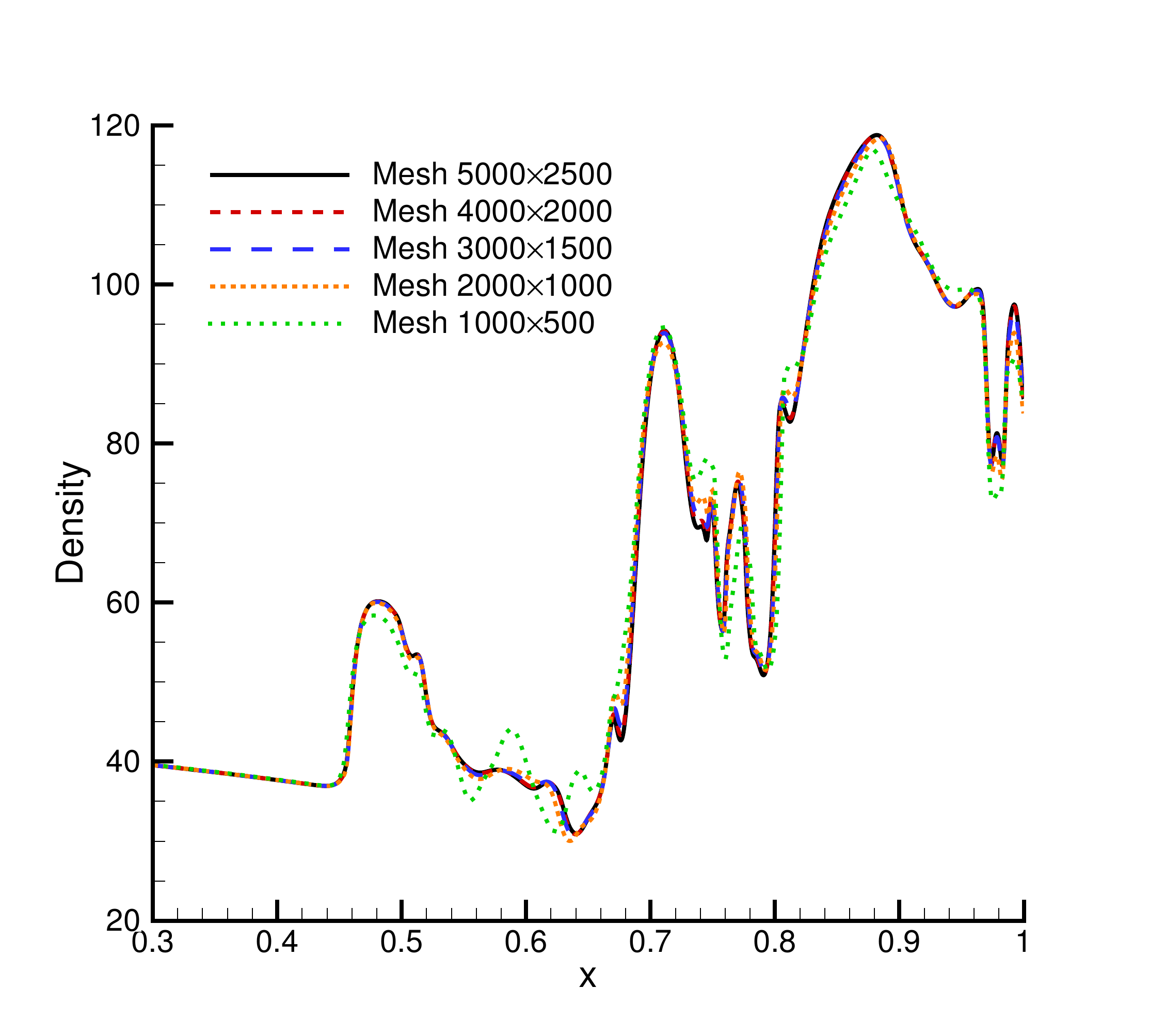}
\centerline{\footnotesize (a)}
\end{minipage}%
\begin{minipage}[t]{0.5\textwidth}
\centering
\includegraphics[width=\textwidth]{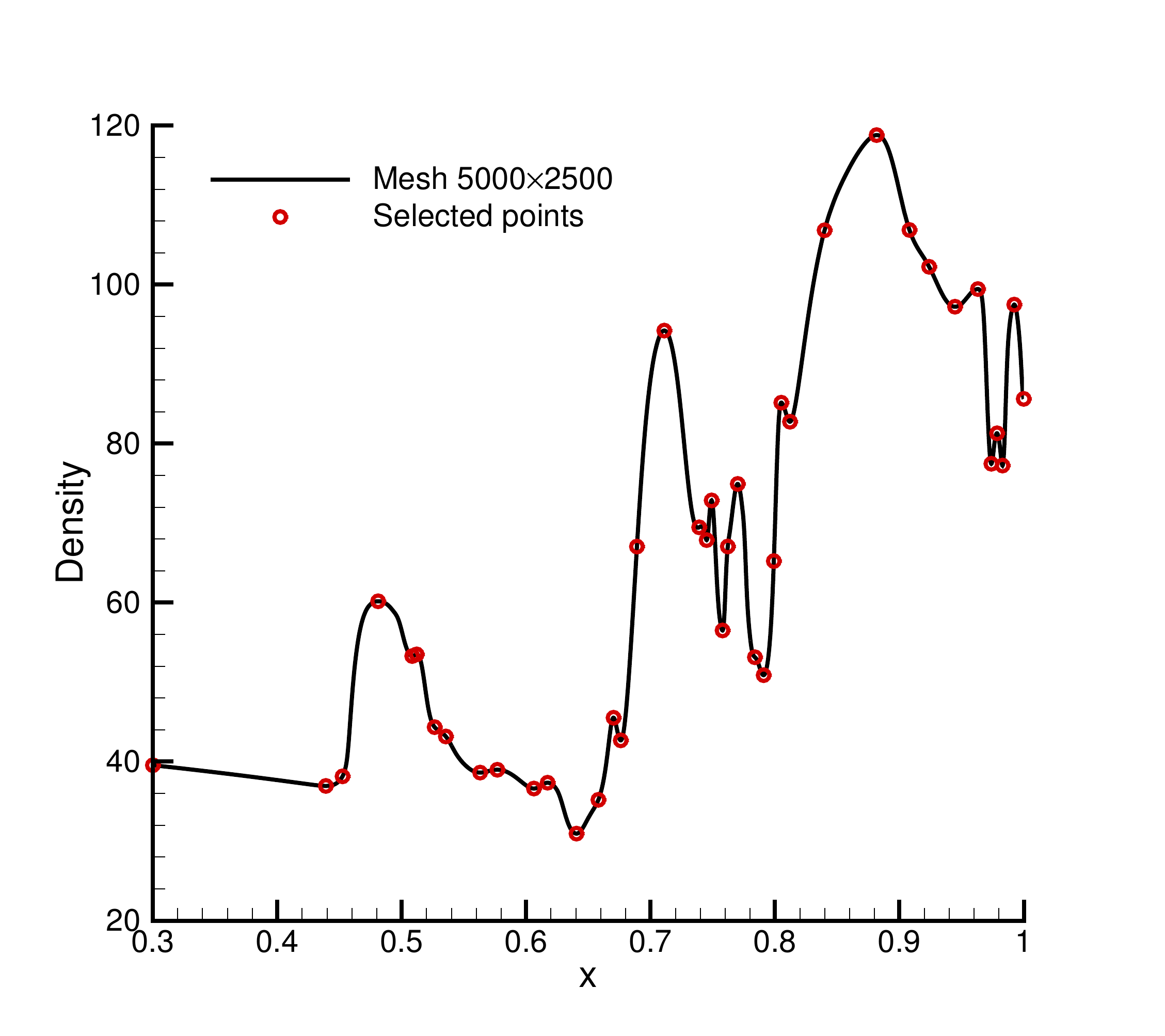}
\centerline{\footnotesize (b)}
\end{minipage}
\begin{minipage}[t]{\textwidth}
\centering
\caption{Density distribution along the bottom wall at $t=1$ for the $\Rey = 1000$ case. \protect\\ (\textit{a}) Comparison between different grids; (\textit{b}) Positions of the selected points in table~\ref{vst_1000_extract}.}\label{vst_1000_bottom}
\end{minipage}
\end{figure}

To perform a quantitative comparison, the density distribution along the bottom wall is shown in figure~\ref{vst_1000_bottom}(a). The difference between the curves on the $2000 \times 1000$ and $3000 \times 1500$ grids is already very small. As in the $\Rey = 200$ case, the critical points of the density distribution obtained on the finest grid are extracted and listed in table~\ref{vst_1000_extract} as a reference. The positions of the selected points are shown in figure~\ref{vst_1000_bottom}(b).

\begin{table}
  \begin{center}
\def~{\hphantom{0}}
    \begin{tabular}{cc|cc|cc|cc}
   $x$  &  $\rho$ & $x$ & $\rho$ & $x$ & $\rho$ & $x$ & $\rho$  \\
   0.3001 & 39.5483 & 0.6063 & 36.6144 & 0.7491 & 72.8602  & 0.8817 & 118.8170     \\
   0.4391 & 36.9422 & 0.6173 & 37.3454 & 0.7579 & 56.5015  & 0.9081 & 106.8818     \\
   0.4525 & 38.1477 & 0.6405 & 30.9455  & 0.7621 & 67.0630    & 0.9239 & 102.2508 \\
   0.4811 & 60.1735  & 0.6581 & 35.1934 & 0.7701 & 74.9344     & 0.9447 & 97.2271 \\
   0.5085 & 53.2823  & 0.6703 & 45.5234  & 0.7839 & 53.1310   & 0.9631 & 99.4473 \\
   0.5121 & 53.4914  & 0.6761 & 42.6753  & 0.7909 & 50.8776  & 0.9739 & 77.4691 \\
   0.5265 & 44.3346  & 0.6891 & 67.0539  & 0.7991 & 65.2257   & 0.9785 & 81.3049 \\
   0.5355 & 43.1639  & 0.7111 & 94.2231  & 0.8051 & 85.1548  & 0.9829 & 77.2446 \\
   0.5631 & 38.6210  & 0.7391 & 69.4755  & 0.8121 & 82.7582  & 0.9923 & 97.4887 \\
   0.5769 & 38.9783   & 0.7451 & 67.8694  & 0.8399 & 106.8413 & 0.9999 & 85.6308 \\
\end{tabular}
\caption{Extracted data of the density along the bottom wall. $\Rey = 1000$.} \label{vst_1000_extract}
\end{center}
\end{table}

\subsection{Grid refinement study with the Grid-Convergence Index approach}
As shown in figure~\ref{vst_1000_density}, we can hardly see any difference in the plot of the density distribution on the grid $3000 \times 1500$, $4000 \times 2000$, and $5000 \times 2500$. Since the flow field is very complex, it is important to develop some quantitative measure on the convergence of the computational solutions to the presumed exact solution as the grid spacing is refined to approach zero. We adopt the Grid-Convergence Index (GCI) approach proposed by \citet{Roache1994, Roache1997}.

Based on the generalized theory of the Richardson Extrapolation \citep{Richardson1911}, the Grid-Convergence Index is defined to uniformly report the grid refinement tests. Assume $f_1$ and $f_2$ are solutions on a fine grid and a coarse grid, respectively, the relative error is expressed as 
\begin{equation} \label{relative_e}
\epsilon = (f_2 - f_1) / f_1.
\end{equation}
Then the GCI of the fine-grid solution is defined by the following formula:
\begin{equation} \label{gci_eq}
GCI = c_s |\epsilon| / (r^p - 1),
\end{equation}
where $r$ is the ratio of the grid spacing between the coarse and fine grids ($ r = h_2/h_1 > 1$), and $p$ is the order of accuracy of the scheme. $c_s = 3$ is a safety factor. As pointed out by \citet{Roache1994}, the GCI gives a conservative estimate of the error relative to the unknown `exact' solution.

The underlying assumption of the GCI approach is the smoothness of the solution. The solution must have a Taylor series expansion at least up to the order of the numerical scheme. Despite the existence of many sharp `discontinuities' in the present shock tube problem, the solution of the Navier-Stokes equations is not strictly discontinuous. Thus, the GCI still serves as a reliable measure on the convergence of our computations when the grids used are sufficiently fine enough.

In detail, the GCI on the $2000 \times 1000$ and finer grids are computed. The calculations are performed on the target grid and the first coarser grid next to it, i.e., to get the GCI of the solution on the $3000 \times 1500$ grid, the solutions on the $3000 \times 1500$ grid and its neighbouring $2000 \times 1000$ grid are used in \eqref{relative_e} and \eqref{gci_eq}.

In particular, we choose the $1000 \times 500$ grid as a standard stencil. The GCI based on the averaged density in each stencil cell is computed. Since the cell numbers of all grids are integer multiples of the stencil cell number in both $x$ and $y$ directions, no interpolation or other approximation is needed. Following the suggestion of \citet{Roache1994}, since a uniform order $p$ can not be found all across the field which contains shocks and other discontinuities, a conservative value $p = 1$ is used. After the GCI on each cell of the stencil is obtained, the average and root mean square of all the GCIs are taken and reported. 

\citet{Roache1994} also proposed a method for checking whether the asymptotic range of convergence is reached by two GCIs on three different grids, on the premise that the order of scheme is known. This is based on the fact that the GCI is essentially an estimate of the error level. Similarly, in the present case that the practical order of the scheme cannot be well defined, we assume that when three GCIs are located in a straight line in the log-log plot against the grid spacing, a conclusion can be drawn that the solution is converging with a constant order.

\begin{figure}
\centering
\includegraphics[width=0.5\textwidth]{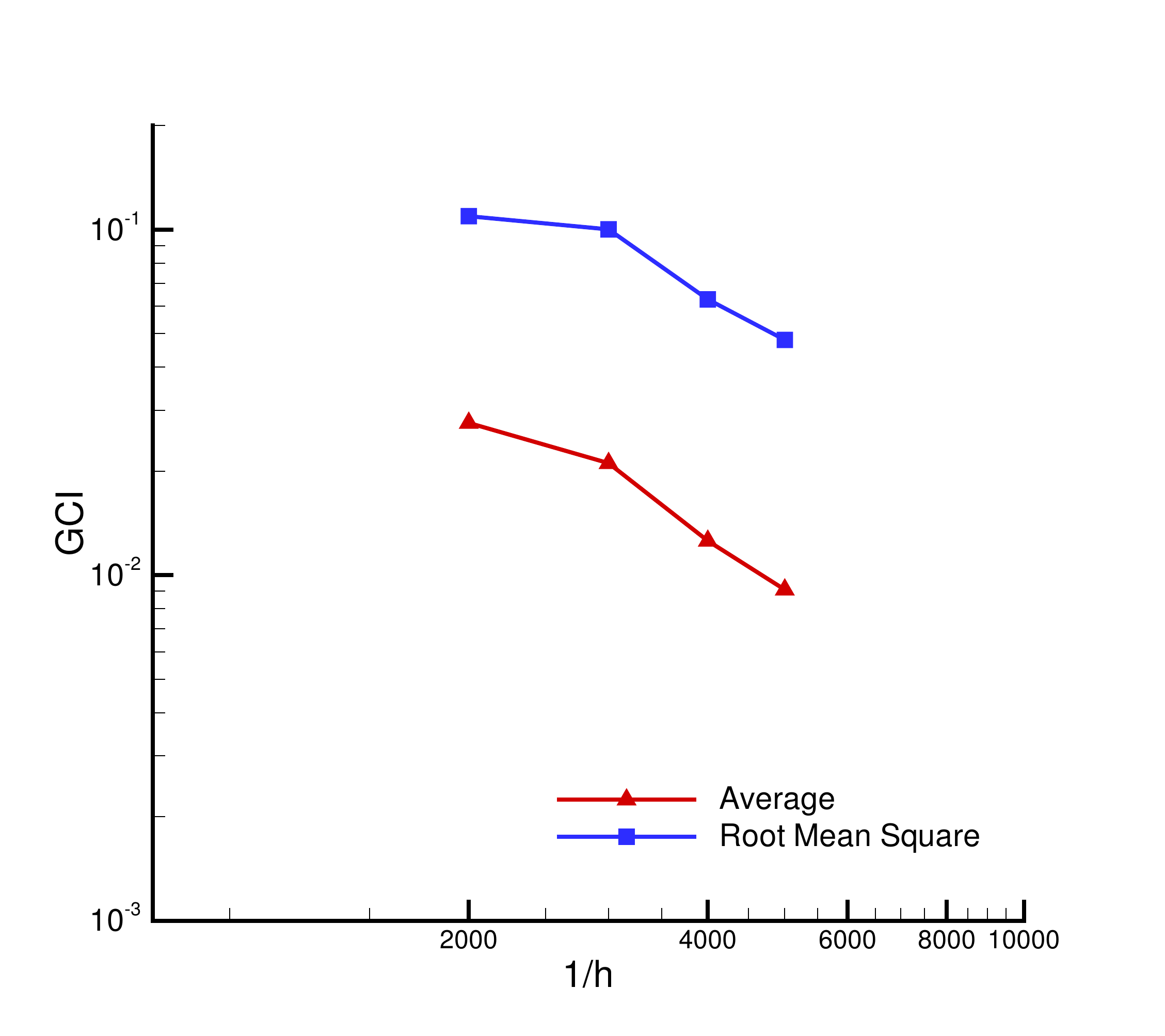}
\caption{Overall Grid-Convergence Index for the viscous shock tube problem at $\Rey = 1000$. $h$ is the grid spacing.}\label{gci}
\end{figure}

The results are shown in figure~\ref{gci}. For both averaging methods, the points corresponding to the $3000 \times 1500$ grid, the $4000 \times 2000$ grid and the $5000 \times 2500$ grid are approximately in a line, indicating that the asymptotic range is achieved on the $3000 \times 1500$ grid, whereas the result of the $2000 \times 1000$ grid is out of the range. This conclusion agrees well with figure~\ref{vst_1000_density}, where the visible details of the density distribution stay unchanged for the $3000 \times 1500$ and finer grids, but not for the $2000 \times 1000$ one.

If we go back to the original meaning of the GCI, it is seen in figure~\ref{gci}, from an overall perspective, that the averaged relative error of the result obtained by the $5000 \times 2500$ grid is less than 1\%, with respect to the exact solution.

The viscous shock tube problem at $\Rey = 1000$ is naively simple in geometry and initial and boundary conditions. Yet, it encompasses the evolution of almost all elementary flow phenomena of a viscous compressible flow and their mutual interactions, resulting in a complex dynamic flow field with a multitude of fine scales. As such it offers a difficult but arguably necessary test case to demonstrate the accuracy and efficiency of modern high-resolution and high-order numerical methods for compressible viscous flows. The grid-converged solution for this problem as well as the rigorous GCI approach presented here provide the research community a useful database and approach in comparing and assessing different numerical methods for their numerical discretization, flux models, shock capture strategies, effect of numerical dissipation, time evolution, and implementation of boundary conditions.

\section{The $\Rey=1000$ Case: Analysis of the Complex Flow Physics} \label{section process}

The dynamic evolution process of the flow field at $\Rey=1000$ is of great significance for understanding the fluid dynamics of the interactions between boundary layers, vortices, and wave systems in supersonic flow. Analysis and discussion of the flow physics of this problem, however, has been rather minimum in previous papers except those by \citet{Daru2004,Daru2009}. (\citet{Chen2015} calculated a slightly different problem and gave some discussions on the flow behaviours at early stages.) This is partly due to the complexity of this problem and partly lack of adequate proof of numerical convergence. With the solution on the $3000\times1500$ grid proven above be grid converged, we proceed to present and analyse the details of the flow field and its time evolution. Important observations during the process are emphasised.

\begin{figure}
\centering
\begin{minipage}[t]{0.33\textwidth}
\centering
\includegraphics[width=\textwidth]{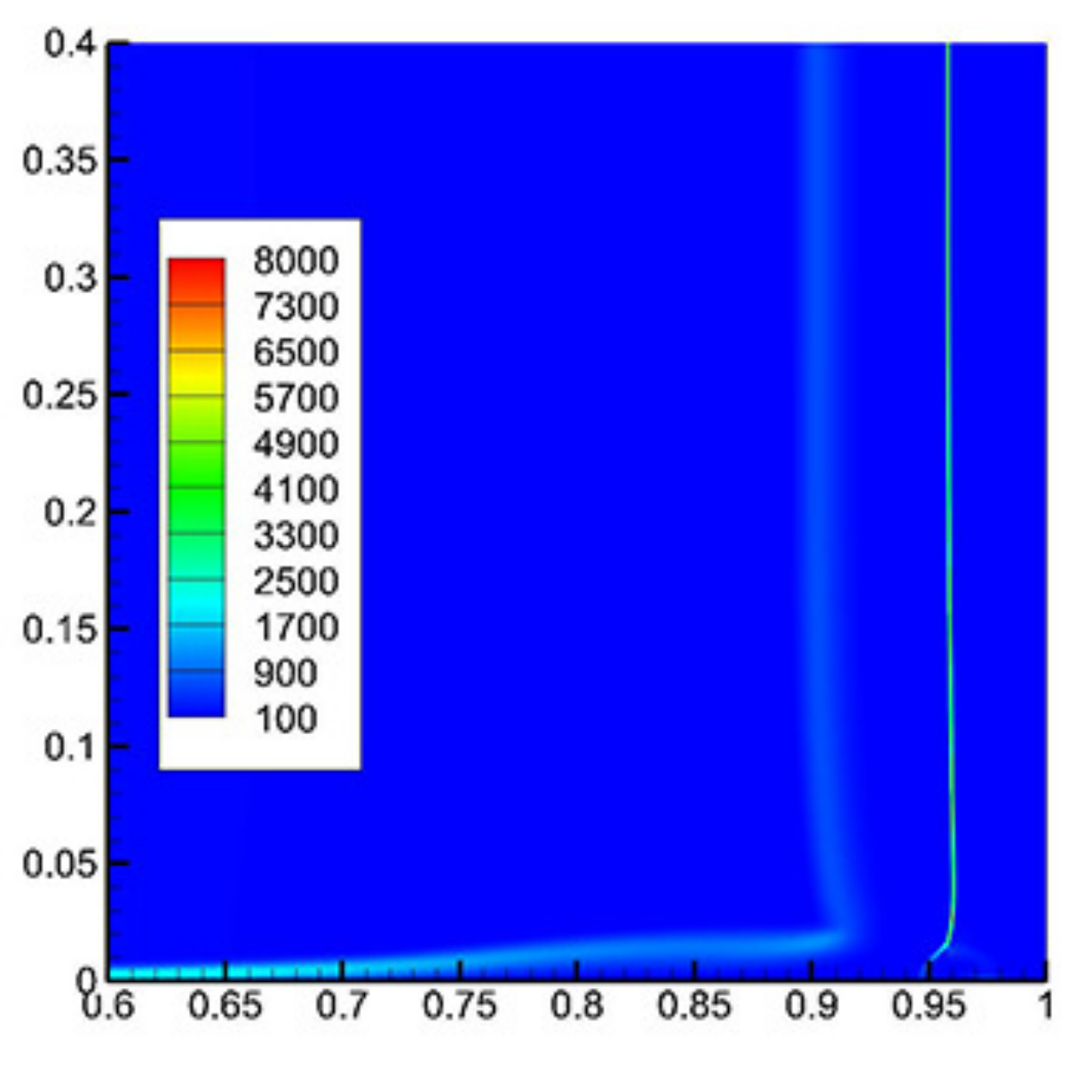}
\centerline{\footnotesize (a)}
\end{minipage}%
\begin{minipage}[t]{0.33\textwidth}
\centering
\includegraphics[width=\textwidth]{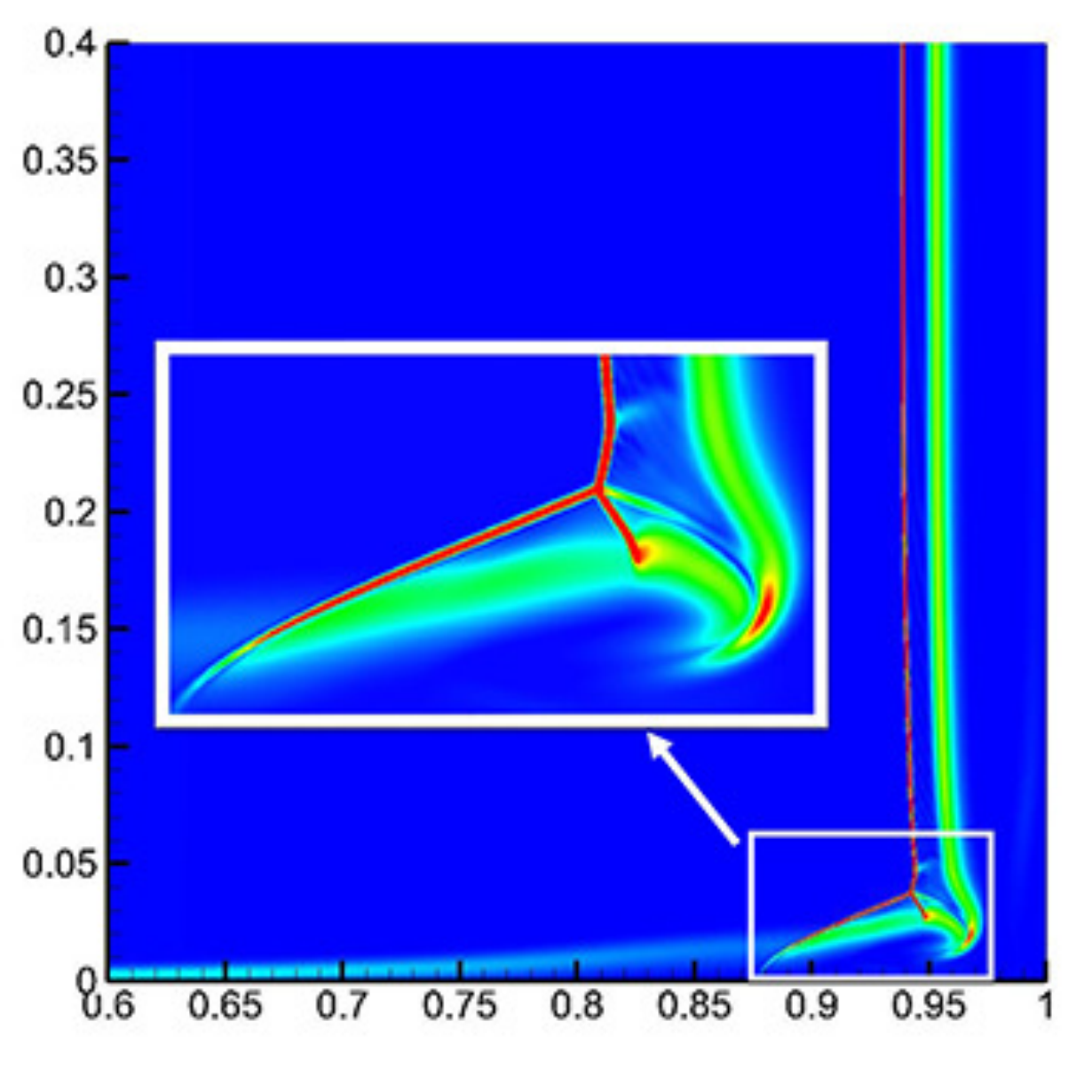}
\centerline{\footnotesize (b)}
\end{minipage}%
\begin{minipage}[t]{0.33\textwidth}
\centering
\includegraphics[width=\textwidth]{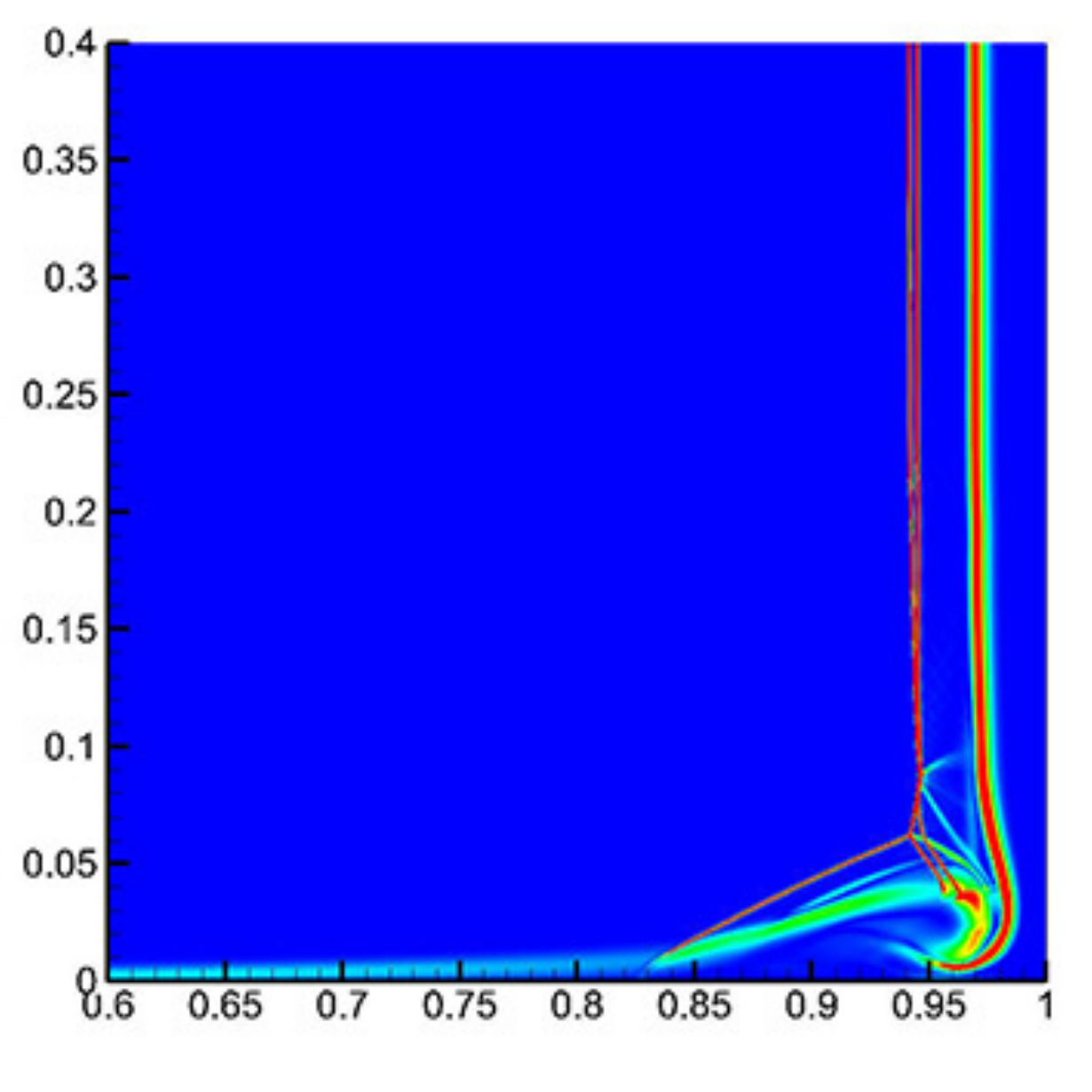}
\centerline{\footnotesize (c)}
\end{minipage}

\begin{minipage}[t]{0.33\textwidth}
\centering
\includegraphics[width=\textwidth]{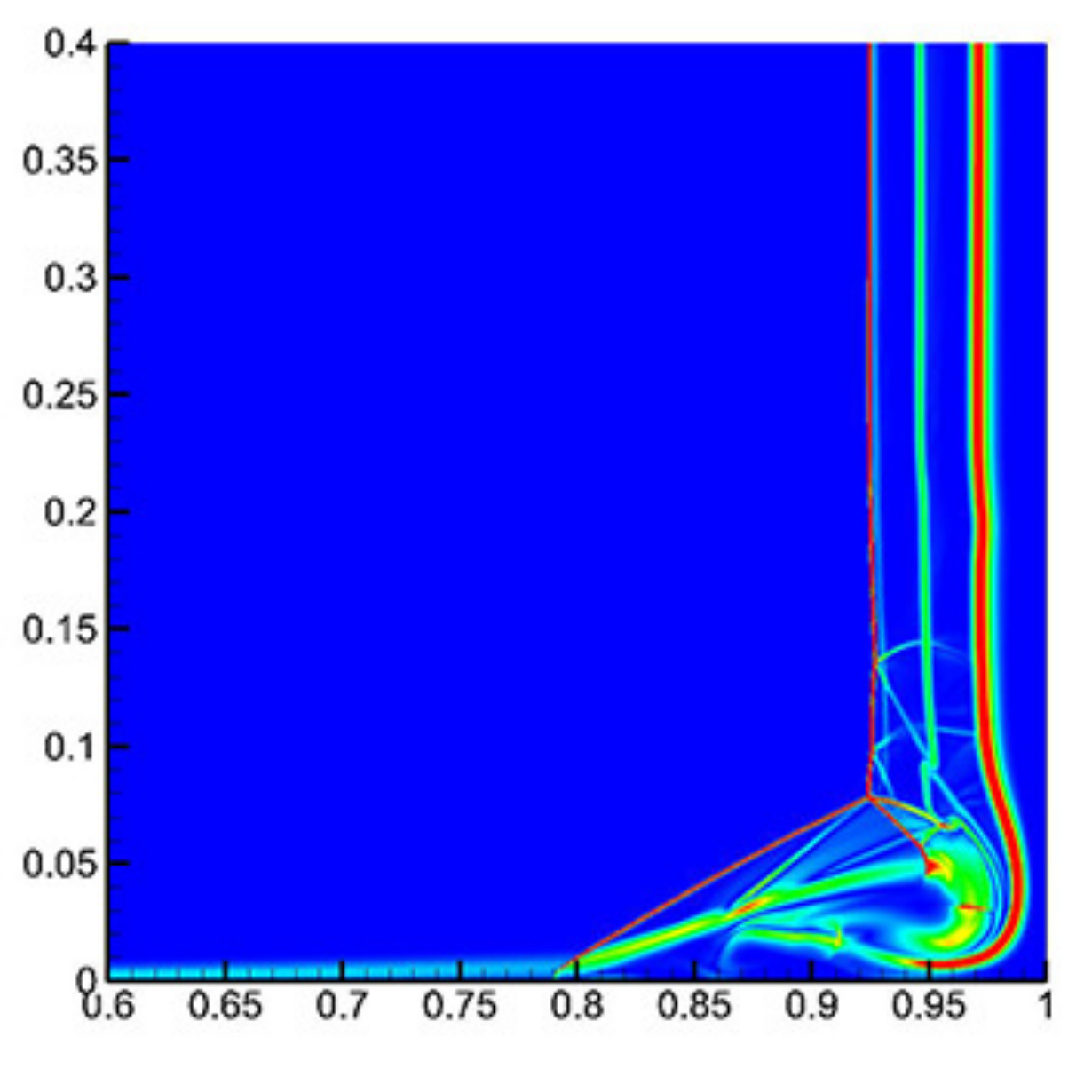}
\centerline{\footnotesize (d)}
\end{minipage}%
\begin{minipage}[t]{0.33\textwidth}
\centering
\includegraphics[width=\textwidth]{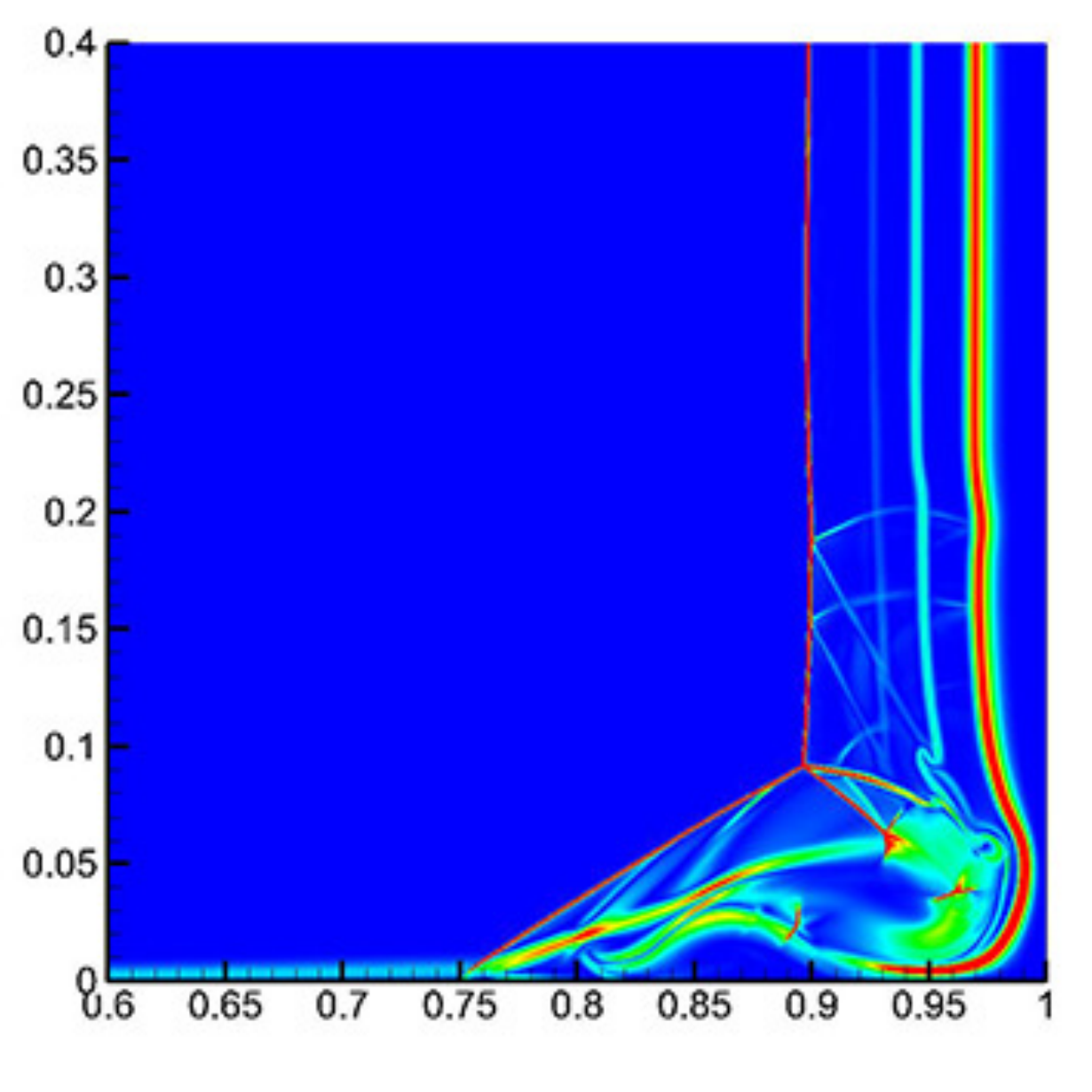}
\centerline{\footnotesize (e)}
\end{minipage}%
\begin{minipage}[t]{0.33\textwidth}
\centering
\includegraphics[width=\textwidth]{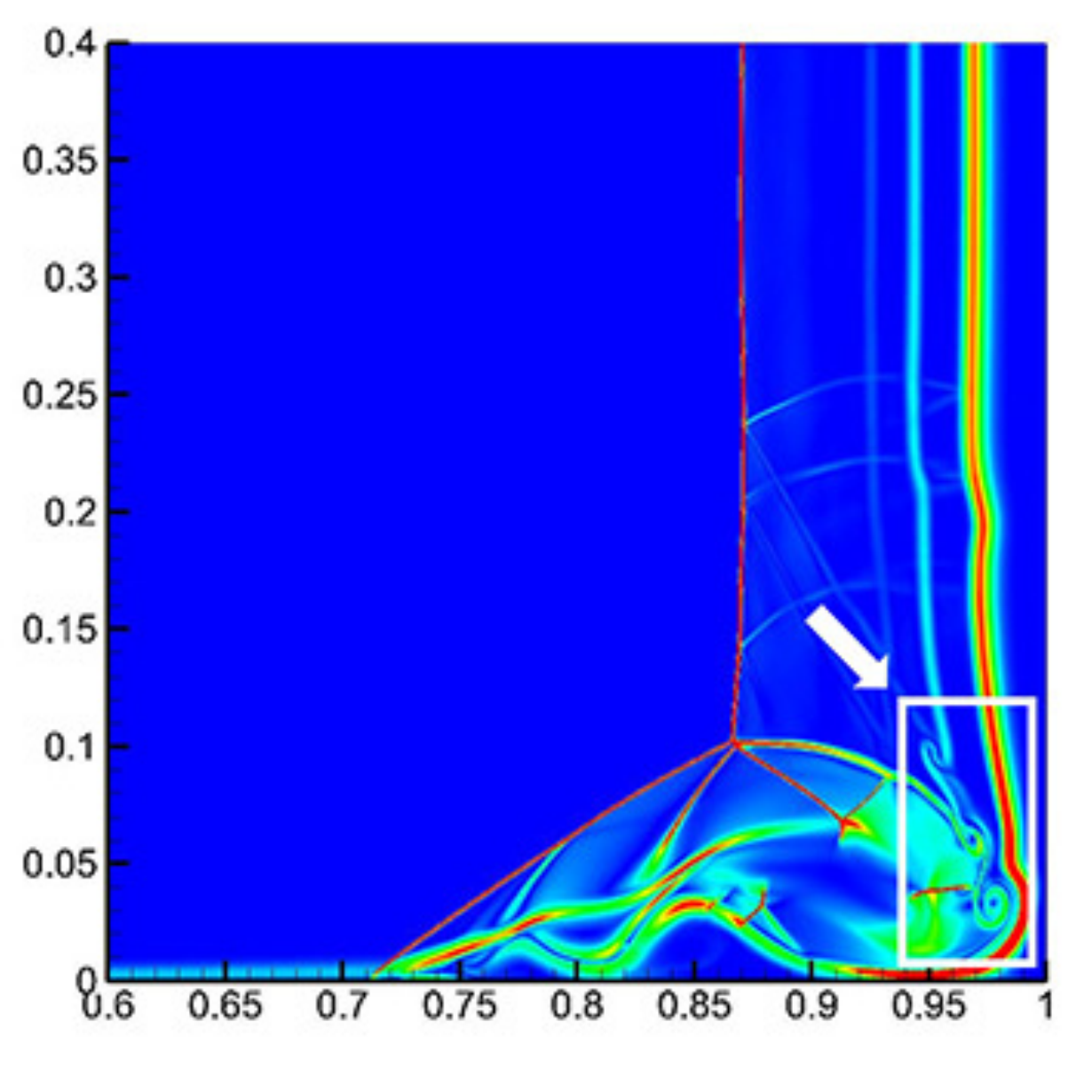}
\centerline{\footnotesize (f)}
\end{minipage}

\begin{minipage}[t]{0.33\textwidth}
\centering
\includegraphics[width=\textwidth]{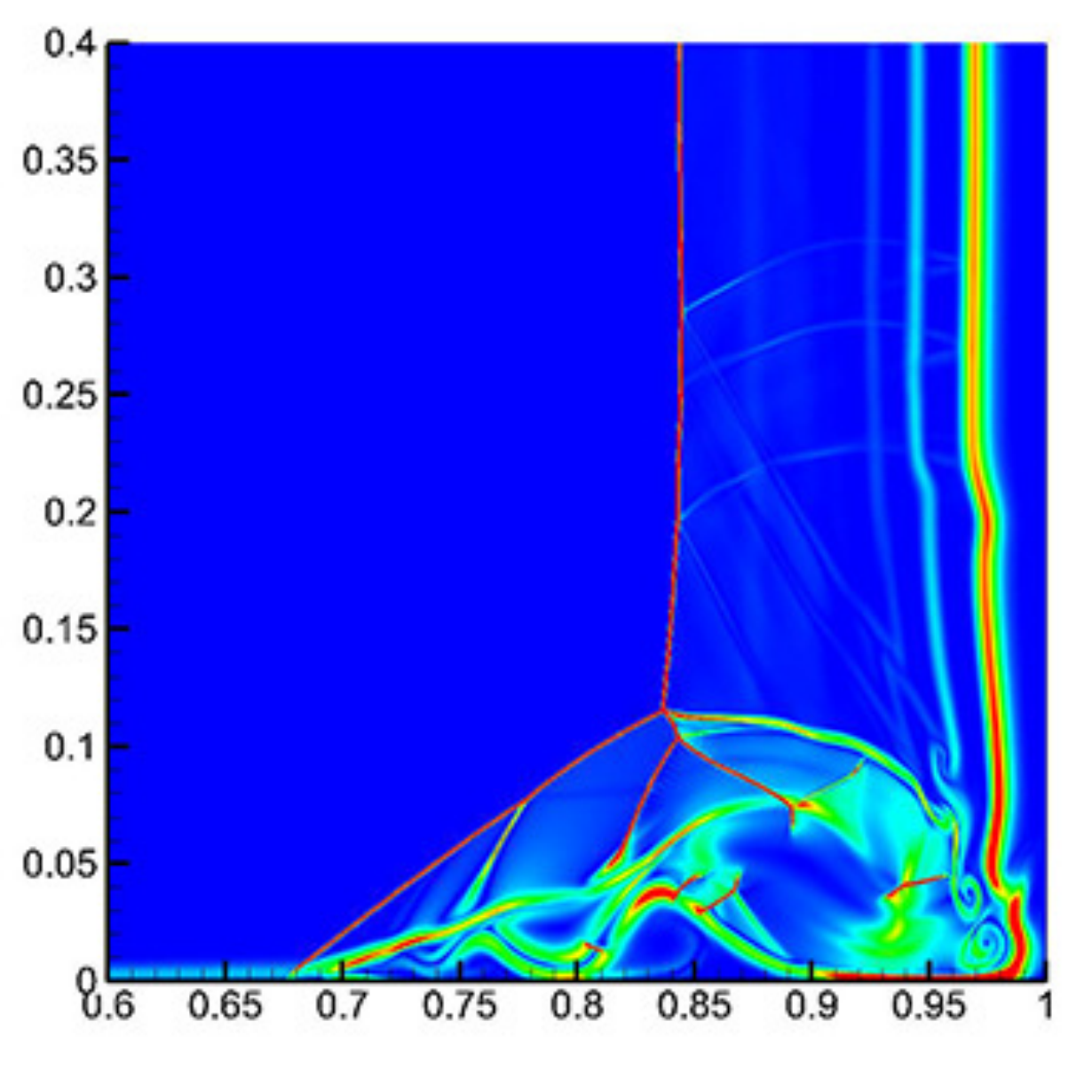}
\centerline{\footnotesize (g)}
\end{minipage}%
\begin{minipage}[t]{0.33\textwidth}
\centering
\includegraphics[width=\textwidth]{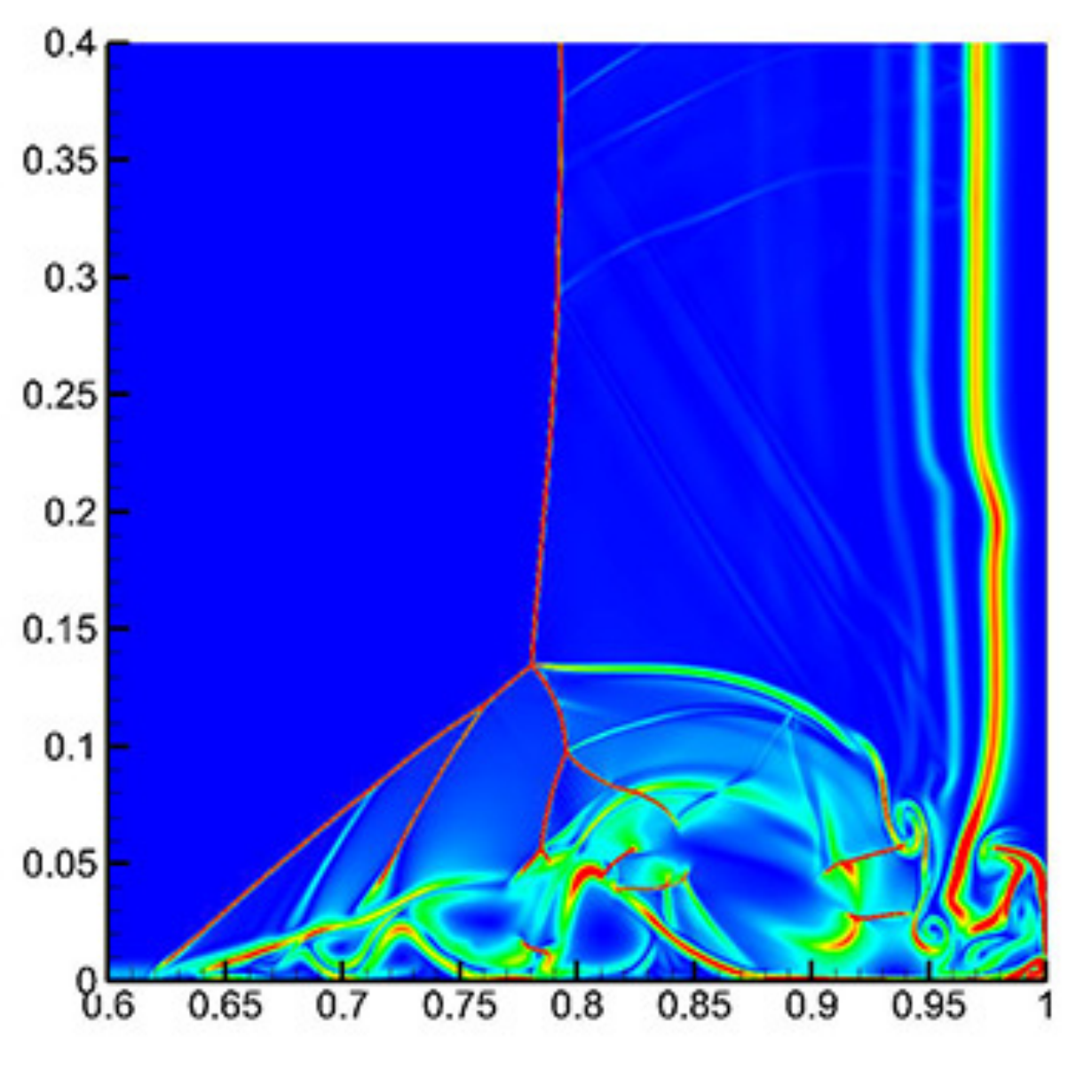}
\centerline{\footnotesize (h)}
\end{minipage}%
\begin{minipage}[t]{0.33\textwidth}
\centering
\includegraphics[width=\textwidth]{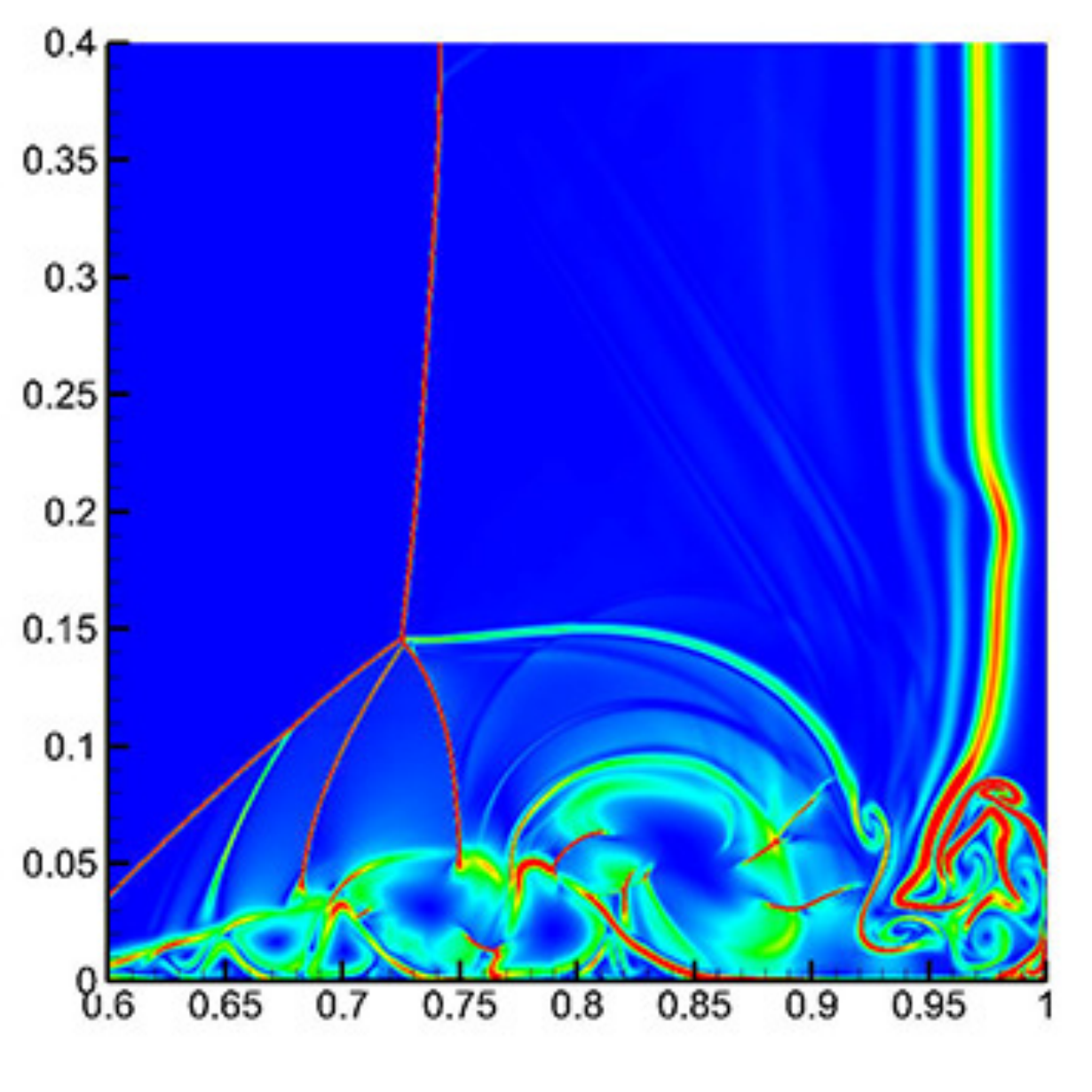}
\centerline{\footnotesize (i)}
\end{minipage}

\begin{minipage}[t]{0.33\textwidth}
\centering
\includegraphics[width=\textwidth]{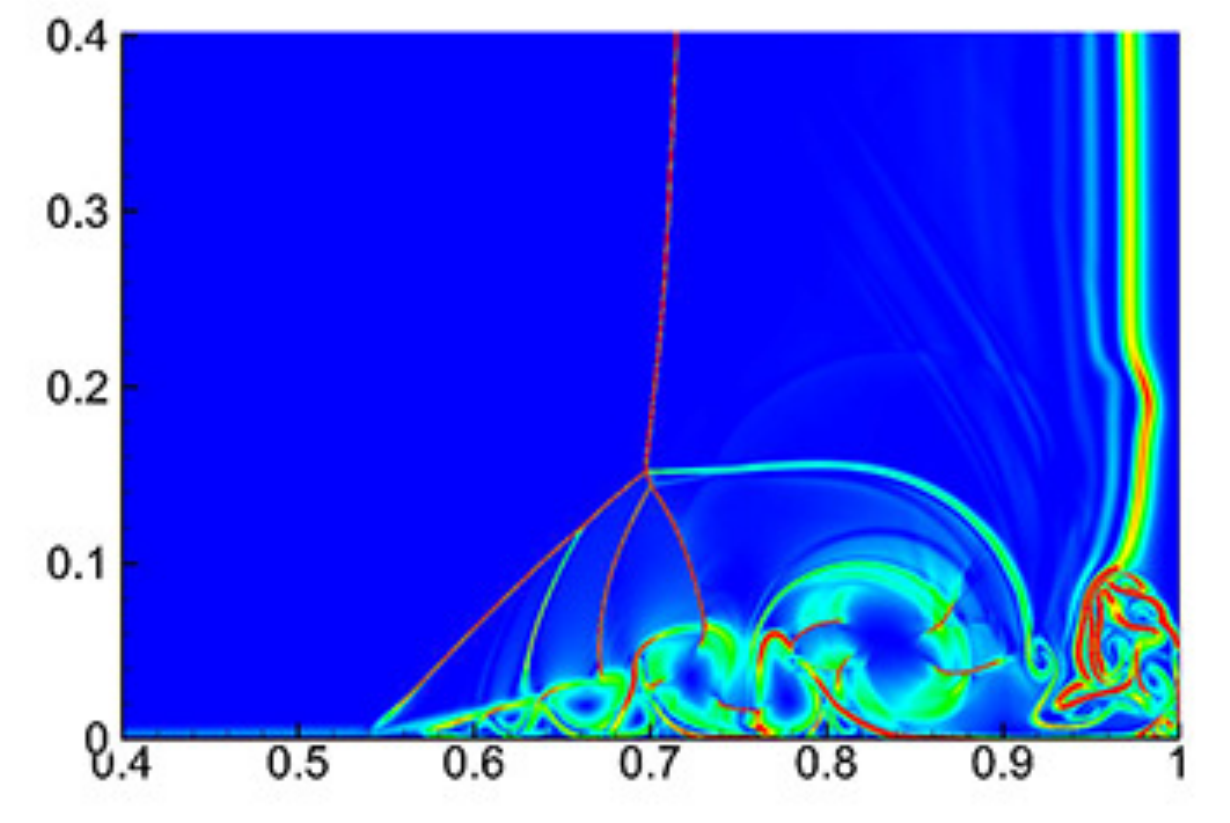}
\centerline{\footnotesize (j)}
\end{minipage}%
\begin{minipage}[t]{0.33\textwidth}
\centering
\includegraphics[width=\textwidth]{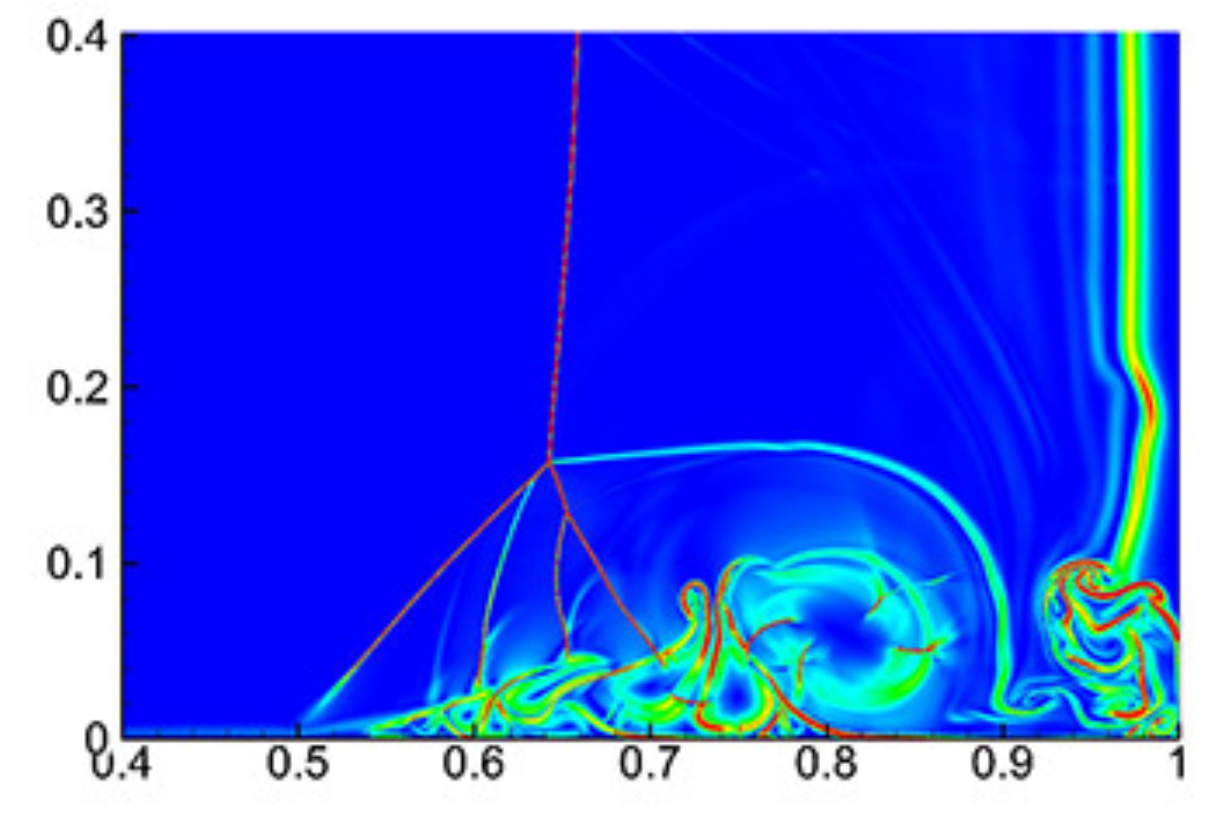}
\centerline{\footnotesize (k)}
\end{minipage}%
\begin{minipage}[t]{0.33\textwidth}
\centering
\includegraphics[width=\textwidth]{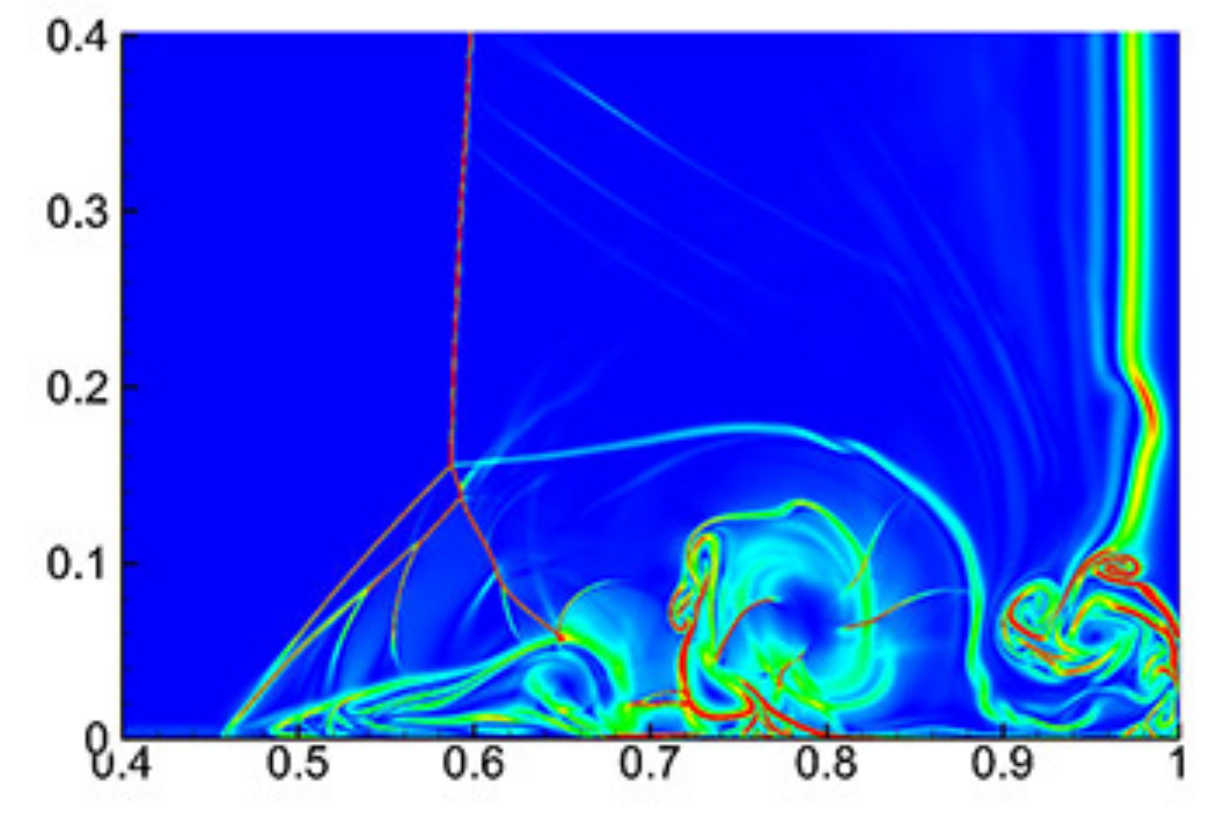}
\centerline{\footnotesize (l)}
\end{minipage}
\centering
\caption{Distribution of density gradient magnitude at   (\textit{a}) $t=0.25$, (\textit{b}) $t=0.30$ (the key structure is enlarged), (\textit{c}) $t=0.35$, (\textit{d}) $t=0.40$, (\textit{e}) $t=0.45$, (\textit{f}) $t=0.50$ (vortical structures are marked out), (\textit{g}) $t=0.55$, (\textit{h}) $t=0.65$, (\textit{i}) $t=0.75$, (\textit{j}) $t=0.80$, (\textit{k}) $t=0.90$ and (\textit{l}) $t=1.00$.\protect\\}\label{dgm}
\end{figure}

Before detailed description, we present the whole history of the physical dynamic process in figure~\ref{dgm}, where the magnitude of the density gradient at different time points of interest are shown in chronological order.

\begin{figure}
\centering
\begin{minipage}[t]{0.5\textwidth}
\centering
\includegraphics[width=\textwidth]{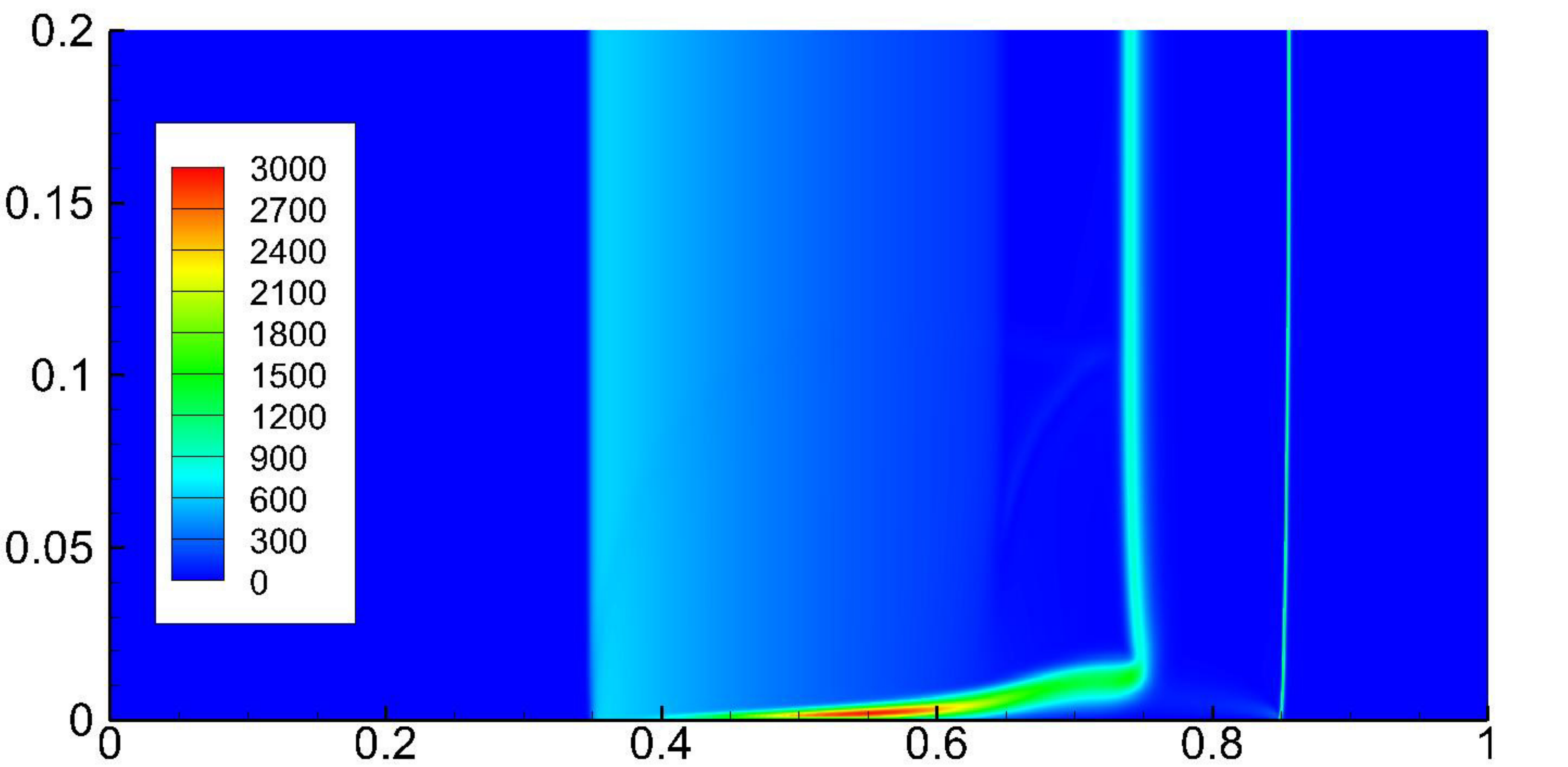}
\centerline{\footnotesize (a)}
\end{minipage}%
\begin{minipage}[t]{0.5\textwidth}
\centering
\includegraphics[width=\textwidth]{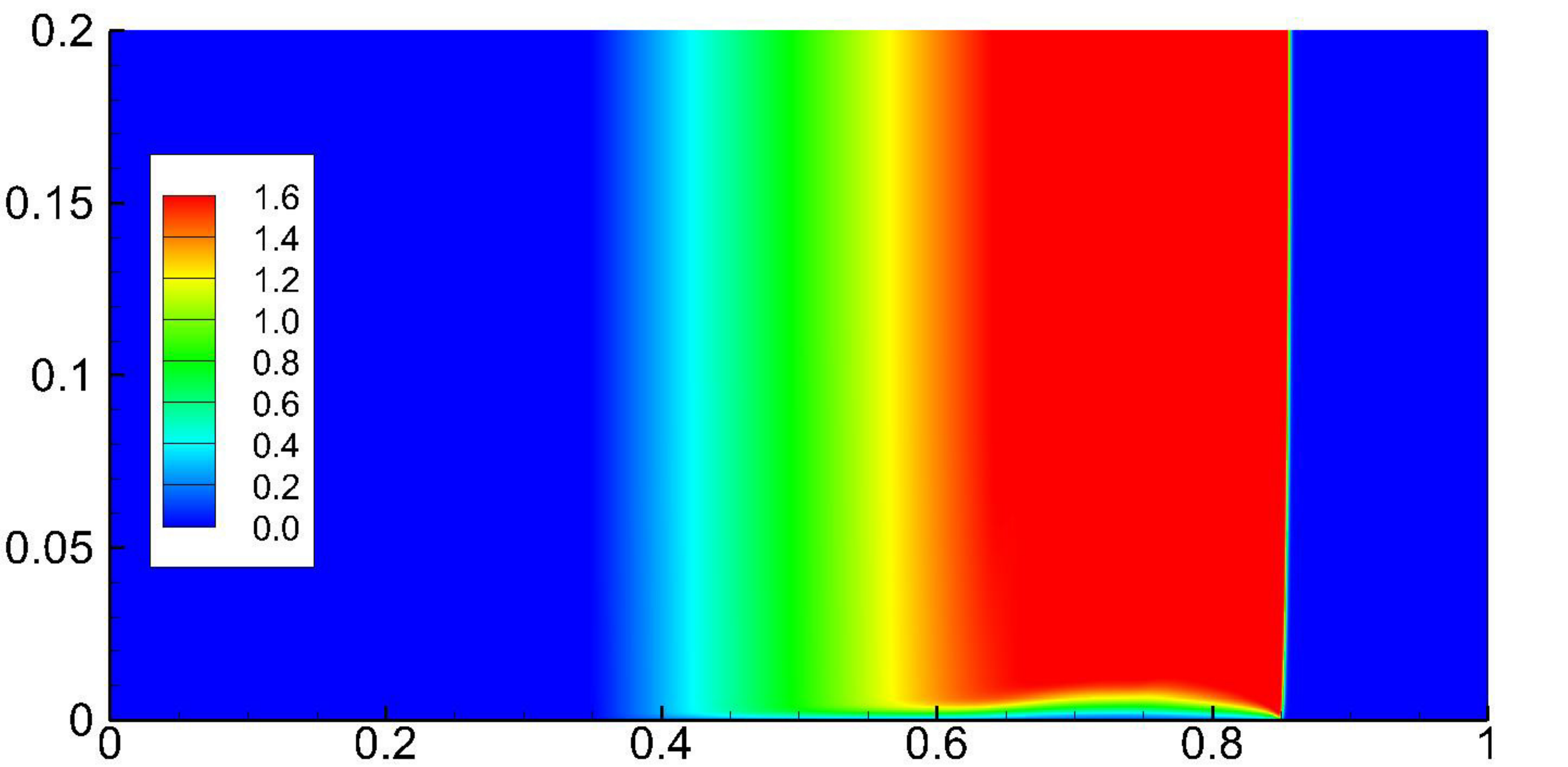}
\centerline{\footnotesize (b)}
\end{minipage}
\centering
\caption{Flow field at $t=0.15$ (the $y$-axis is stretched for clarity): (\textit{a}) Distribution of density gradient magnitude; (\textit{b}) Distribution of the velocity in $x$-direction.}\label{dgm_u_30}
\end{figure}

At $t=0$, break of the diaphragm results in three different waves: a right-moving shock wave, a contact discontinuity following the shock, and an expansion wave propagating in both directions. The waves travel freely into the undisturbed region creating a boundary layer on the bottom wall behind. See figure~\ref{dgm_u_30}. This configuration is similar to the inviscid case in figure~\ref{xt}, except for the creation of the wall boundary layer and thickening of the two discontinuities (especially the contact discontinuity) due to viscous effect.

The boundary layer is attached to and dragged by the right-moving shock wave, as can be seen in figure~\ref{dgm_u_30}(b), where the distribution of the velocity in the $x$-direction is shown. The boundary layer thickens as one moves away from its initiation point at the foot of the shock much like a usual boundary layer over a flat plate until $x=0.75$ where the contact discontinuity is located. The effective Reynolds number is increased due to the high density in the freestream flow behind the contact surface, resulting in a decrease of the boundary layer thickness.

At this stage the boundary layer is behind the shock wave and is theoretically of zero thickness at the foot of the shock. Therefore, the shock front remains effectively straight across the channel and curves only slightly as it touches the wall. On the contrary, the contact discontinuity, being a material wave front that moves with the fluid, is dramatically bent over the boundary layer because of the no-slip condition on the wall. It is seen from figure~\ref{dgm_u_30}(a) that a very oblique contact discontinuity is stretched along the horizontal wall and it connects with the vertical one outside the boundary layer.

\begin{figure}
\centering
\begin{minipage}[t]{0.4\textwidth}
\centering
\includegraphics[width=0.8\textwidth]{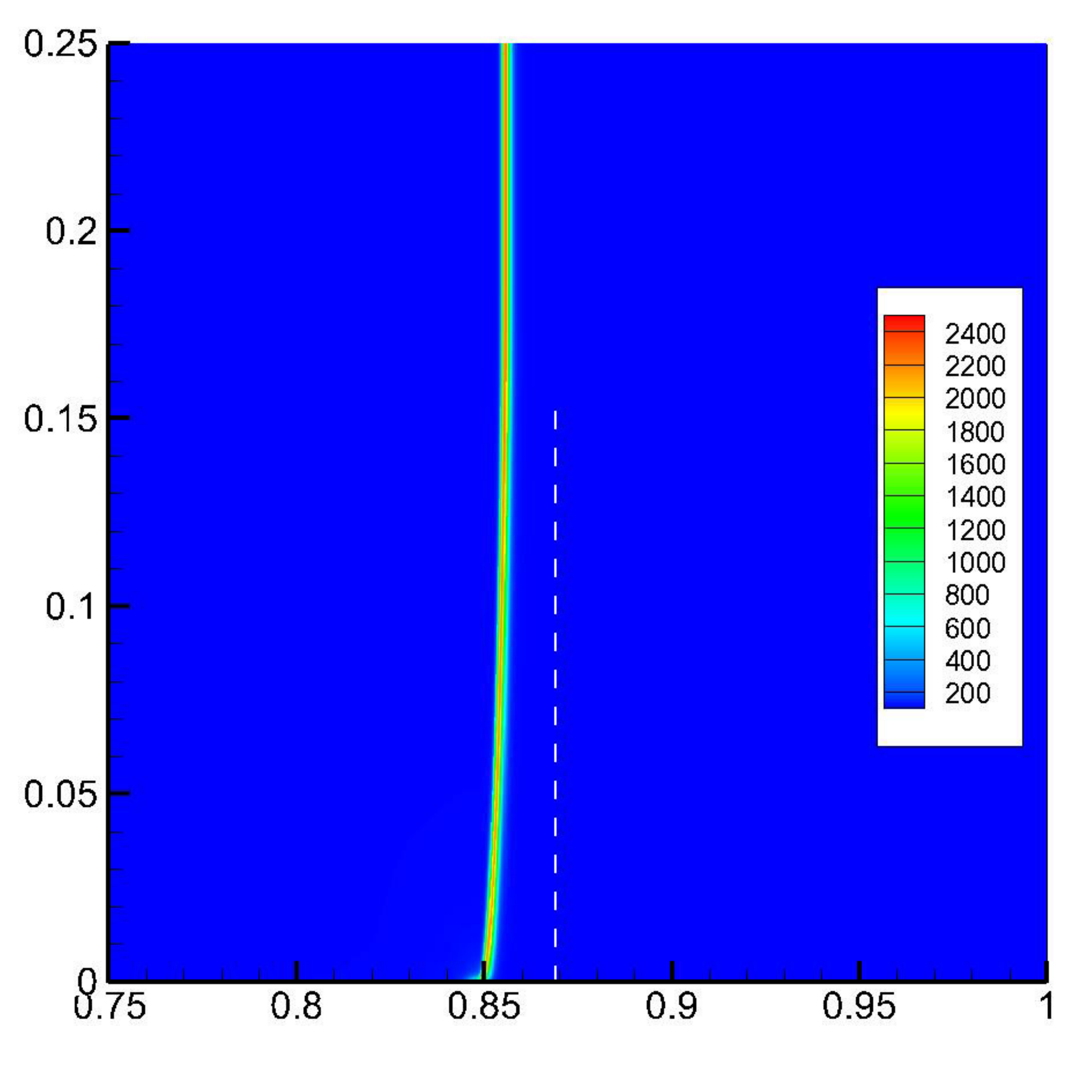}
\centerline{\footnotesize (a)}
\end{minipage}%
\begin{minipage}[t]{0.4\textwidth}
\centering
\includegraphics[width=0.8\textwidth]{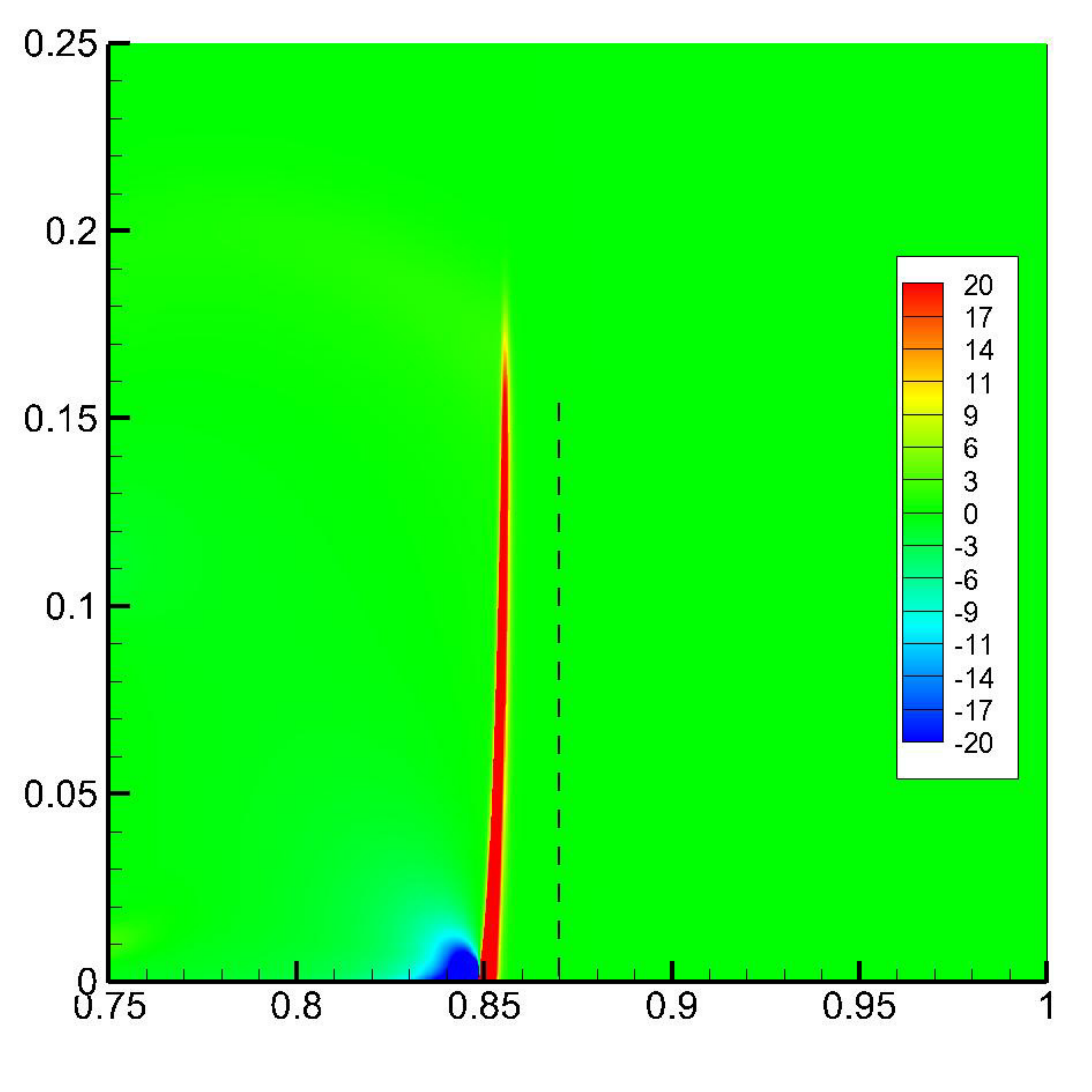}
\centerline{\footnotesize (b)}
\end{minipage}
\centering
\caption{Distribution of the pressure gradient at $t=0.15$: (\textit{a}) pressure gradient magnitude; (\textit{b}) $y$-component of the pressure gradient. Dashed lines indicate the vertical direction.}\label{pg_ypg_30}
\end{figure}

The curved near-wall section of the shock wave gets enlarged with time. Since the pressure gradient is perpendicular to the shock surface, the curving of the shock generates a non-zero $y$-direction component of the pressure gradient. Figure~\ref{pg_ypg_30}(a) shows the distribution of the magnitude of the pressure gradient at $t =  0.15$. The shock is more curved at locations closer to the wall. The $y$-component of the pressure gradient is shown in figure~\ref{pg_ypg_30}(b). Obviously this quantity is closely related to the curvature of the shock. As a consequence, the fluid will experience a sudden acceleration when it flows across the narrow shock, obtaining a downward velocity. Although this velocity is very small and nearly invisible because of the much larger flow velocity in the $x$-direction, later we will see that it is of great importance in the following dynamic process.

\begin{figure}
\centering
\begin{minipage}[t]{0.33\textwidth}
\centering
\includegraphics[width=\textwidth]{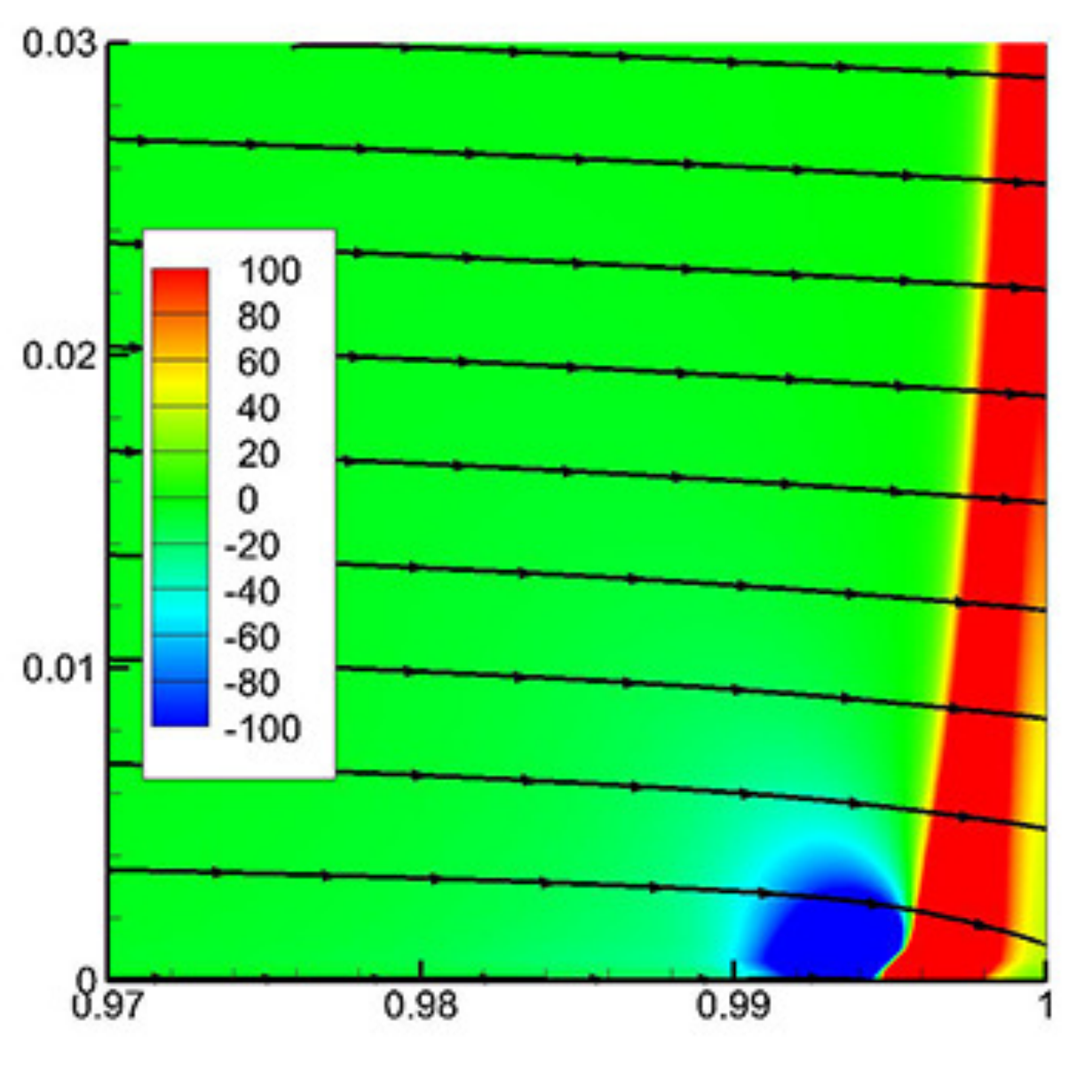}
\centerline{\footnotesize (a)}
\end{minipage}%
\begin{minipage}[t]{0.33\textwidth}
\centering
\includegraphics[width=\textwidth]{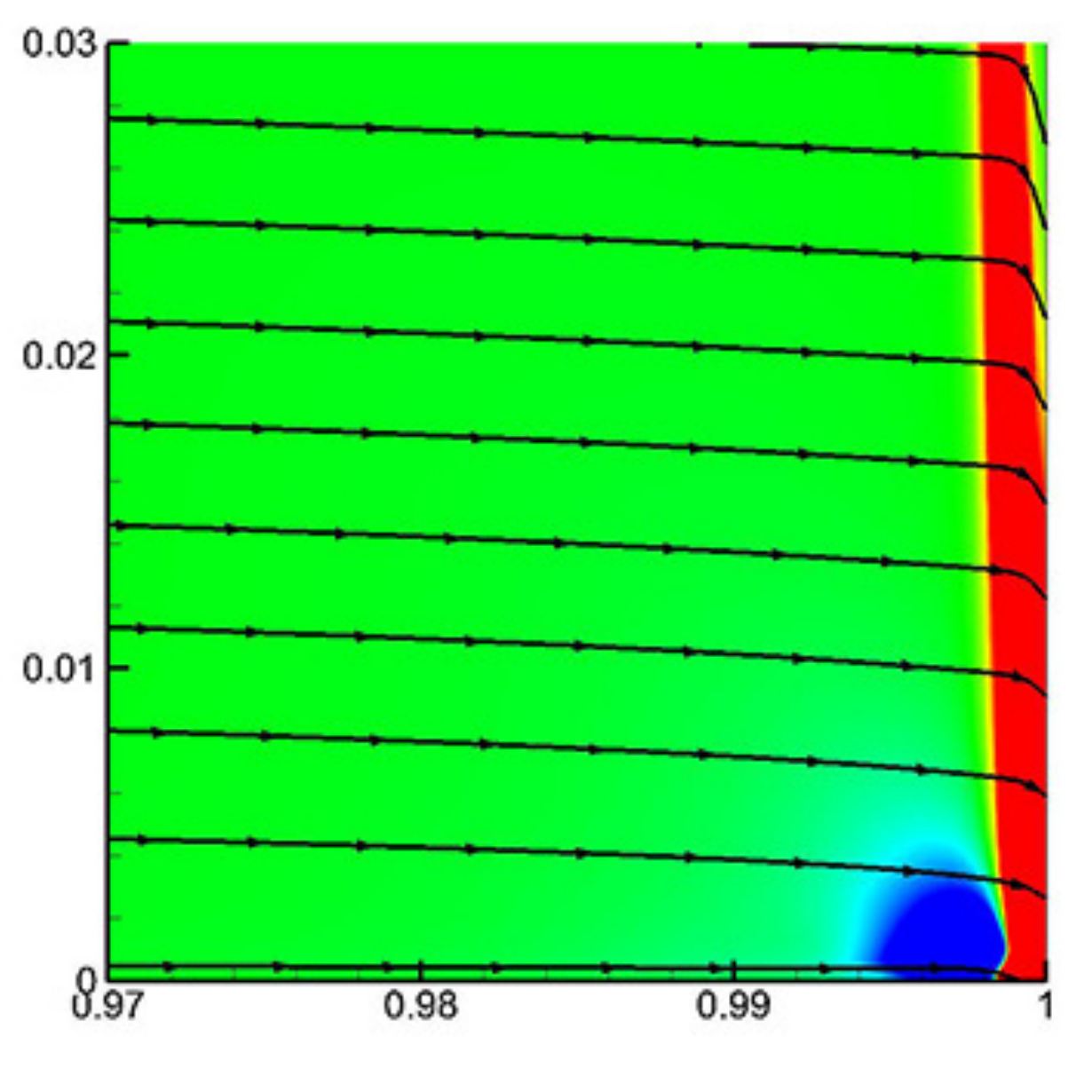}
\centerline{\footnotesize (b)}
\end{minipage}%
\begin{minipage}[t]{0.33\textwidth}
\centering
\includegraphics[width=\textwidth]{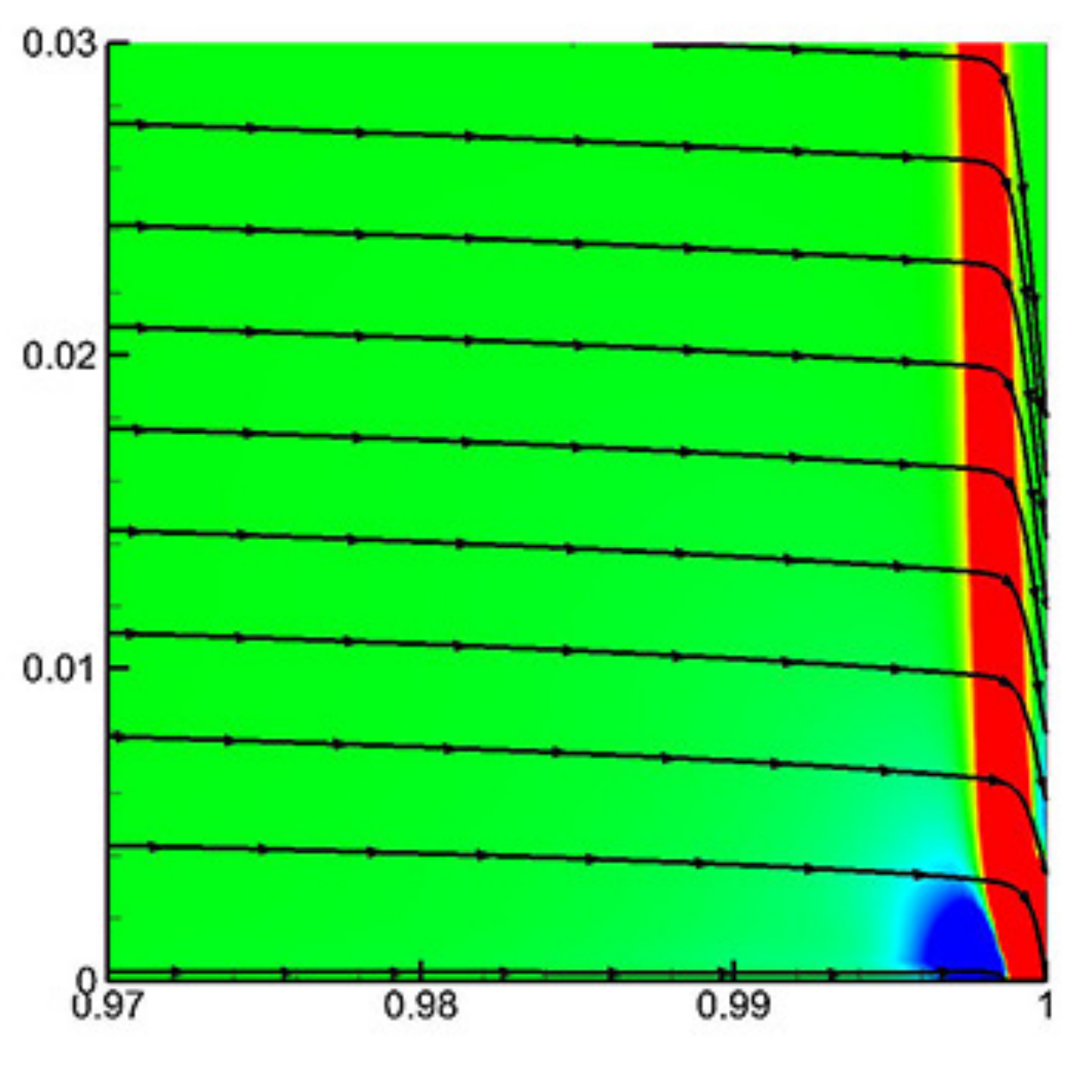}
\centerline{\footnotesize (c)}
\end{minipage}
\centering
\caption{Streamlines and the distribution of the $y$-component of the pressure gradient at (\textit{a}) $t = 0.2124$, (\textit{b}) $t = 0.2140$ and (\textit{c}) $t = 0.2146$.}\label{ypg_reflect}
\end{figure}

\begin{figure}
\centering
\begin{minipage}[t]{0.33\textwidth}
\centering
\includegraphics[width=\textwidth]{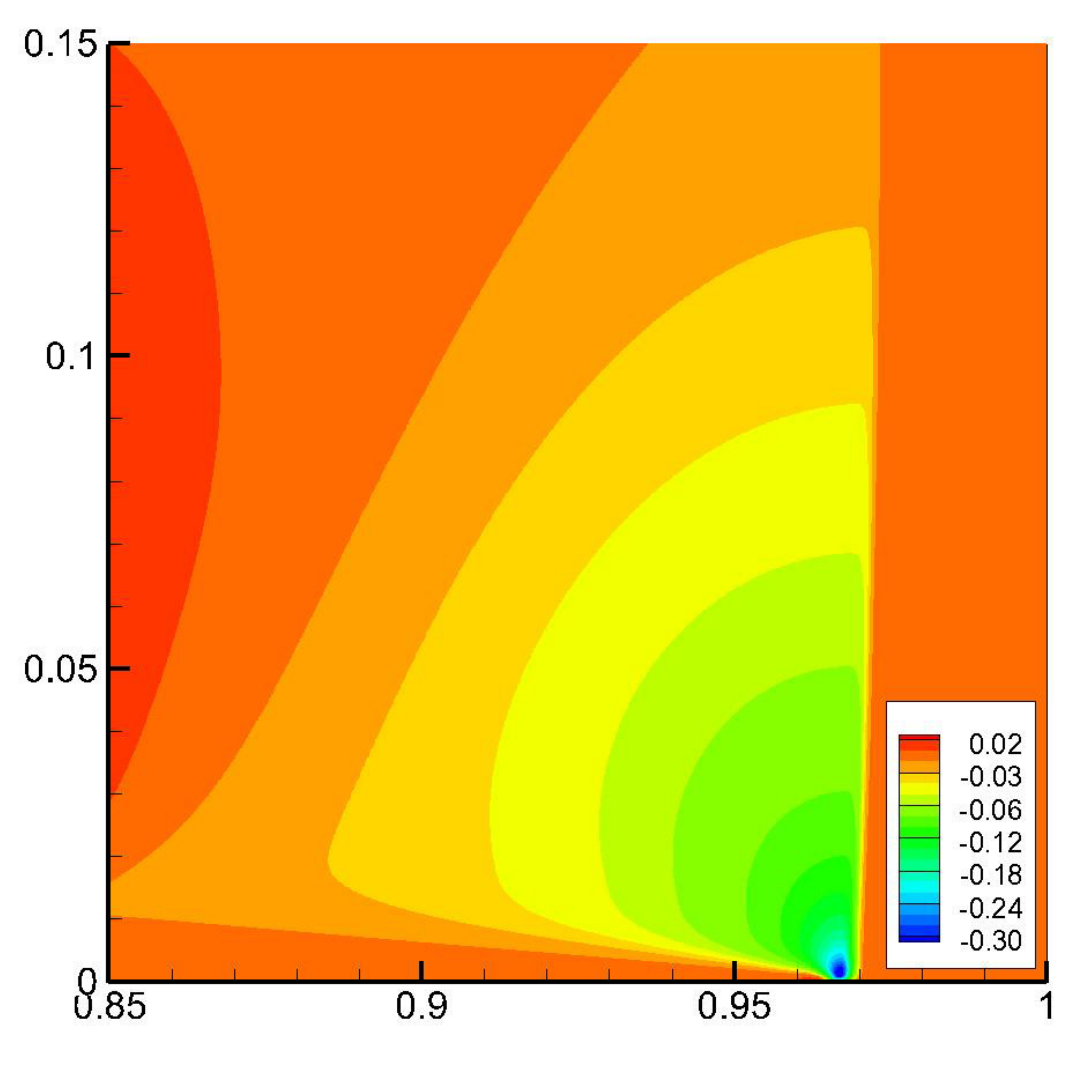}
\centerline{\footnotesize (a)}
\end{minipage}%
\begin{minipage}[t]{0.33\textwidth}
\centering
\includegraphics[width=\textwidth]{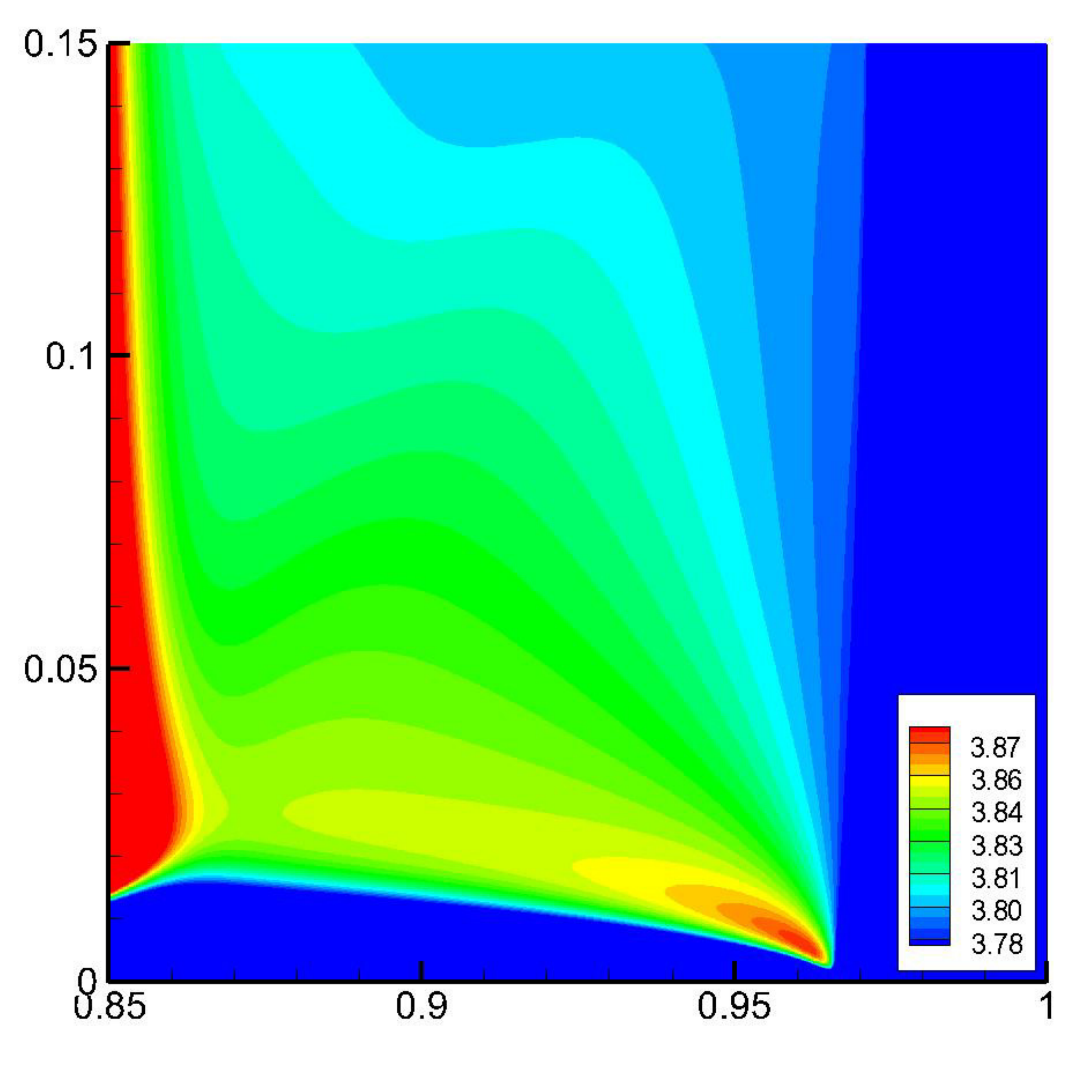}
\centerline{\footnotesize (b)}
\end{minipage}%
\begin{minipage}[t]{0.33\textwidth}
\centering
\includegraphics[width=\textwidth]{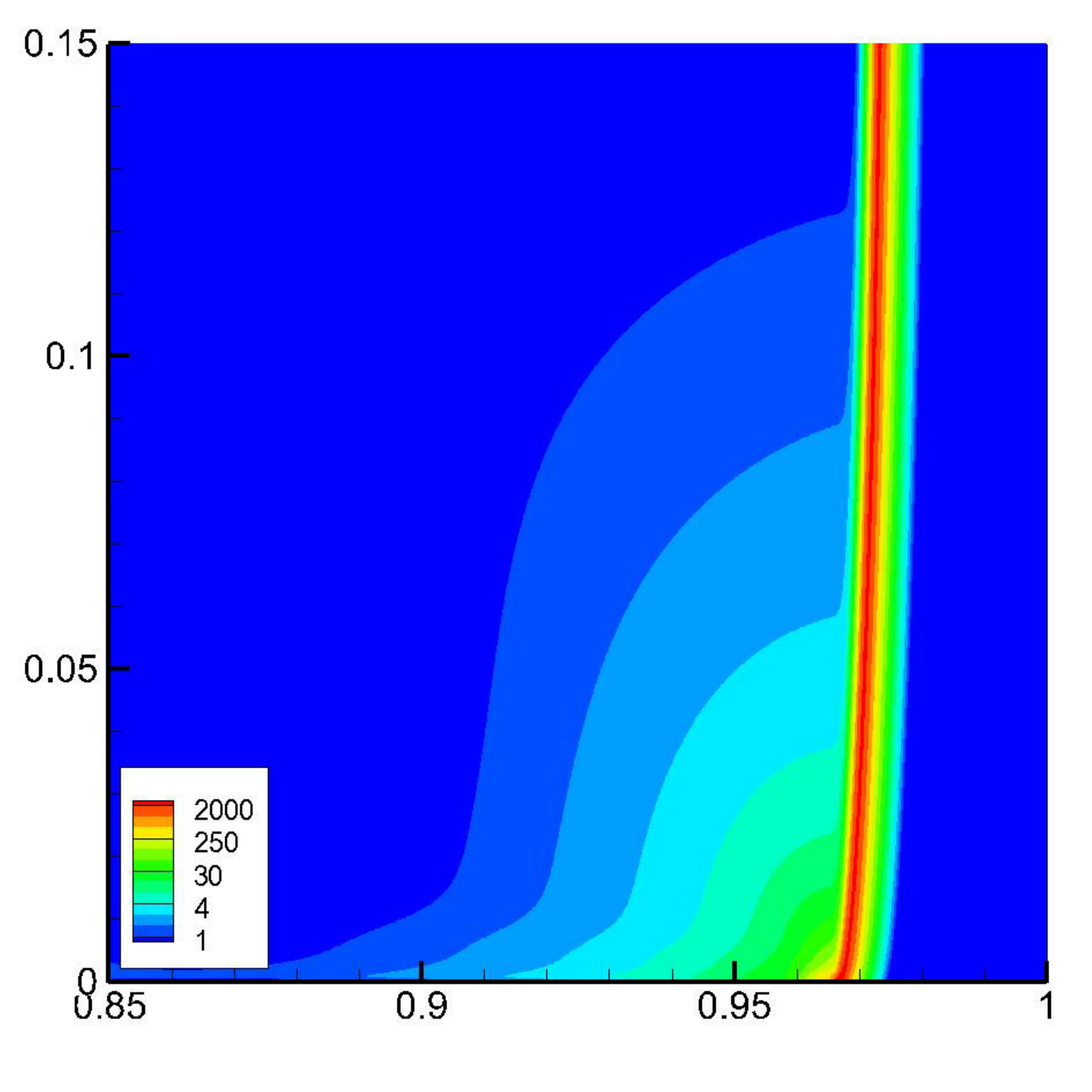}
\centerline{\footnotesize (c)}
\end{minipage}
\centering
\caption{Flow field at $t=0.2$: (\textit{a}) $V$-velocity; (\textit{b}) density; (\textit{c}) pressure gradient magnitude.}\label{downward}
\end{figure}

\begin{figure}
\centering
\includegraphics[width=0.4\textwidth]{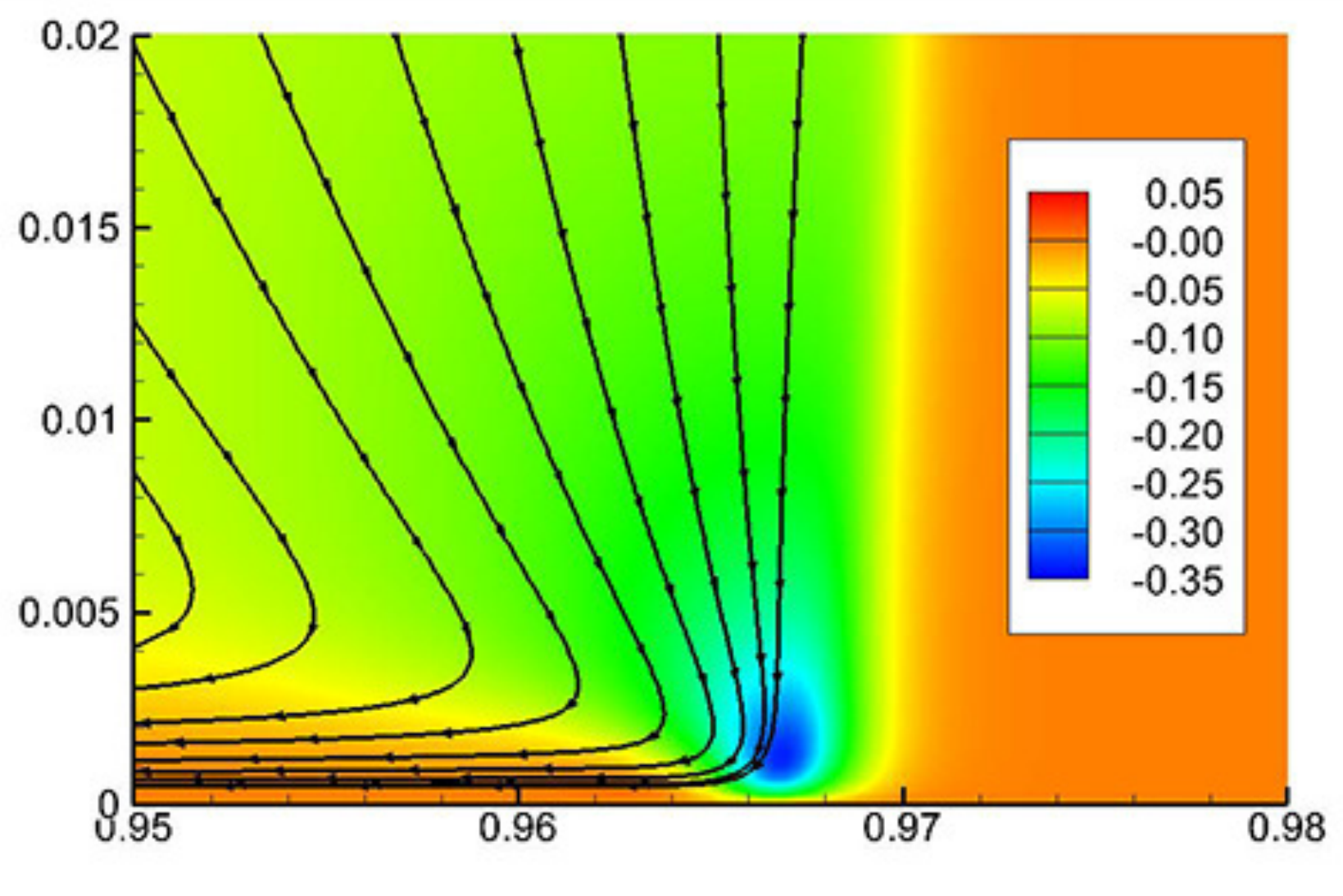}
\centering
\caption{Streamlines under a Galilean transform and distribution of $V$-velocity at $t=0.2$.}\label{Galileo}
\end{figure}

At about $t = 0.21$, the right-travelling shock wave encounters the end wall and is then reflected by it. As the shock is curved, it reaches the wall successively from upper parts to lower parts. Figure~\ref{ypg_reflect} presents three snapshots around the time of reflection. In figure~\ref{ypg_reflect}(a), the upper part shown in the plot has just moved to the wall; in figure~\ref{ypg_reflect}(b), the upper part has been reflected back while the lower part just touches the wall; in figure~\ref{ypg_reflect}(c), the lower part has also completed the reflection. Since theoretically the horizontal velocity of the flow in the region behind a reflected normal shock is zero, the downward-concentrating effect of the curved shock can be observed very obviously in figure~\ref{ypg_reflect}(b) and \ref{ypg_reflect}(c). It is clear from the streamlines that the fluid flows to the lower-right corner from upper regions behind the reflected shock wave. However, we emphasise that this process started from the very beginning: A region with negative velocity in the $y$-direction always exists after the shock wave is generated, see figure~\ref{downward}(a). The gathering of flow near the root of the shock makes the density there larger, as shown in figure~\ref{downward}(b). To get a better view, a Galilean transform is made at $t=0.2$: A constant is subjected from the $U$ velocity in the flow field, so that the $V$ velocity is shown more clearly. The streamlines after transformation are presented in figure~\ref{Galileo}. It demonstrates how the fluid is moving to the bottom wall. This process has no essential difference with the phenomenon  behind the reflected shock shown in figure~\ref{ypg_reflect}.

\begin{figure}
\centering
\begin{minipage}[t]{0.25\textwidth}
\centering
\includegraphics[width=\textwidth]{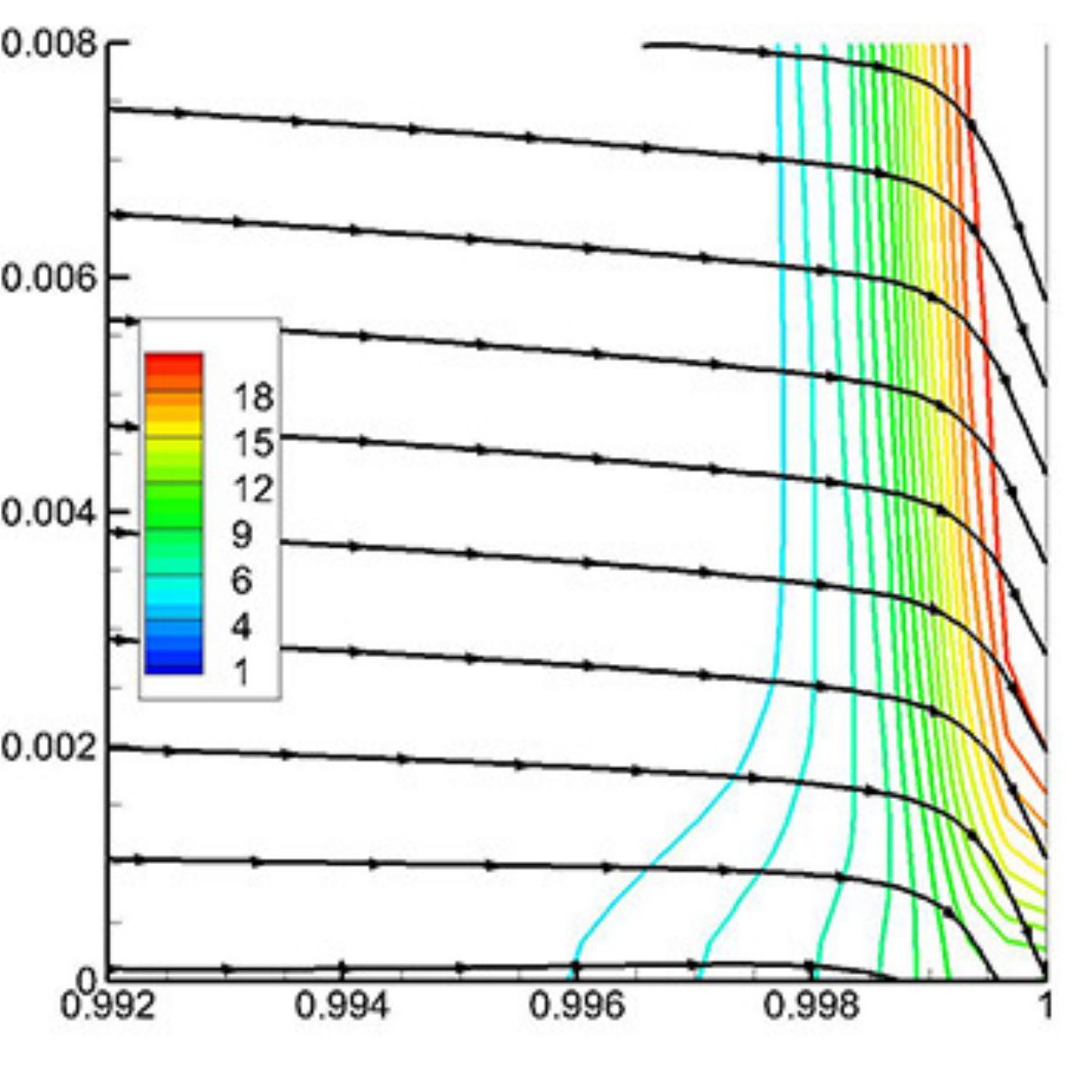}
\centerline{\footnotesize (a)}
\end{minipage}%
\begin{minipage}[t]{0.25\textwidth}
\centering
\includegraphics[width=\textwidth]{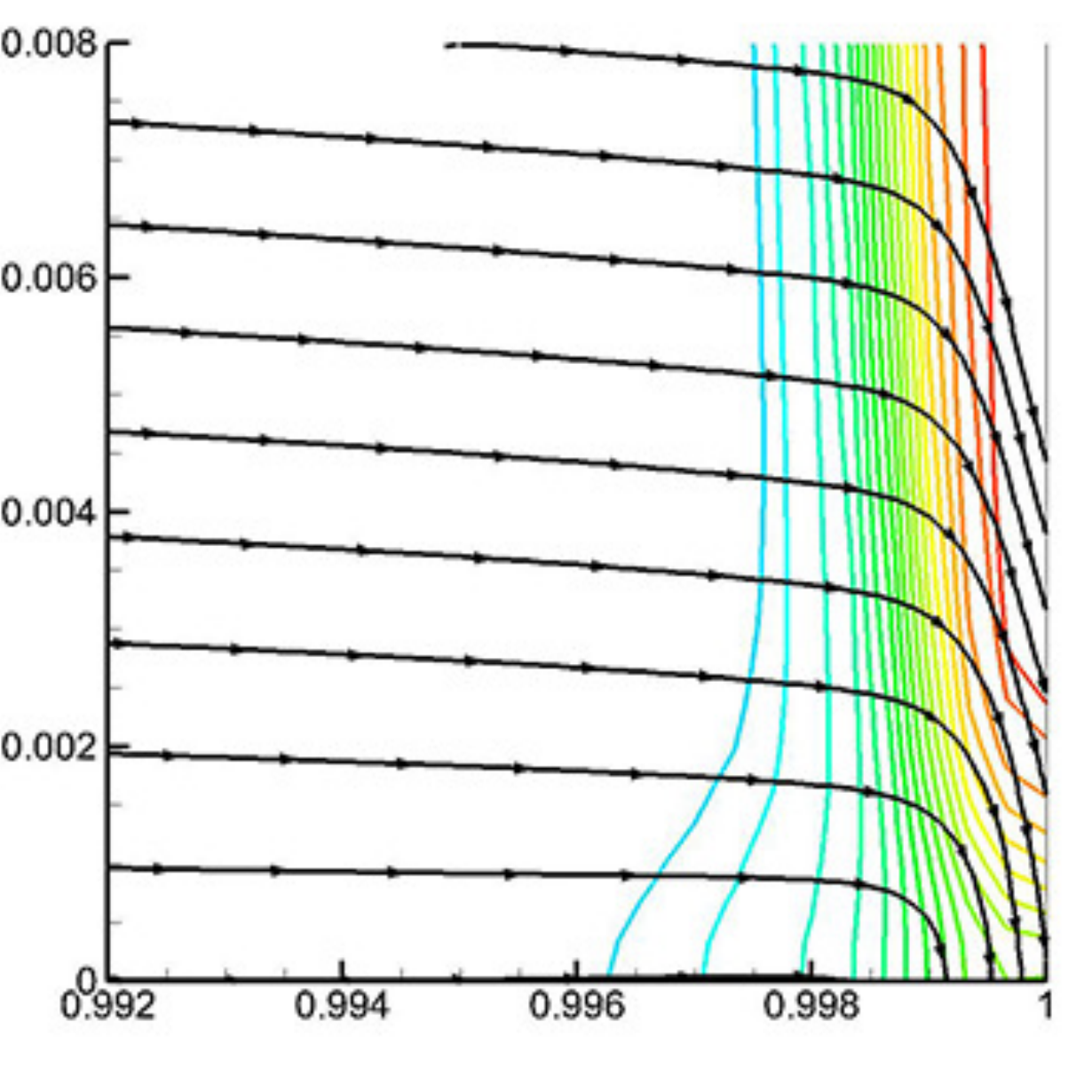}
\centerline{\footnotesize (b)}
\end{minipage}%
\begin{minipage}[t]{0.25\textwidth}
\centering
\includegraphics[width=\textwidth]{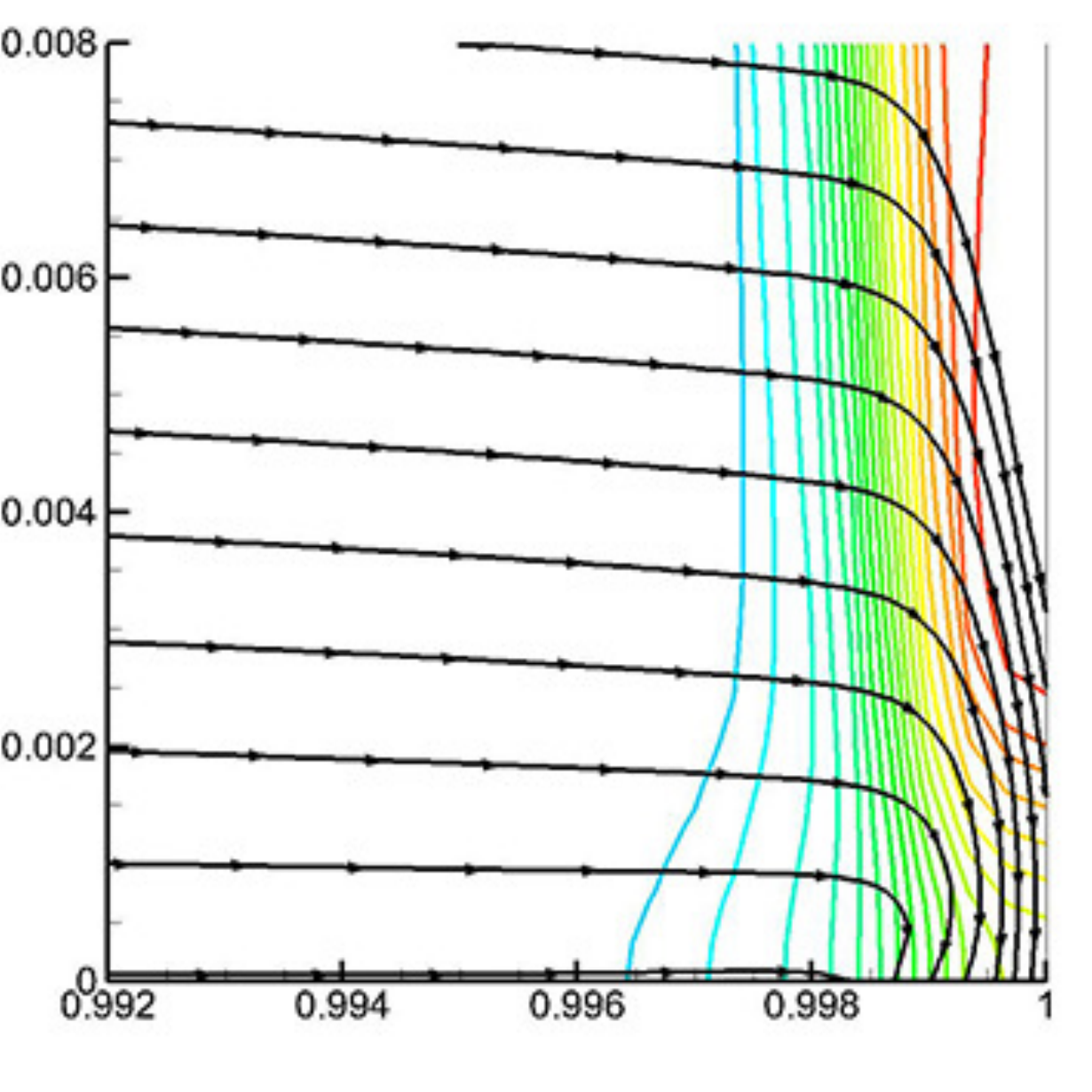}
\centerline{\footnotesize (c)}
\end{minipage}%
\begin{minipage}[t]{0.25\textwidth}
\centering
\includegraphics[width=\textwidth]{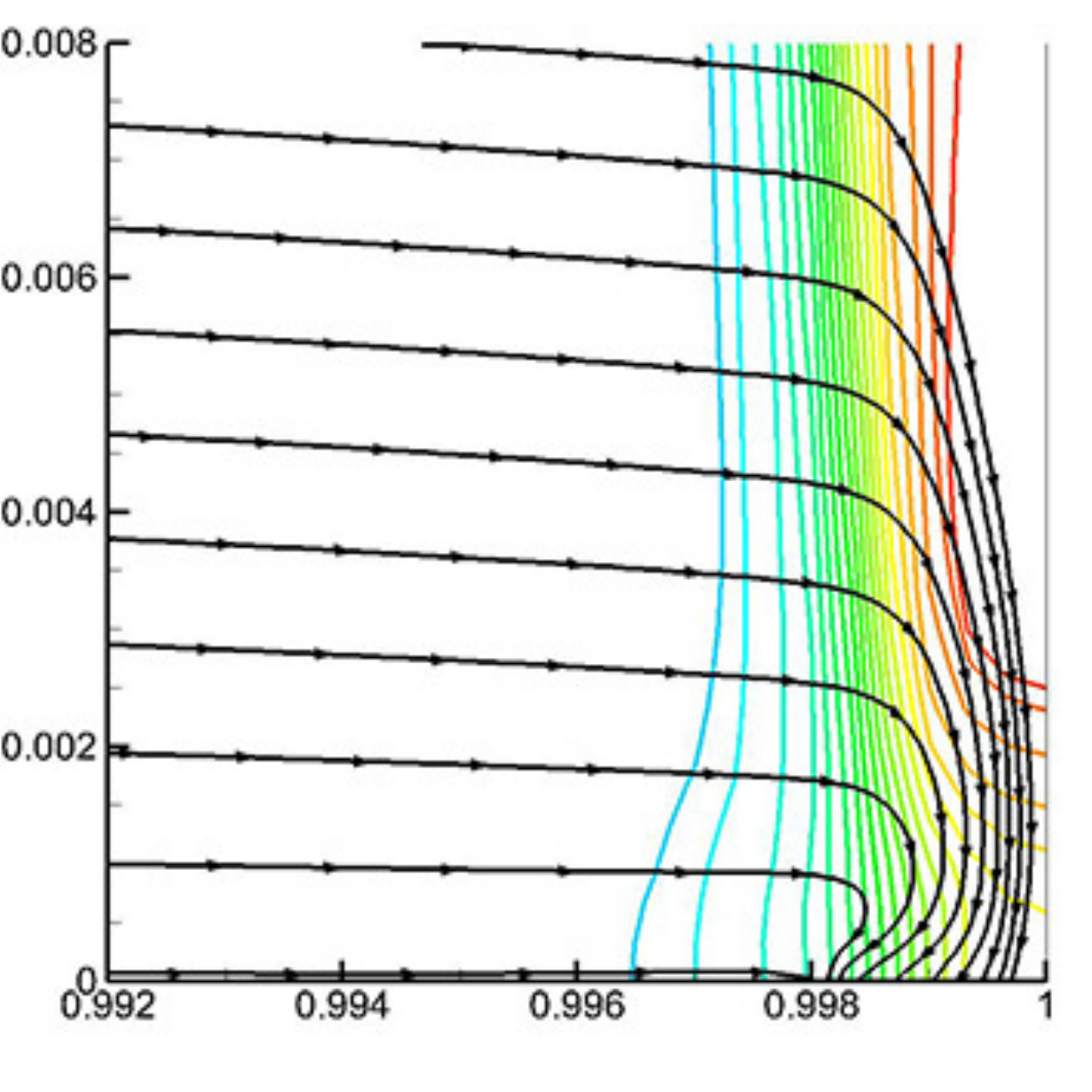}
\centerline{\footnotesize (d)}
\end{minipage}
\centering
\caption{Streamlines and pressure contour lines at (\textit{a}) $t = 0.2144$,  (\textit{b}) $t = 0.2146$, \protect\\ (\textit{c}) $t = 0.2148$ and (\textit{d}) $t = 0.2150$.}\label{p_reflect}
\end{figure}

We will then focus on the flow in the lower right corner. It is seen from figure~\ref{downward}(c) that the shock wave disperses near the bottom wall due to viscous effect. Hence it is more like a sequence of compressible waves in this region. In addition, the shock is very curved there and the strength in the $x$-direction is then weakened. As a consequence, the reflected wave in the near-wall region is not as strong as that in the upper region where the incident shock is thin and normal to the right wall. This effect creates a pressure gradient pointing to the lower left direction, see figure~\ref{p_reflect}. Driven by such a pressure gradient, the downward flow alters its direction to the left. Figures~\ref{p_reflect}(a) to \ref{p_reflect}(d) display the process how the streamlines adjust to the perpendicular direction to the pressure contour lines.

\begin{figure}
\centering
\begin{minipage}[t]{0.33\textwidth}
\centering
\includegraphics[width=\textwidth]{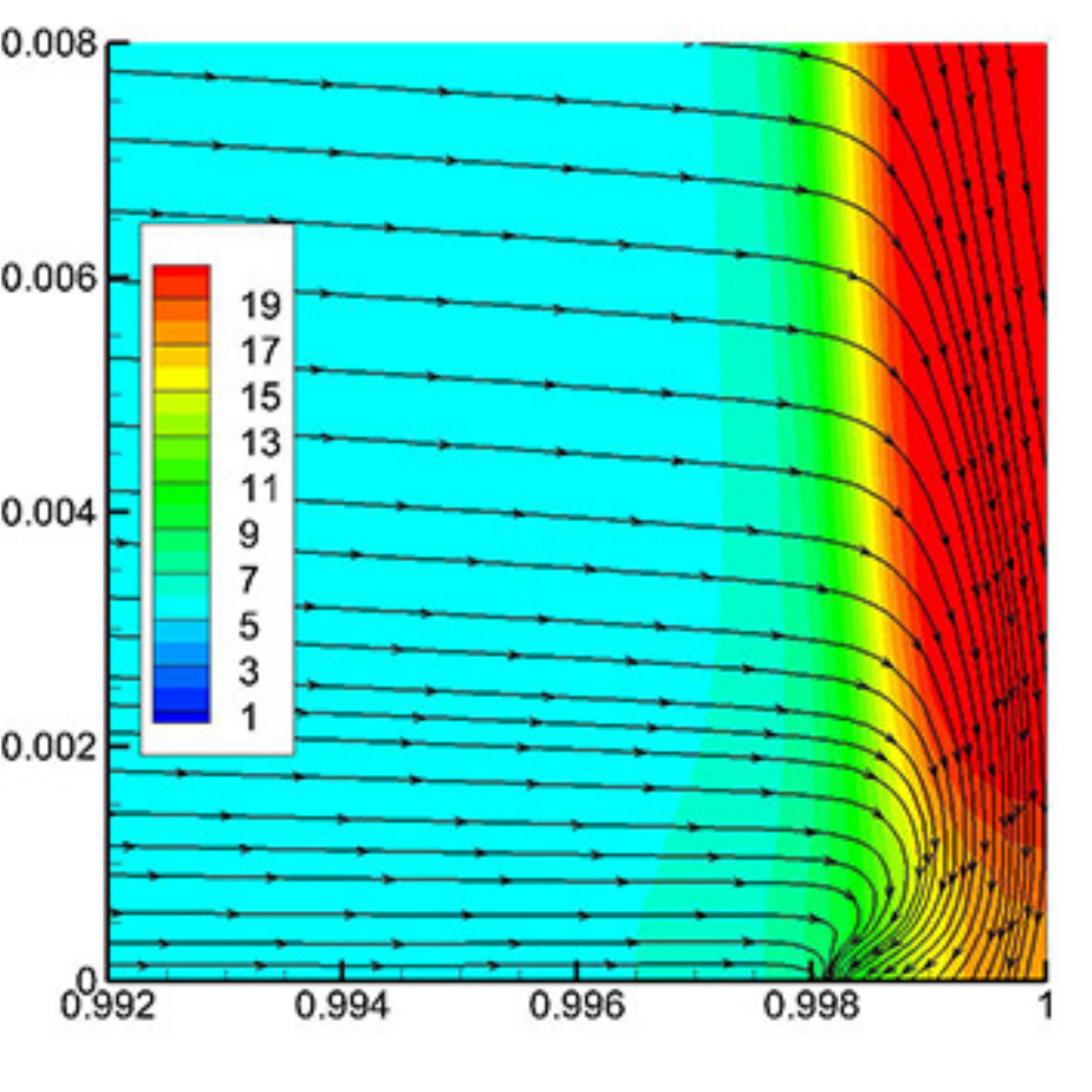}
\centerline{\footnotesize (a)}
\end{minipage}%
\begin{minipage}[t]{0.33\textwidth}
\centering
\includegraphics[width=\textwidth]{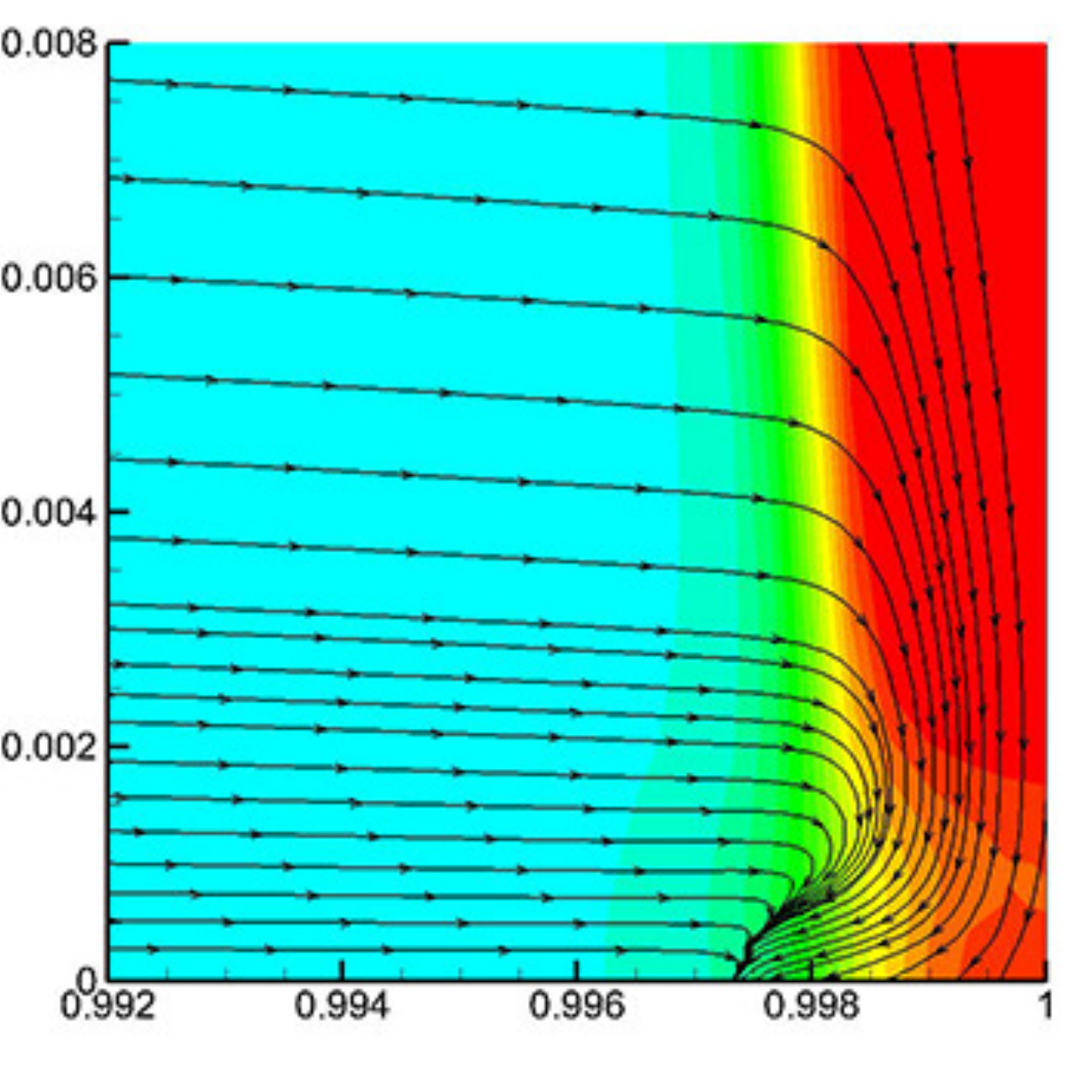}
\centerline{\footnotesize (b)}
\end{minipage}%
\begin{minipage}[t]{0.33\textwidth}
\centering
\includegraphics[width=\textwidth]{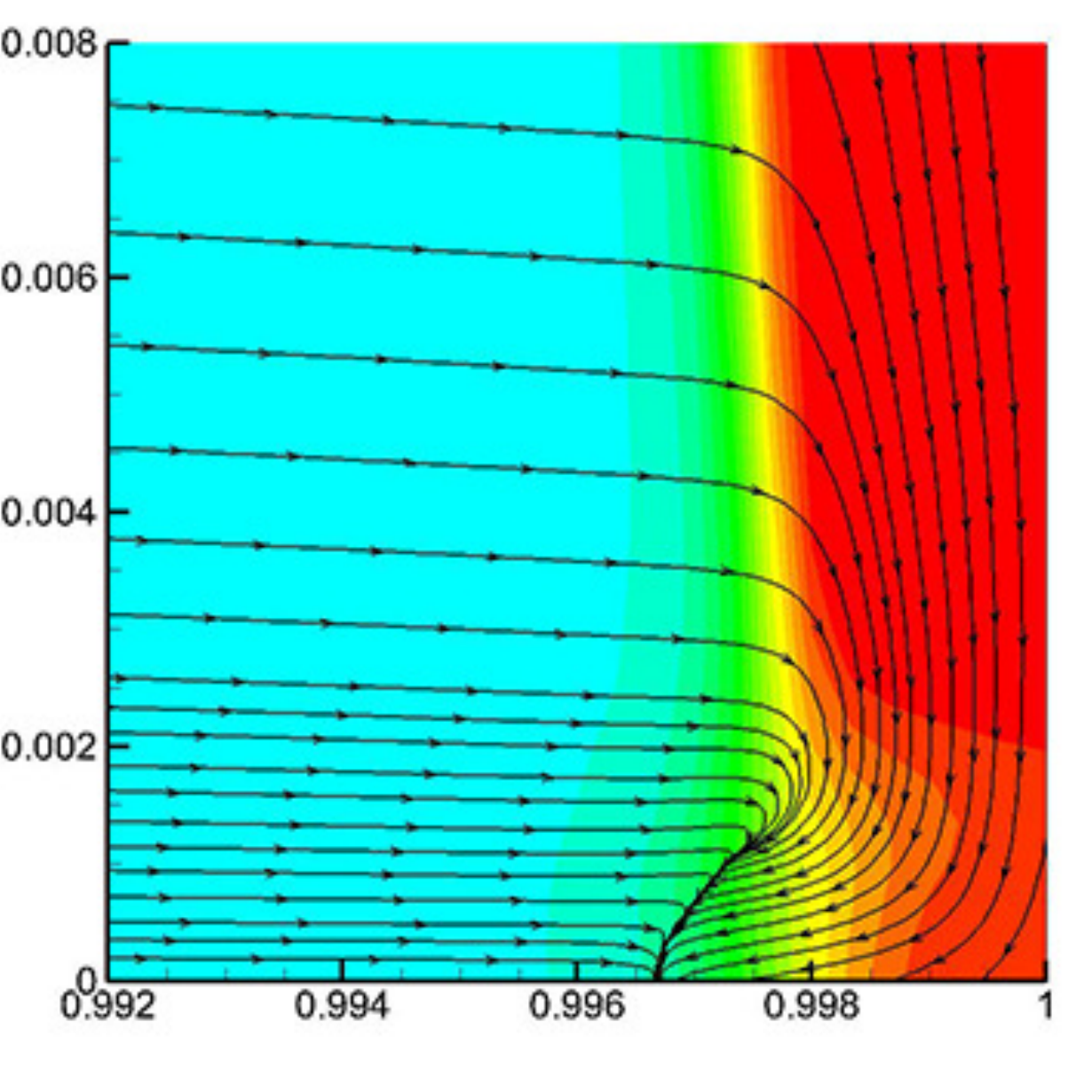}
\centerline{\footnotesize (c)}
\end{minipage}
\begin{minipage}[t]{0.33\textwidth}
\centering
\includegraphics[width=\textwidth]{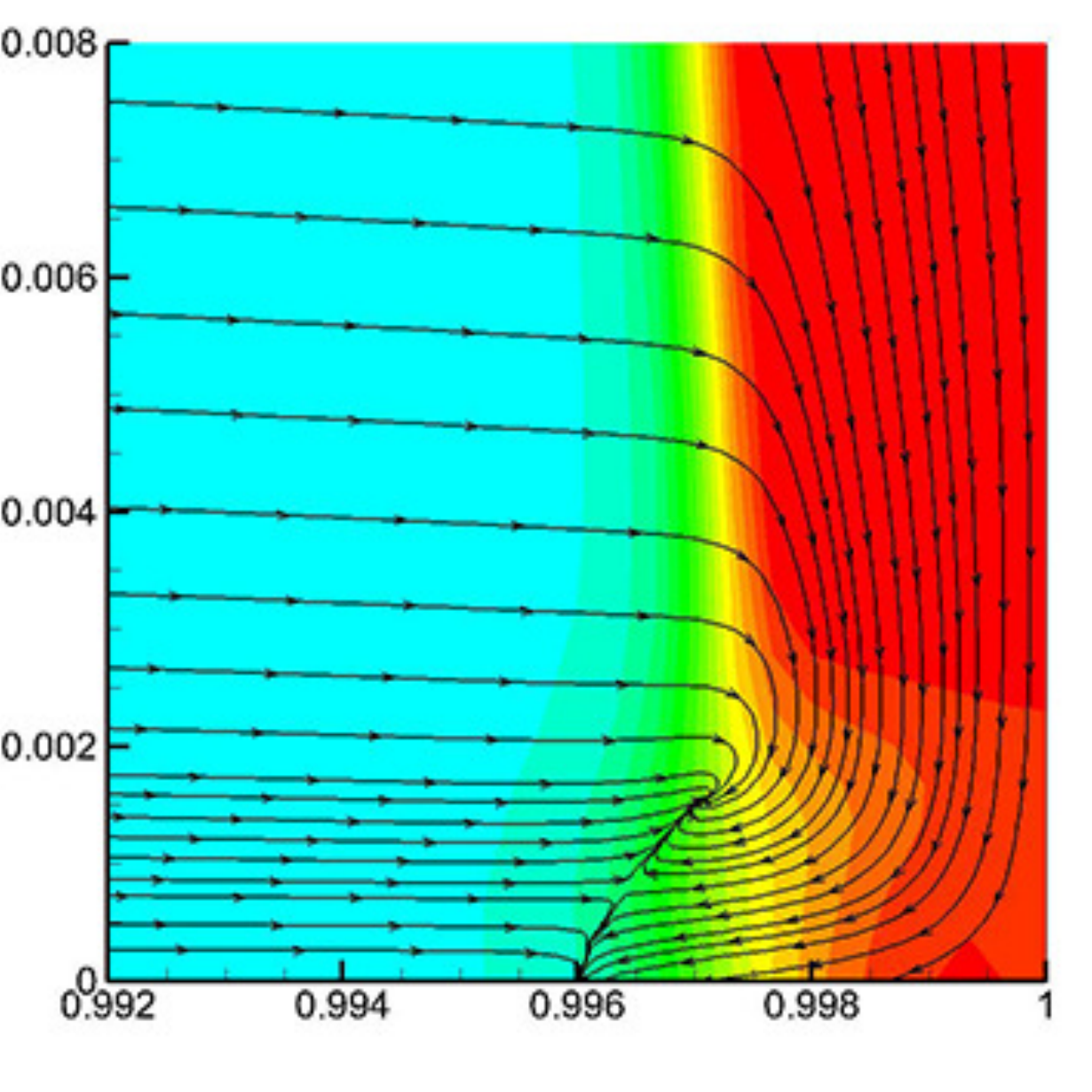}
\centerline{\footnotesize (d)}
\end{minipage}%
\begin{minipage}[t]{0.33\textwidth}
\centering
\includegraphics[width=\textwidth]{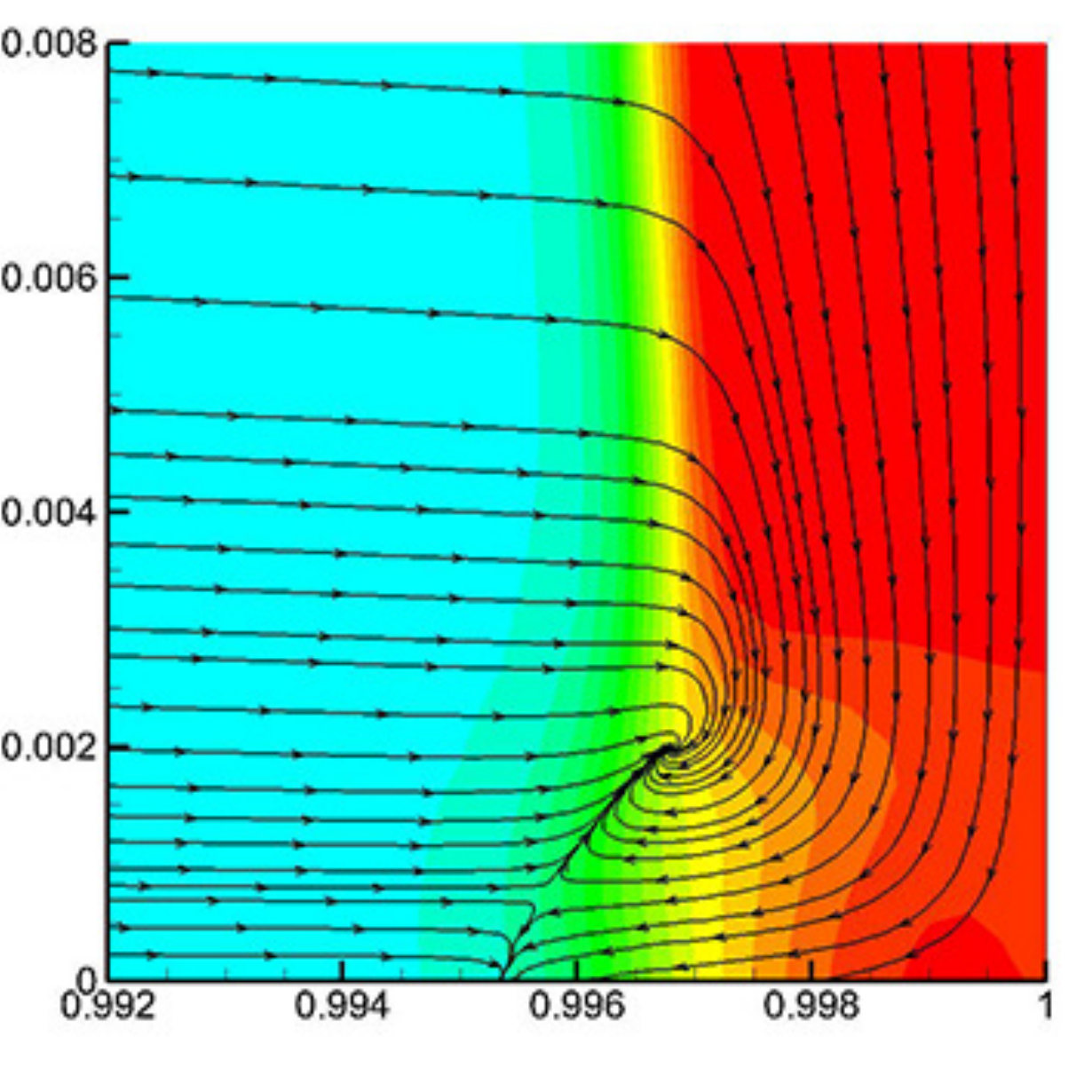}
\centerline{\footnotesize (e)}
\end{minipage}%
\begin{minipage}[t]{0.33\textwidth}
\centering
\includegraphics[width=\textwidth]{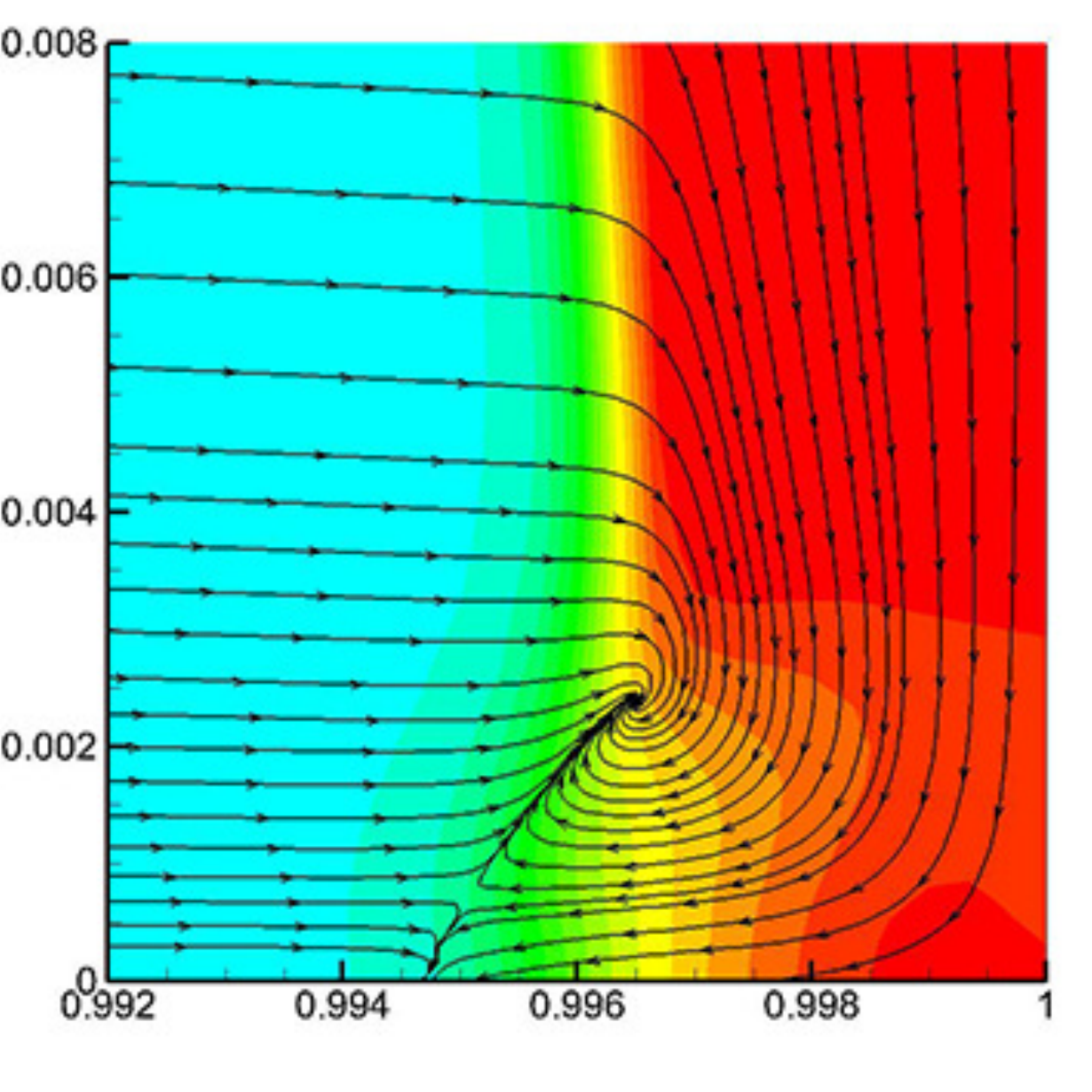}
\centerline{\footnotesize (f)}
\end{minipage}
\centering
\caption{Streamlines and pressure distribution at (\textit{a}) $t=0.2150$, (\textit{b}) $t=0.2154$, \protect\\ (\textit{c}) $t=0.2158$, (\textit{d}) $t=0.2162$, (\textit{e}) $t=0.2166$ and (\textit{f}) $t=0.2170$.}\label{vortex}
\end{figure}

The reversed flow at the corner shown in figure~\ref{p_reflect}(d) soon encounters the incident flow around the position of the left edge of the reflected shock. With continuous supply of fluid, an oblique separation line forms and gets longer between the two parts of the fluid. This process is shown in figure~\ref{vortex}. In the last three snapshots of figure~\ref{vortex} we can see that the fluid beside the separation line is forced to flow downward or upward, generating two sink points at the ends of the separation line and a saddle point in the middle.

\begin{figure}
\centering
\begin{minipage}[t]{0.33\textwidth}
\centering
\includegraphics[width=\textwidth]{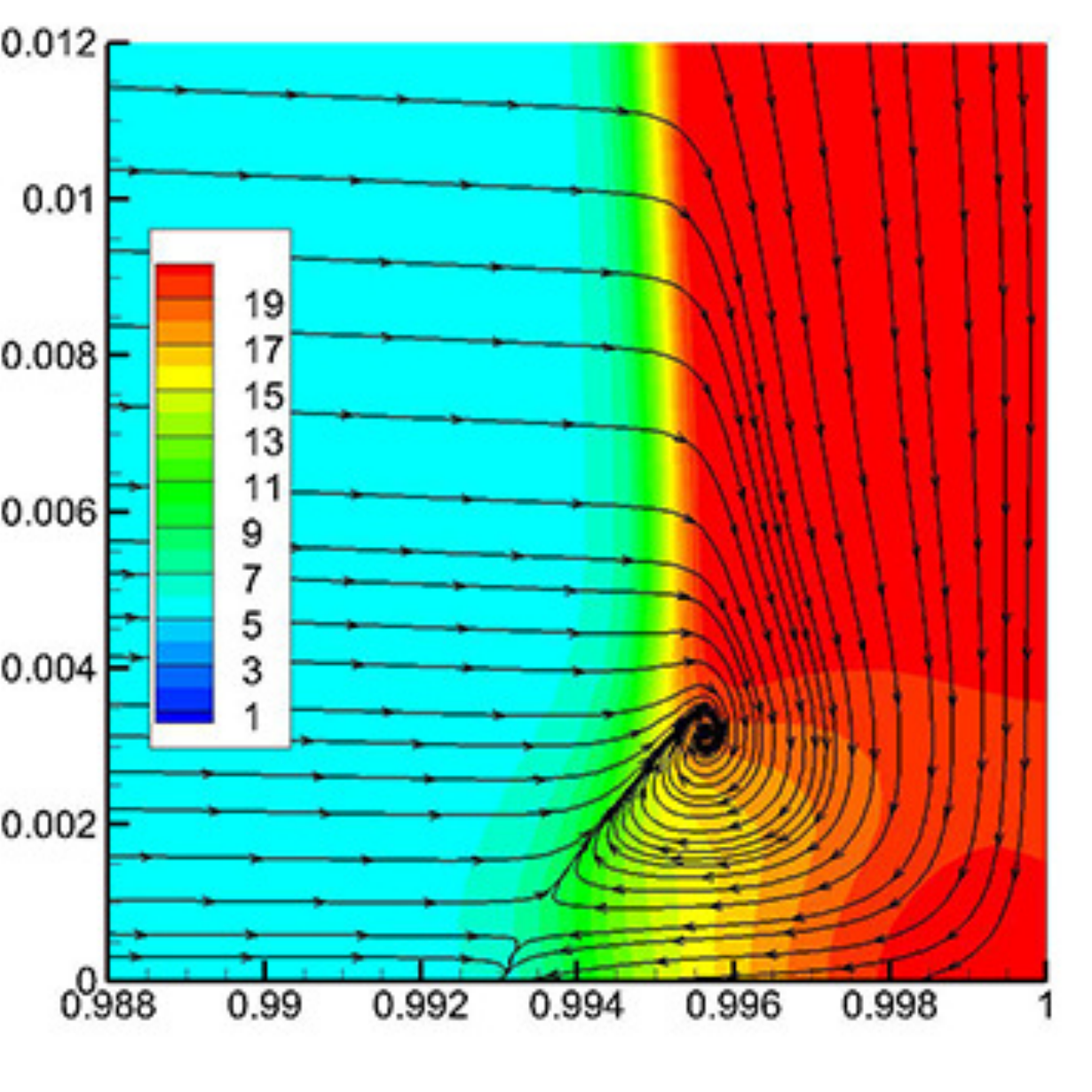}
\centerline{\footnotesize (a)}
\end{minipage}%
\begin{minipage}[t]{0.33\textwidth}
\centering
\includegraphics[width=\textwidth]{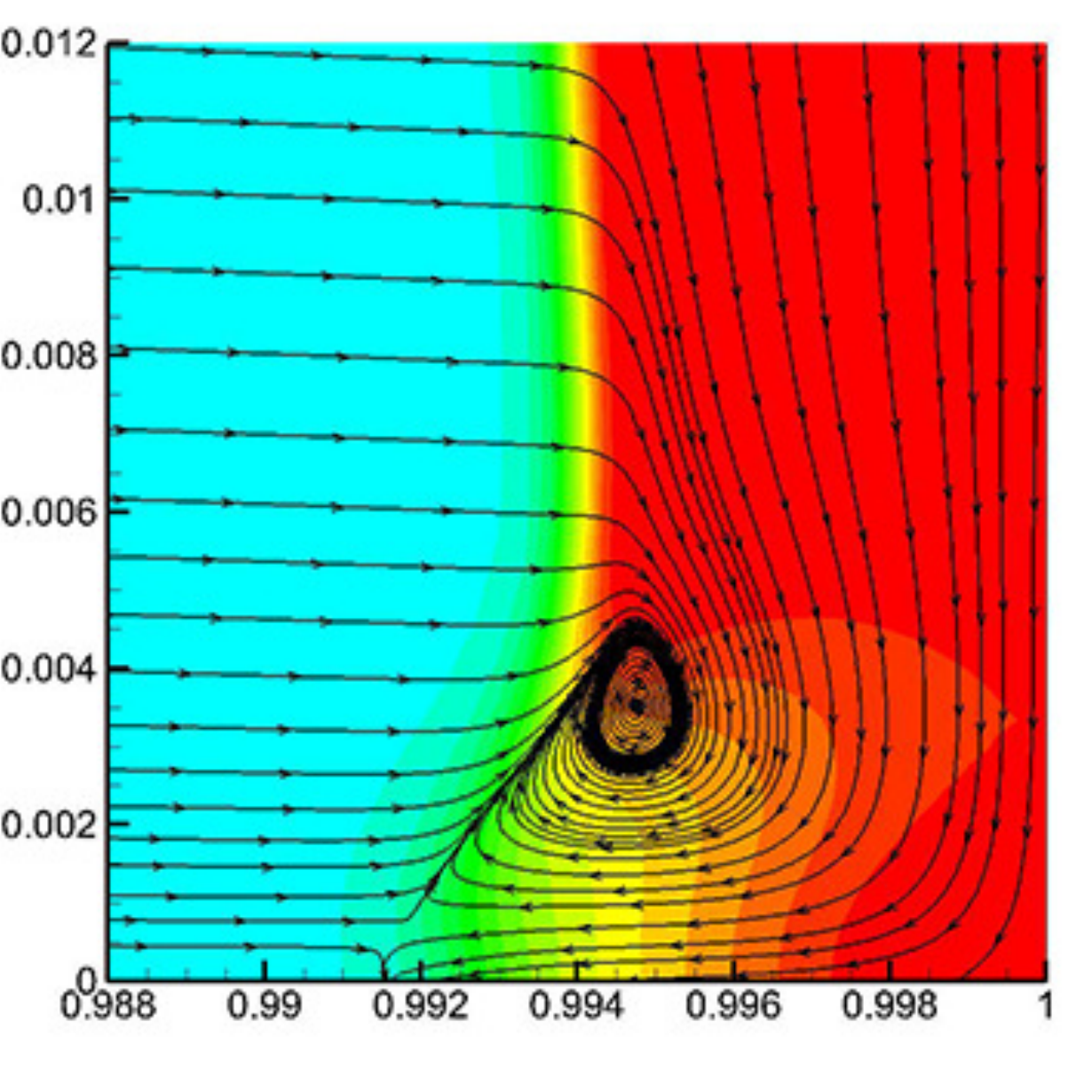}
\centerline{\footnotesize (b)}
\end{minipage}%
\begin{minipage}[t]{0.33\textwidth}
\centering
\includegraphics[width=\textwidth]{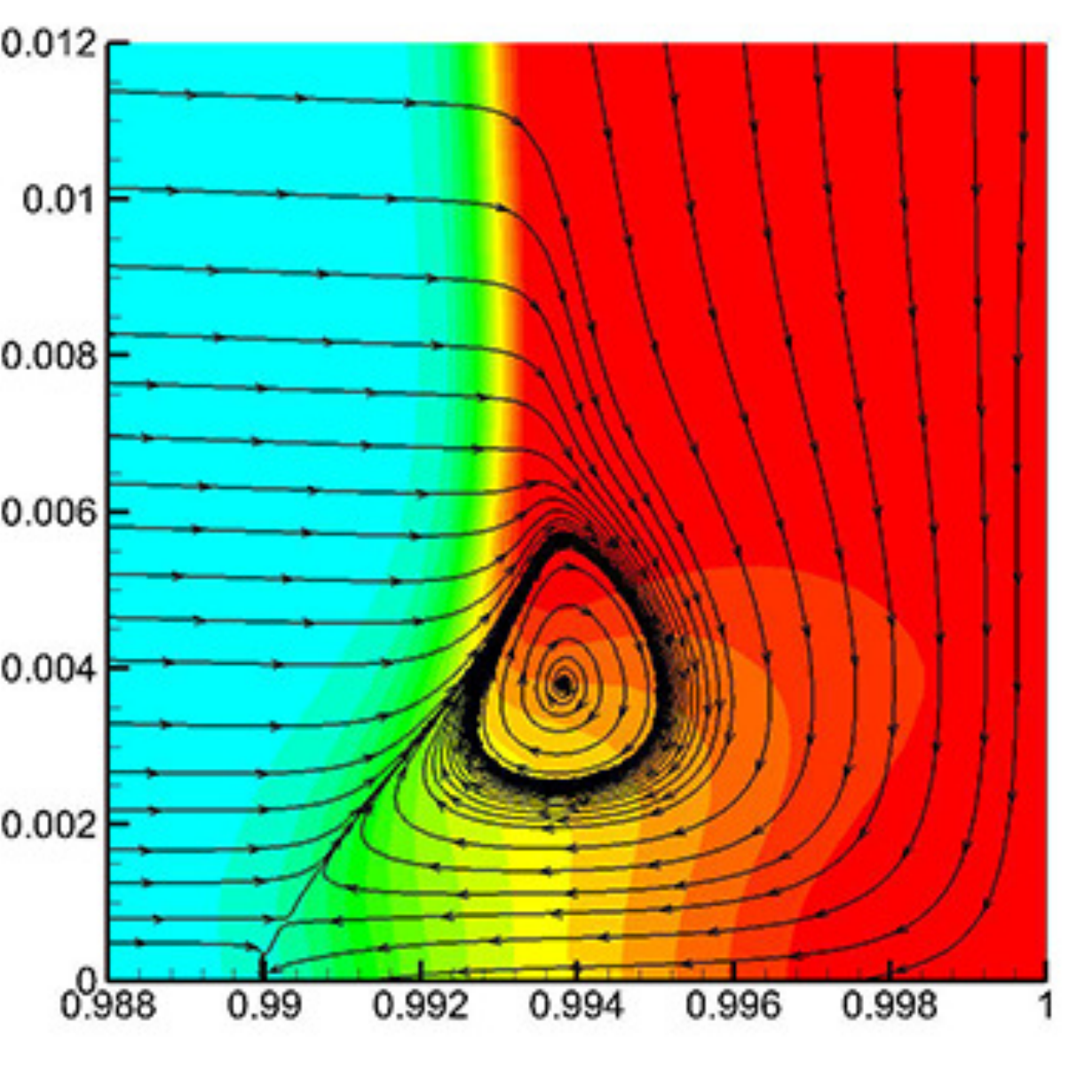}
\centerline{\footnotesize (c)}
\end{minipage}
\centering
\caption{Streamlines and pressure distribution at (\textit{a}) $t=0.218$, (\textit{b}) $t=0.219$ and \protect\\ (\textit{c}) $t=0.220$.}\label{vortex2}
\end{figure}

With the lifting of the upper sink point, its distance to the bottom wall increases, hence the fluid around the sink point has larger velocity and momentum. In this situation, the streamlines roll up forming a vortex around the point, which gets larger in size with entrainment of more fluids, see figure~\ref{vortex2}. Notice that the streamlines and the pressure contour lines finally adjust to be orthogonal to each other.

\begin{figure}
\centering
\begin{minipage}[t]{0.4\textwidth}
\centering
\includegraphics[width=0.8\textwidth]{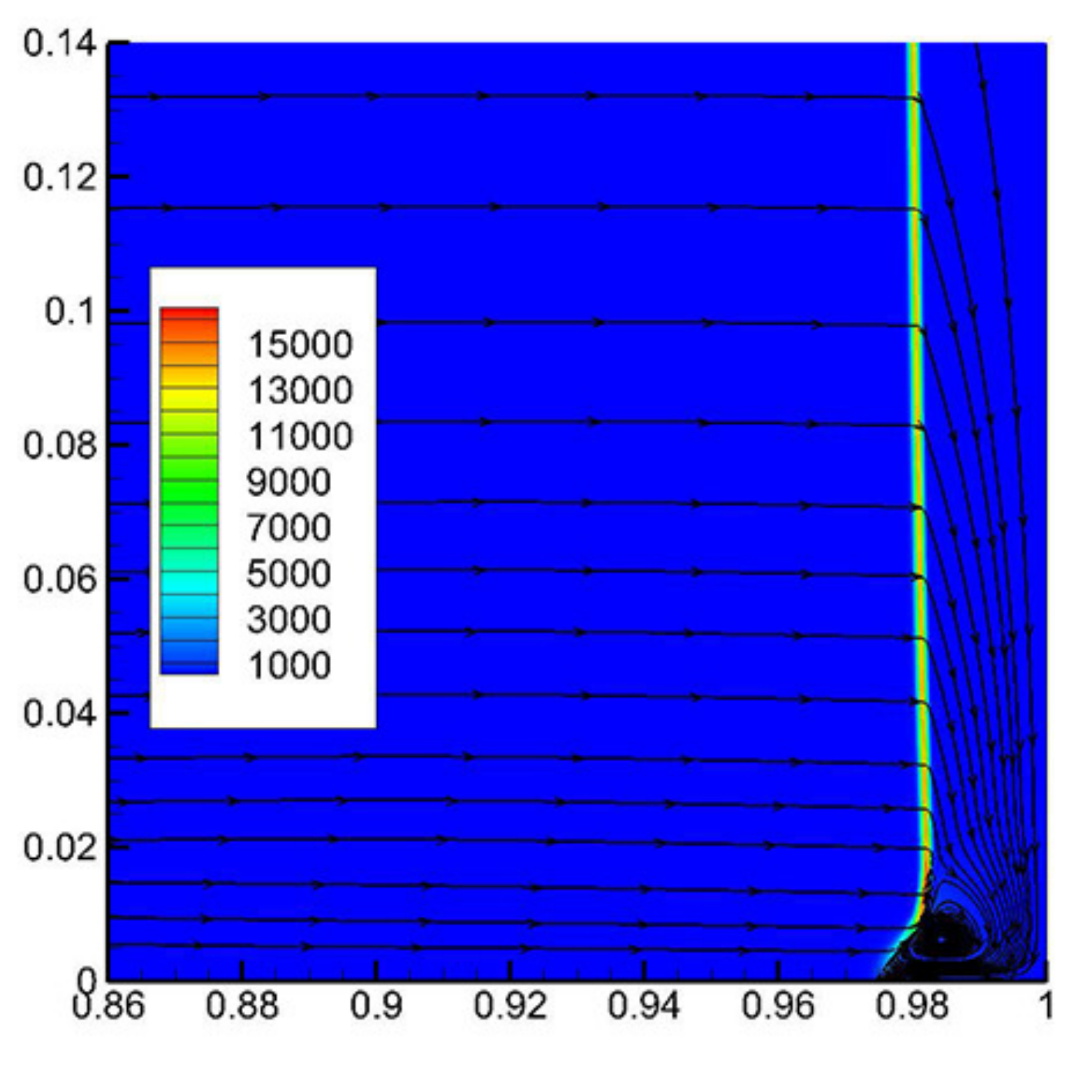}
\centerline{\footnotesize (a)}
\end{minipage}%
\begin{minipage}[t]{0.4\textwidth}
\centering
\includegraphics[width=0.8\textwidth]{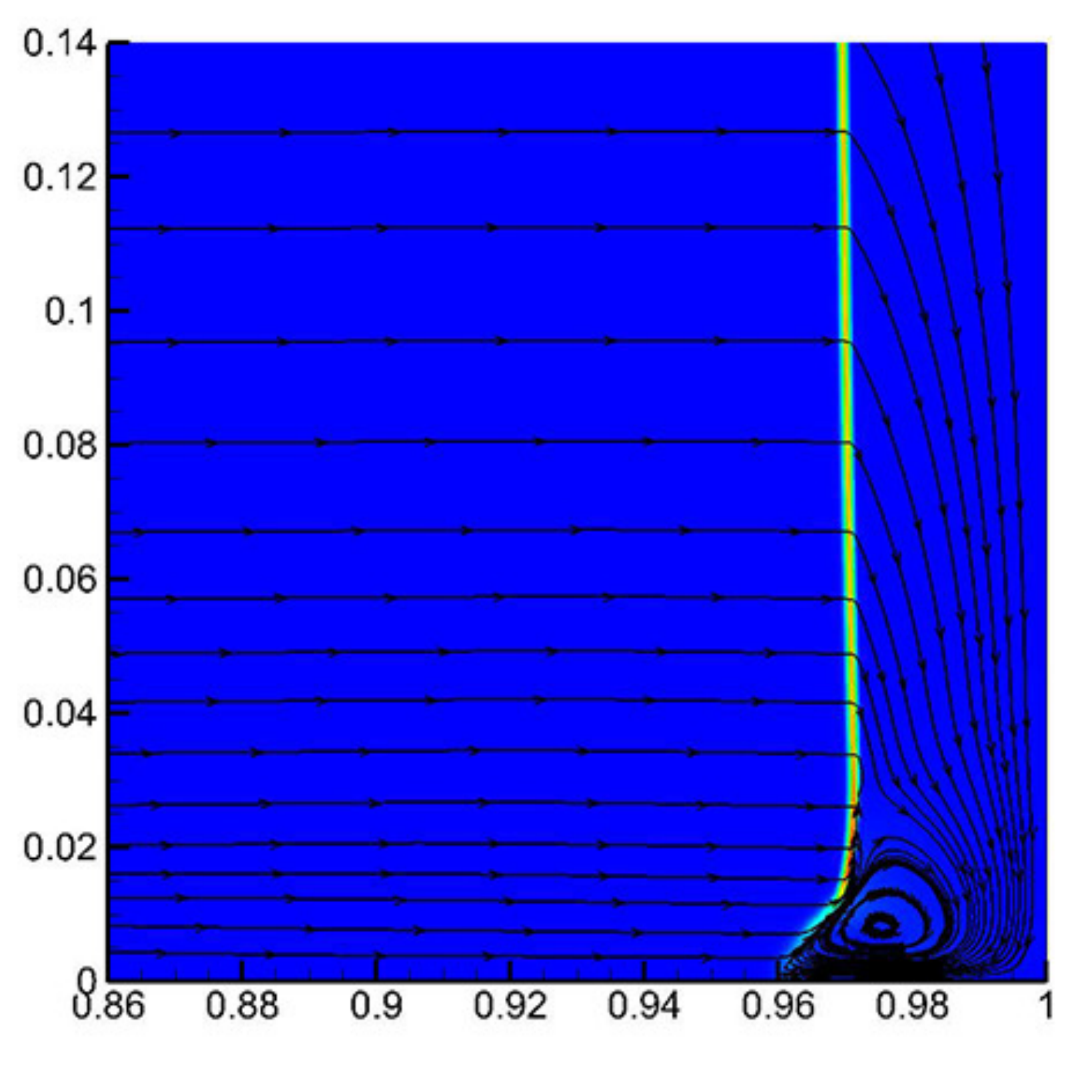}
\centerline{\footnotesize (b)}
\end{minipage}
\centering
\caption{Streamlines and distribution of the pressure gradient magnitude at (\textit{a}) $t = 0.23$ and (\textit{b}) $t = 0.24$.}\label{pgm_vortex}
\end{figure}

It is interesting that there is a close connection between the vortex and the oblique reflected shock wave. Notice that the left edge of the vortex is aligned with the oblique shock. The rotation of the vortex makes the difference on the left and right sides of the oblique shock larger so that the strength of the shock is enhanced. And the asymmetric pressure distribution in the direction parallel to the oblique shock caused by the vortex rotation makes the shock more oblique, as shown in figure~\ref{pgm_vortex}. On the other hand, after the flow passes the oblique shock, the normal component of the velocity decreases to near zero, while the tangential component remains unchanged. Therefore, the fluid behind the oblique reflected shock flows upwards along it, which is right in the same direction with the rotating flow in the vortex. This means that the oblique shock provides a momentum injection mechanism to the vortex and makes it larger and stronger.

\begin{figure}
\centering
\begin{minipage}[t]{0.25\textwidth}
\centering
\includegraphics[width=\textwidth]{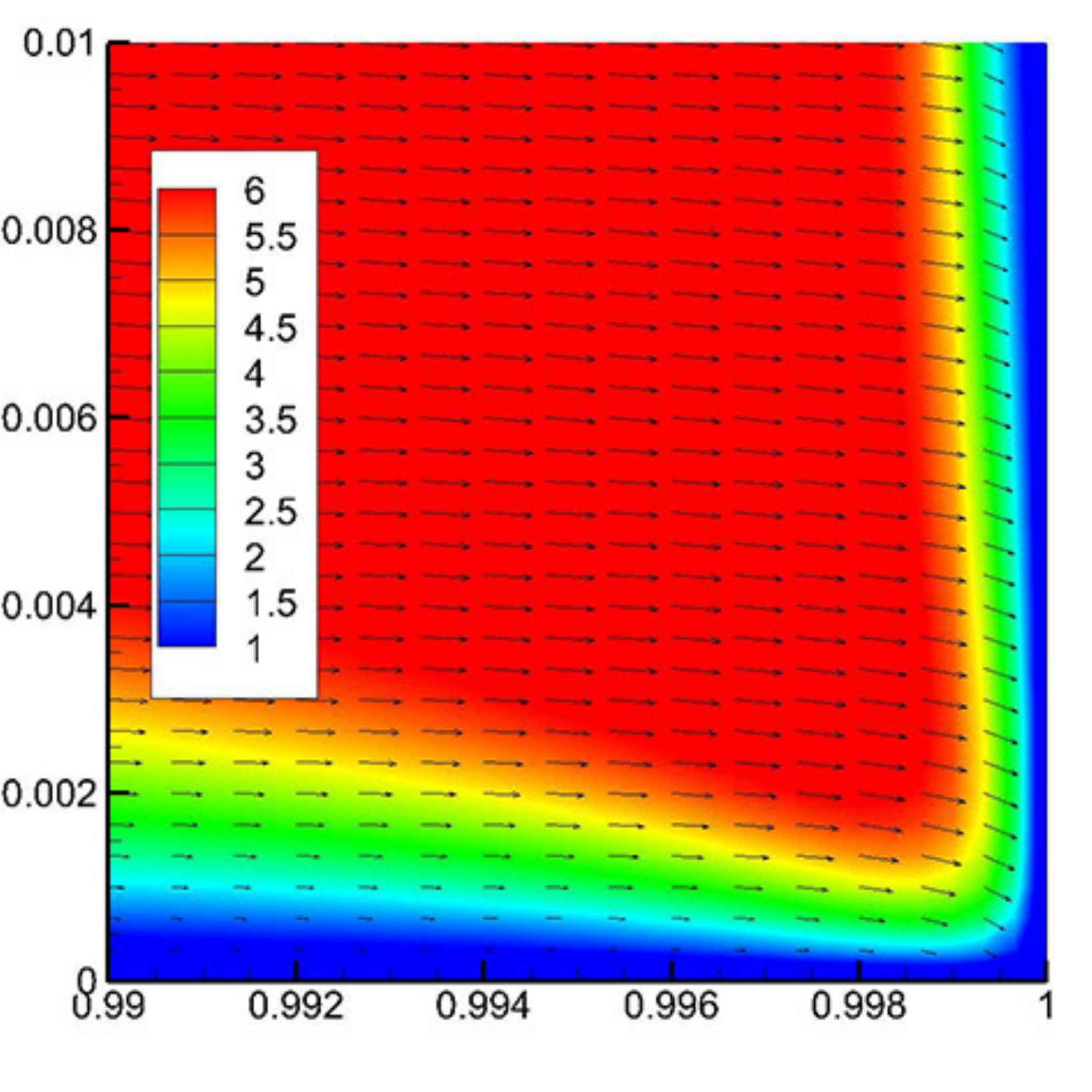}
\centerline{\footnotesize (a)}
\end{minipage}%
\begin{minipage}[t]{0.25\textwidth}
\centering
\includegraphics[width=\textwidth]{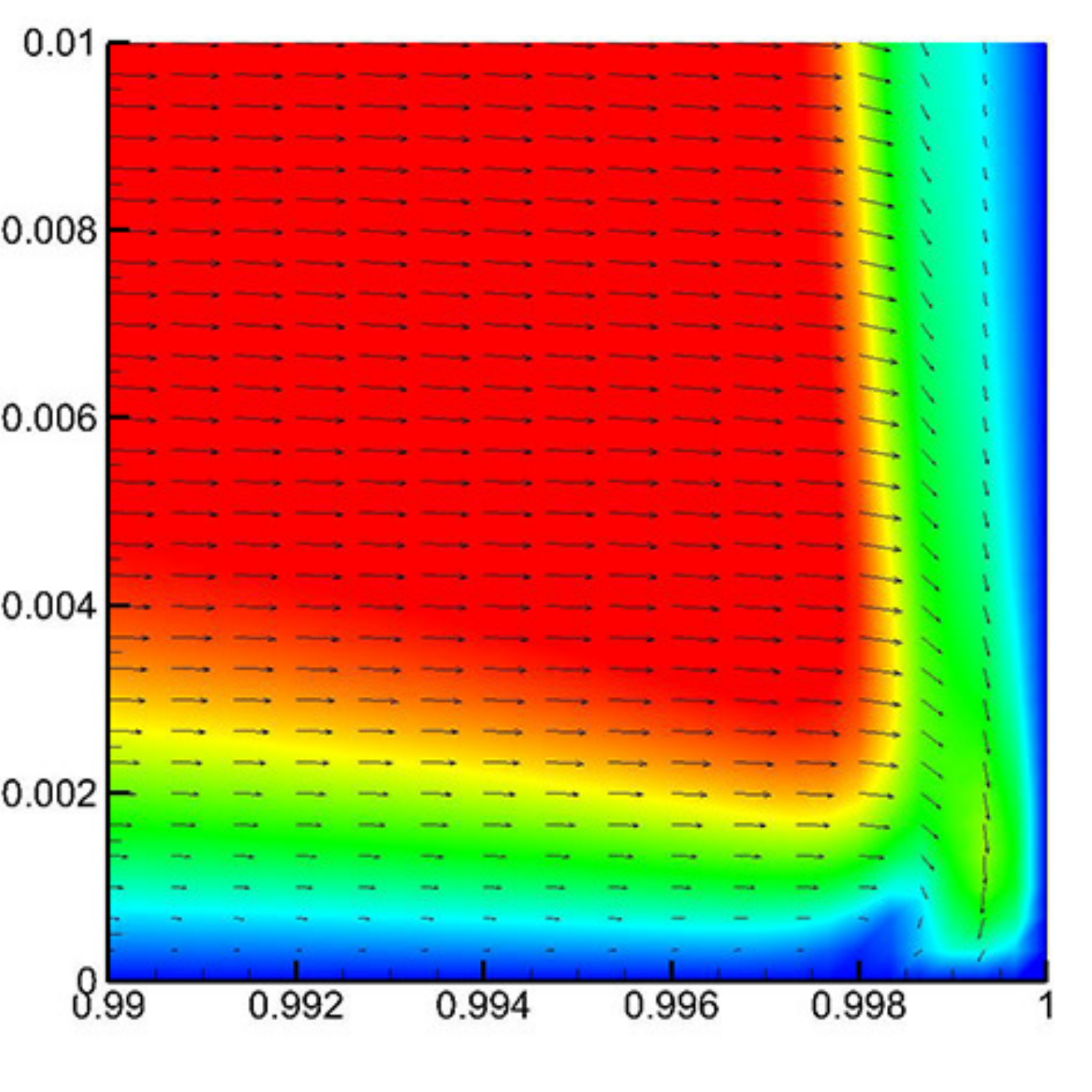}
\centerline{\footnotesize (b)}
\end{minipage}%
\begin{minipage}[t]{0.25\textwidth}
\centering
\includegraphics[width=\textwidth]{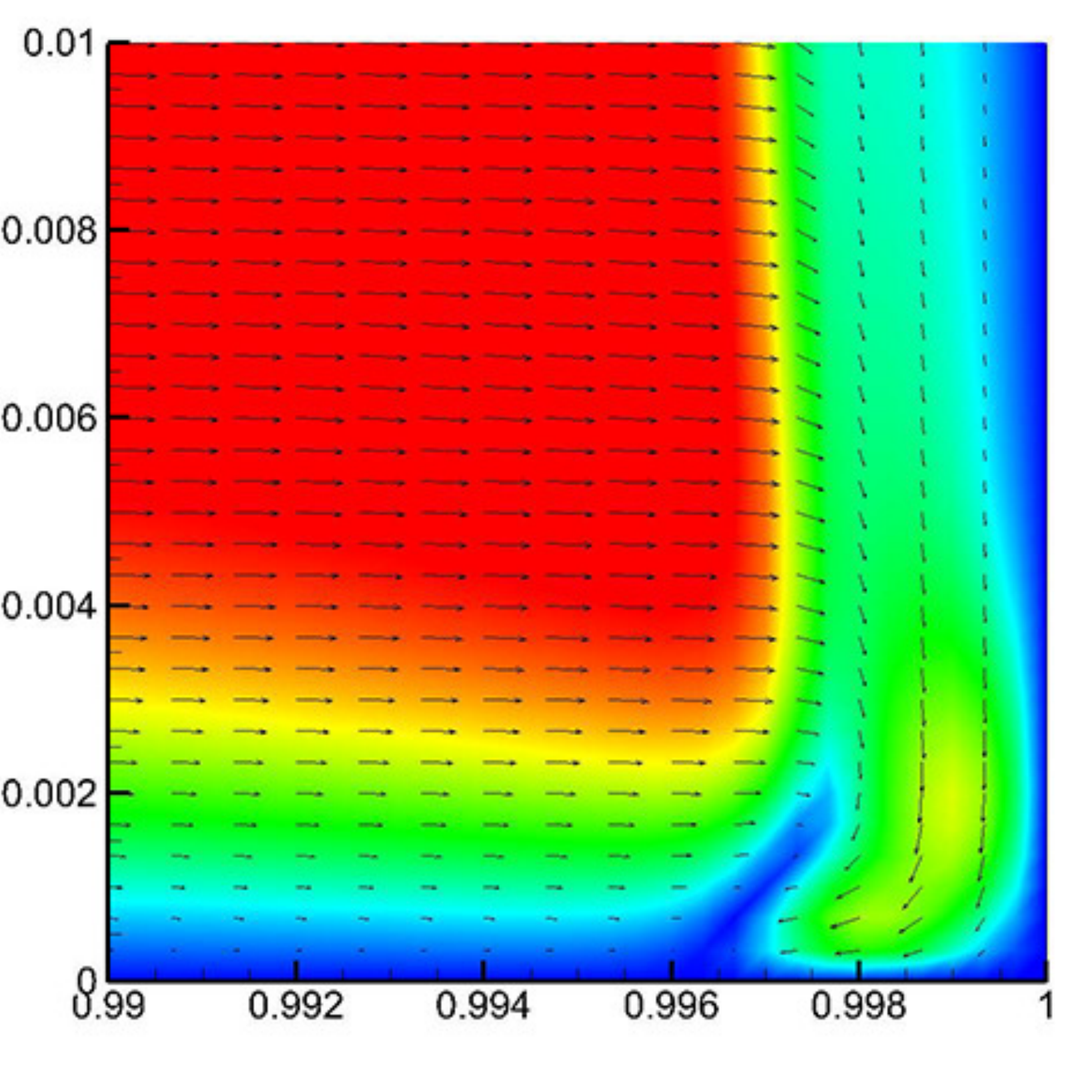}
\centerline{\footnotesize (c)}
\end{minipage}%
\begin{minipage}[t]{0.25\textwidth}
\centering
\includegraphics[width=\textwidth]{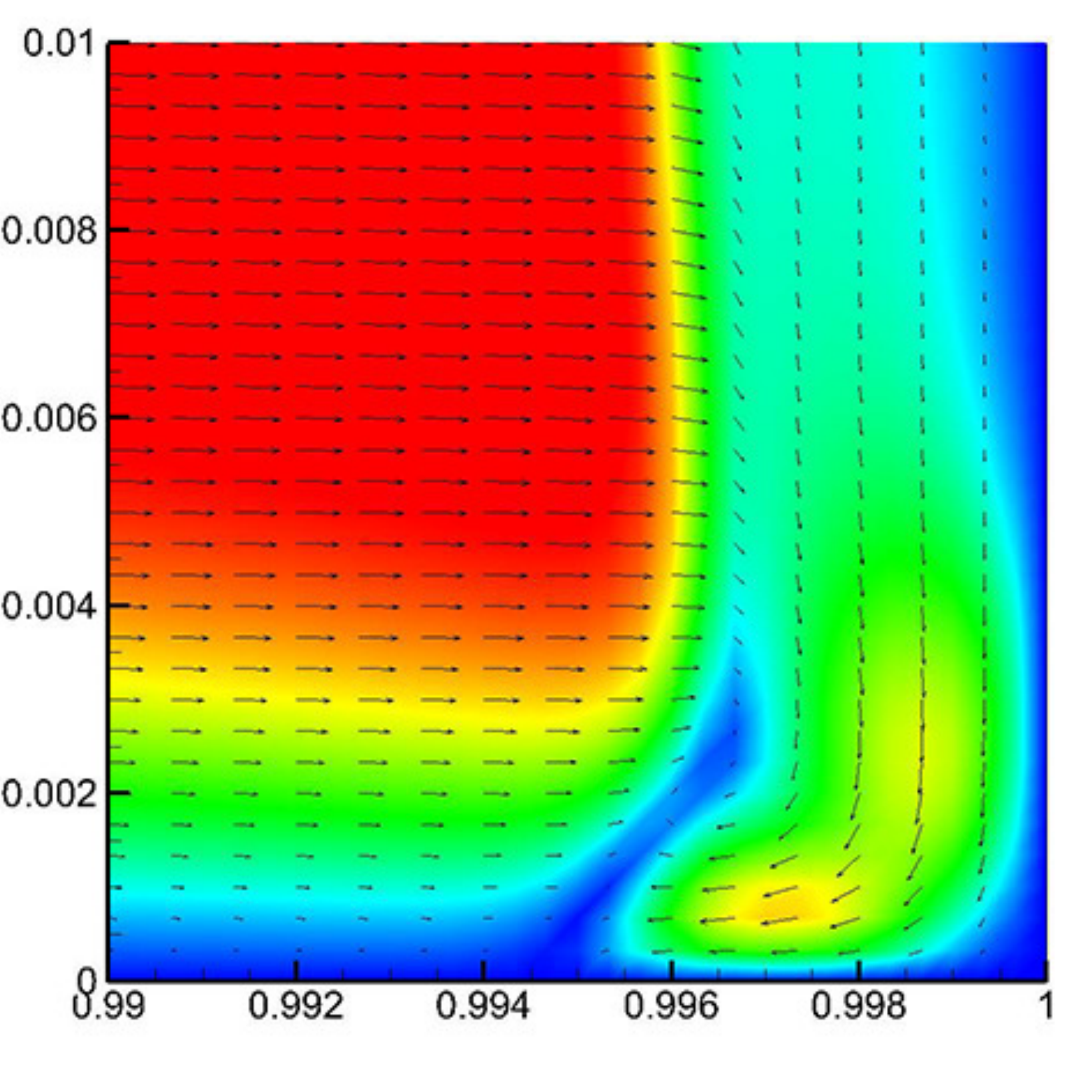}
\centerline{\footnotesize (d)}
\end{minipage}
\centering
\caption{Momentum vectors (every other point is plotted in the $x$-direction) and distribution of the momentum magnitude at (\textit{a}) $t=0.214$, (\textit{b}) $t=0.215$, (\textit{c}) $t=0.216$ and (\textit{d}) $t=0.217$.}\label{momentum}
\end{figure}

The process in this stage can also be interpreted in another view: The downward moving fluid behind the reflected shock wave carries higher momentum than the fluid in the boundary layer. Then it is easy for the former to insert inside the boundary layer, as shown in figure~\ref{momentum}, where the momentum vectors and the distribution of the momentum magnitude are plotted. 

\begin{figure}
\centering
\begin{minipage}[t]{0.33\textwidth}
\centering
\includegraphics[width=\textwidth]{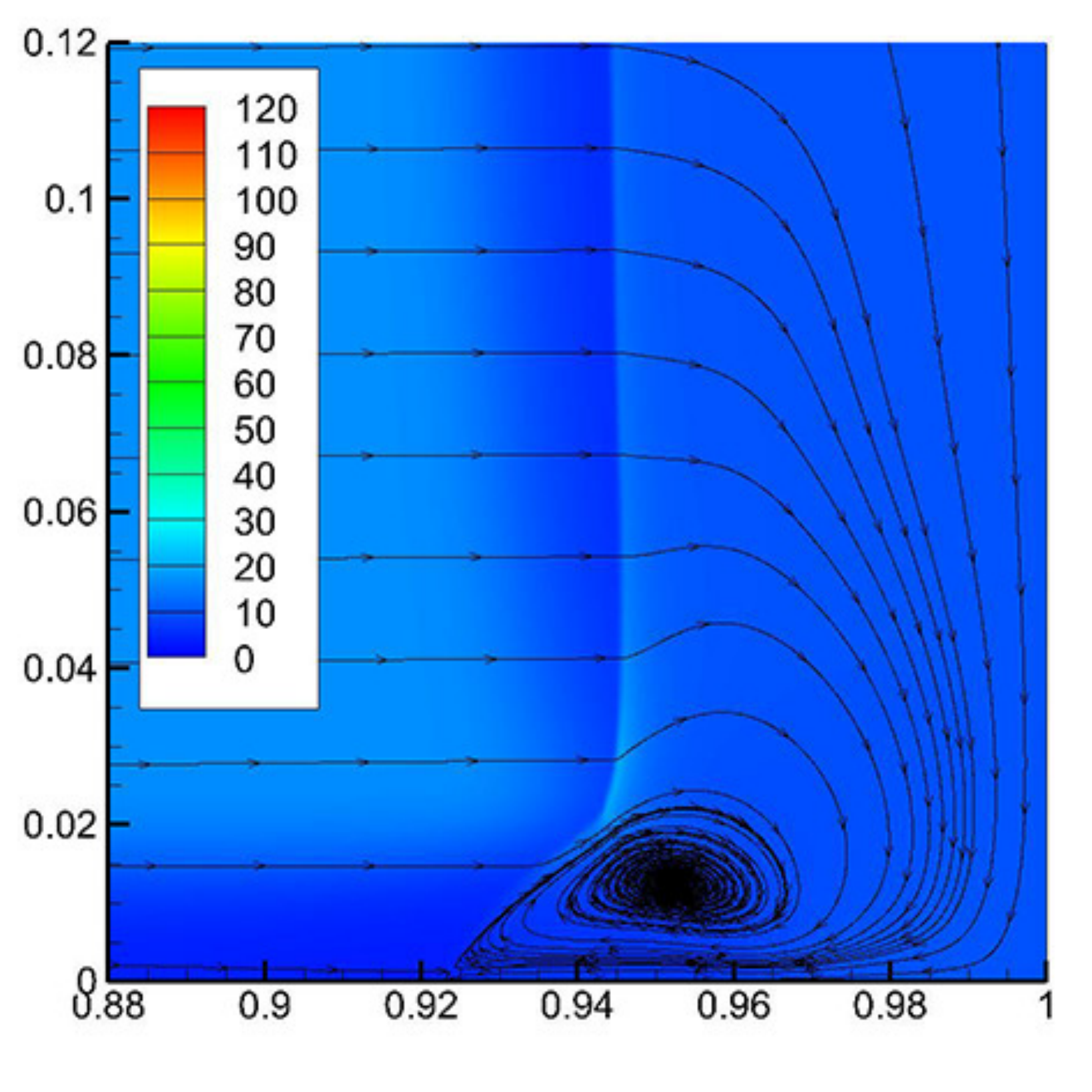}
\centerline{\footnotesize (a)}
\end{minipage}%
\begin{minipage}[t]{0.33\textwidth}
\centering
\includegraphics[width=\textwidth]{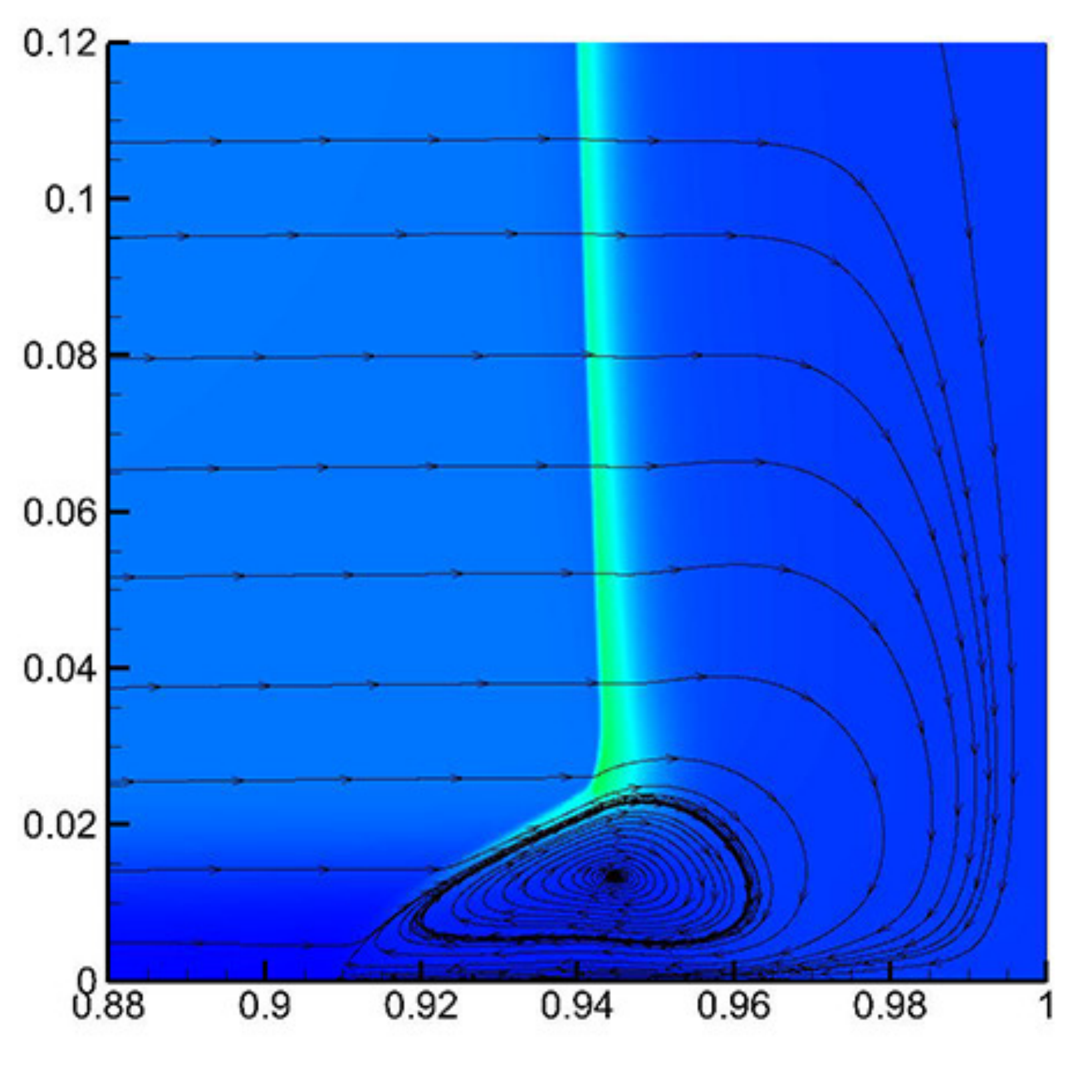}
\centerline{\footnotesize (b)}
\end{minipage}%
\begin{minipage}[t]{0.33\textwidth}
\centering
\includegraphics[width=\textwidth]{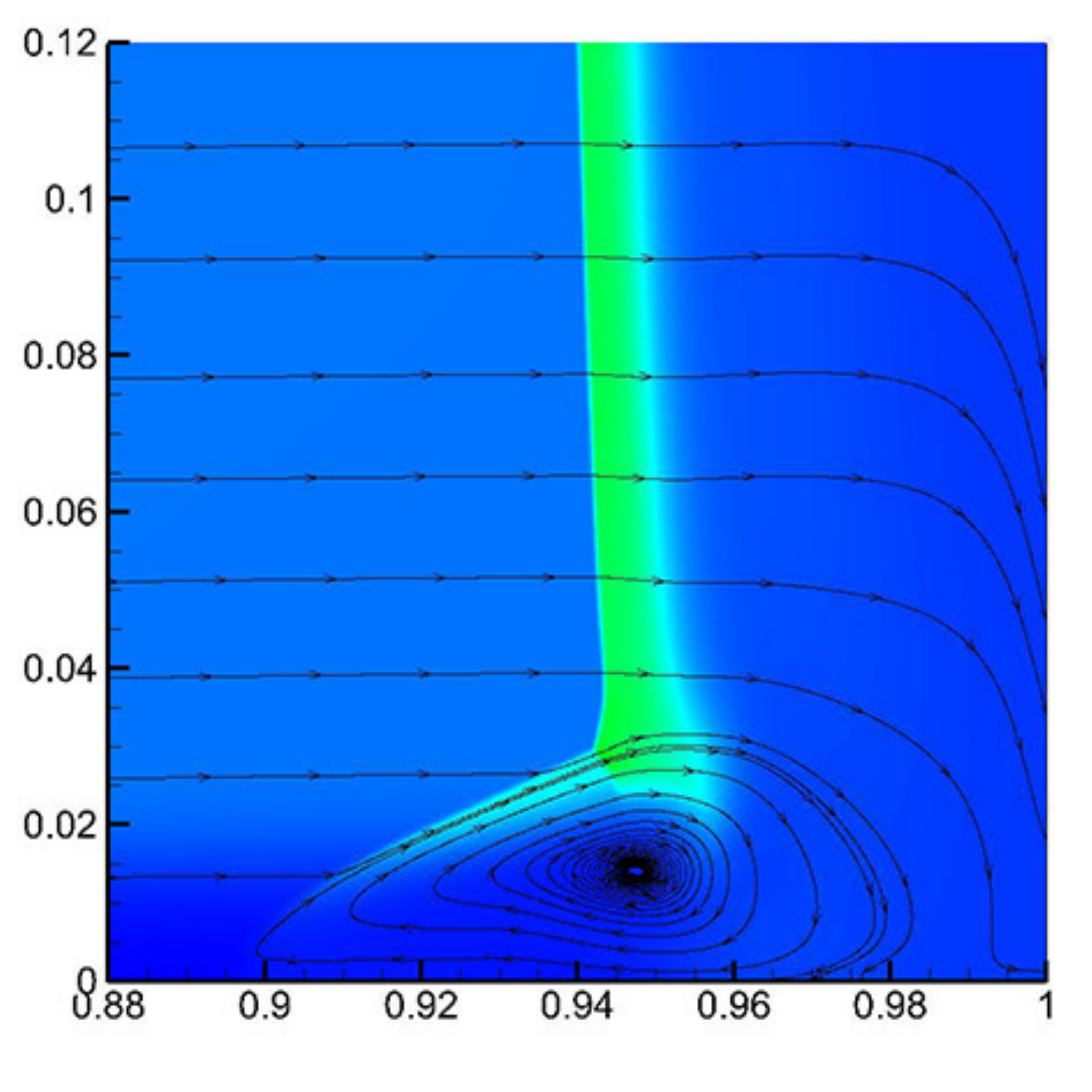}
\centerline{\footnotesize (c)}
\end{minipage}
\centering
\caption{Streamlines and density distribution at (\textit{a}) $t=0.265$, (\textit{b}) $t=0.275$ and \protect\\ (\textit{c}) $t=0.285$.}\label{density}
\end{figure}

At about $t=0.27$, the reflected shock wave encounters the right-travelling contact discontinuity and is nearly stopped by it. The contact discontinuity then moves on with a lower speed. Simultaneously, a new shock wave is formed and propagates to the right. The interaction process is presented in figure~\ref{density}. Notice that the contact discontinuity has not reached the reflected shock wave in figure~\ref{density}(a).

\begin{figure}
\centering
\begin{minipage}[t]{0.33\textwidth}
\centering
\includegraphics[width=\textwidth]{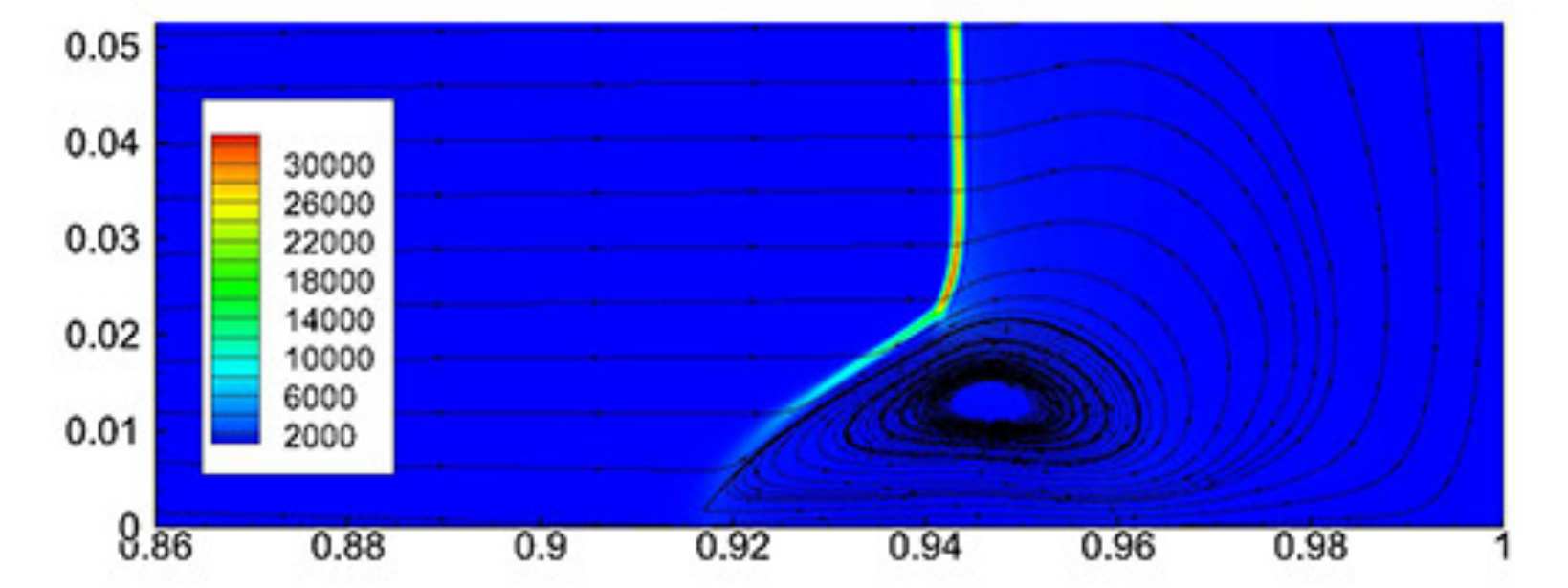}
\centerline{\footnotesize (a)}
\end{minipage}%
\begin{minipage}[t]{0.33\textwidth}
\centering
\includegraphics[width=\textwidth]{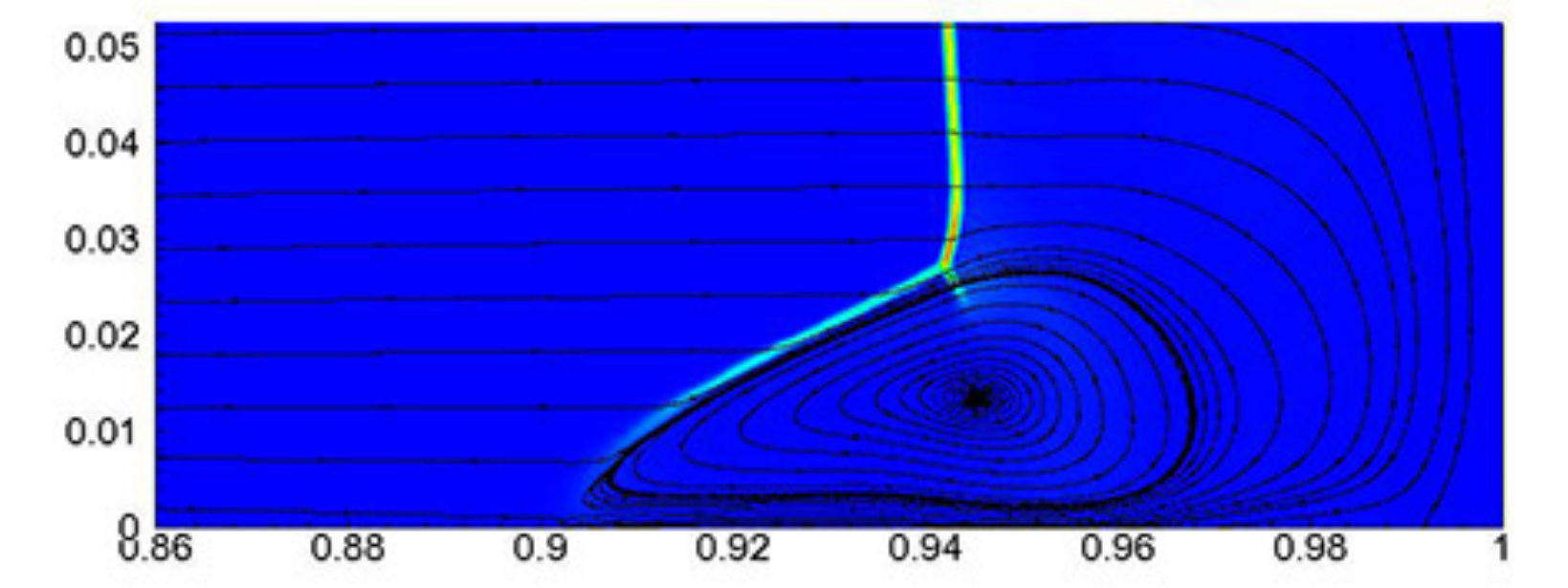}
\centerline{\footnotesize (b)}
\end{minipage}%
\begin{minipage}[t]{0.33\textwidth}
\centering
\includegraphics[width=\textwidth]{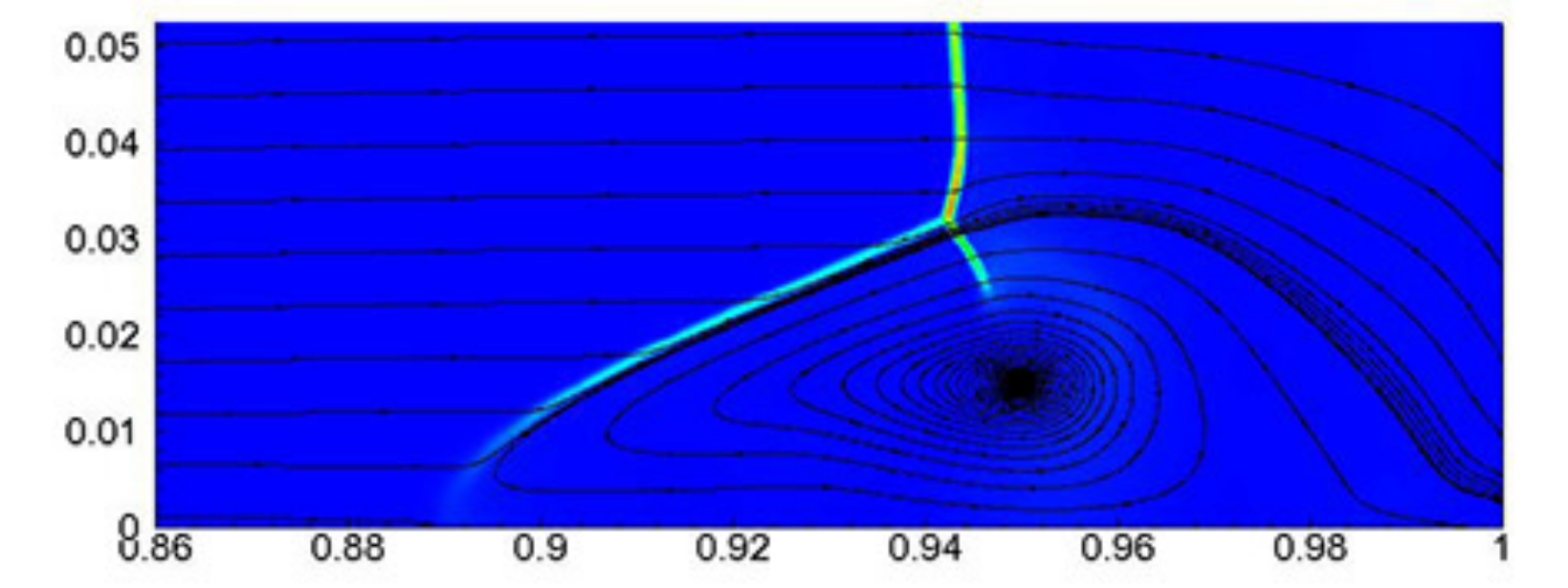}
\centerline{\footnotesize (c)}
\end{minipage}
\centering
\caption{Streamlines and distribution of pressure gradient magnitude at (\textit{a}) $t=0.27$, \protect\\ (\textit{b}) $t=0.28$ and (\textit{c}) $t=0.29$.}\label{lambda}
\end{figure}

The flow in the bulk region outside the viscous boundary layer is similar to the one-dimensional inviscid case. The viscous flow in the near-wall region has a very different behaviour. Since the shock wave becomes oblique in the lower region, the status change of the flow passing the normal shock and the oblique shock is different. This difference of the two regions behind the reflected shock becomes extremely distinct after the contact discontinuity brings the large-density and high-momentum fluid behind it. Remember that the vortex is carrying fluid along the oblique shock from the lower region to the upper region. To accommodate the huge difference of the fluid property, a shock appears at the interface between the two regions, i.e., bifurcation occurs at the junction point of the normal shock and the oblique shock. See figure~\ref{density}(c). This process is more clearly presented in figure~\ref{lambda}, where we can see a lambda-shaped structure around the triple point.

At about $t=0.32$, the new shock wave produced by the shock/contact-discontinuity interaction has been reflected back by the right wall. It then crosses the right-moving contact discontinuity and is slowed down by it. After that, the shock interacts with the vortex and then with the stationary shock, making it start to move again to the left, along with the triple point of the lambda-shaped shock. There are also many other secondary waves and a number of interactions between them at this stage. But they are relatively weak hence do not affect the primary picture much.

\begin{figure}
\centering
\begin{minipage}[t]{0.33\textwidth}
\centering
\includegraphics[width=\textwidth]{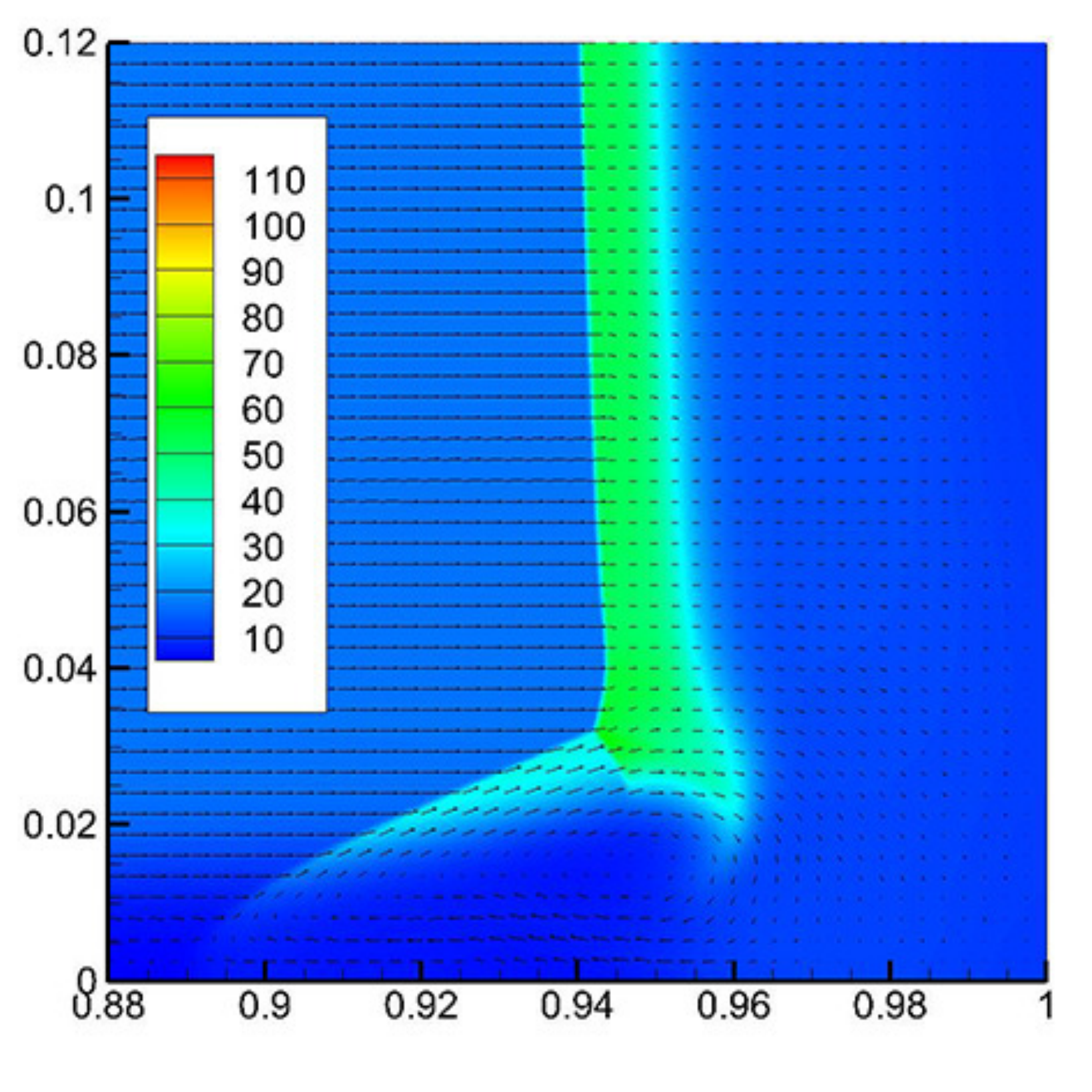}
\centerline{\footnotesize (a)}
\end{minipage}%
\begin{minipage}[t]{0.33\textwidth}
\centering
\includegraphics[width=\textwidth]{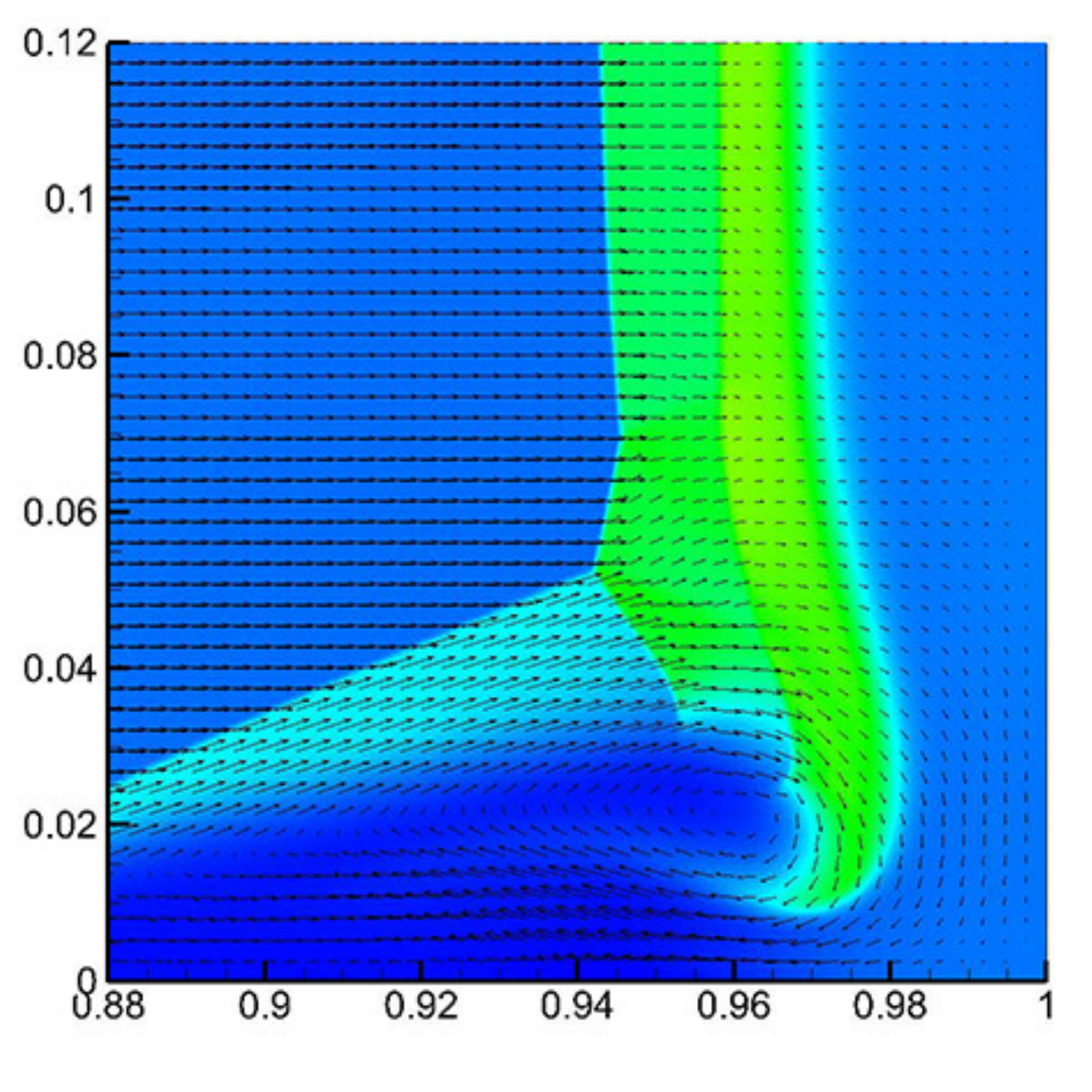}
\centerline{\footnotesize (b)}
\end{minipage}%
\begin{minipage}[t]{0.33\textwidth}
\centering
\includegraphics[width=\textwidth]{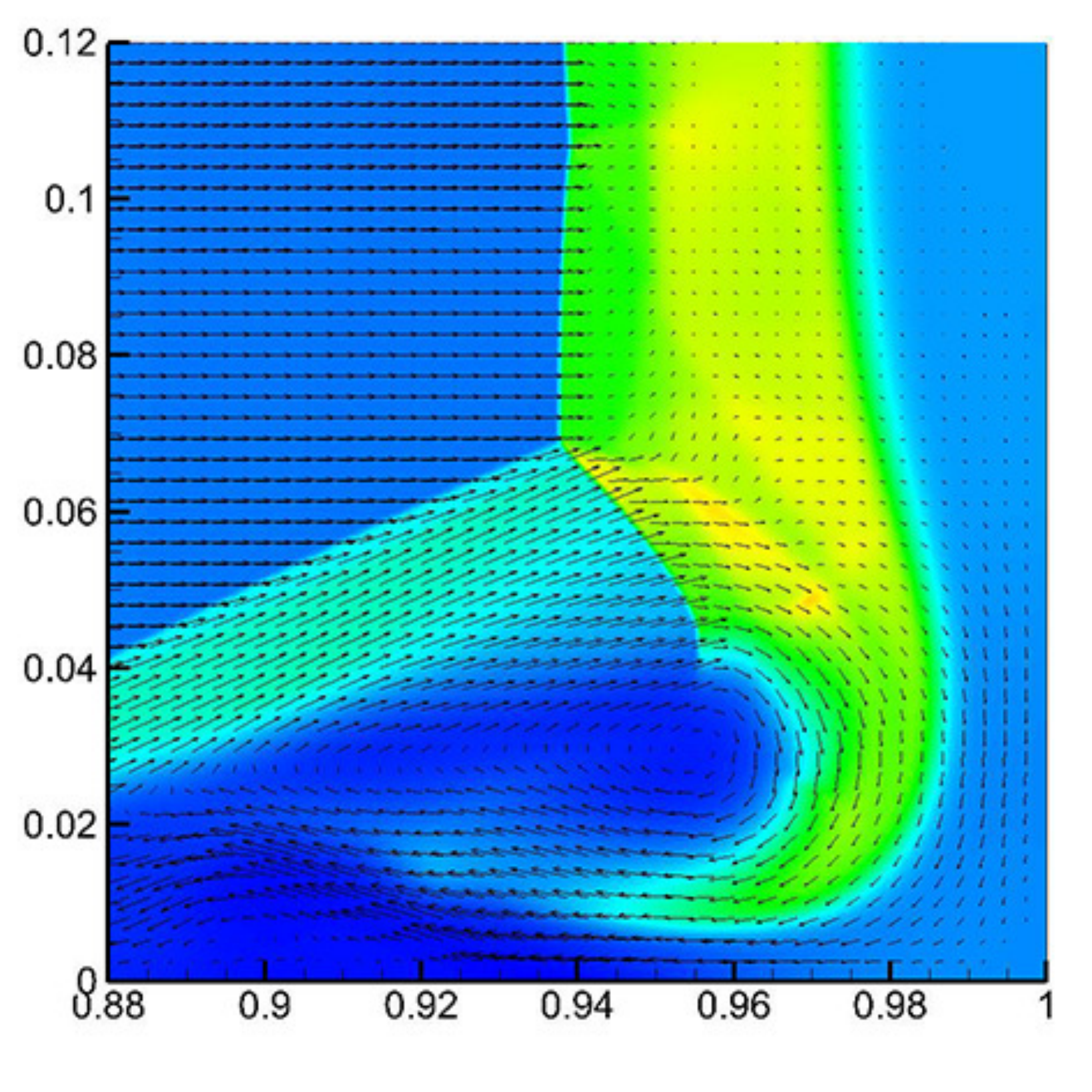}
\centerline{\footnotesize (c)}
\end{minipage}
\centering
\caption{Velocity vectors (every 8 point is plotted in both directions) and density distribution at (\textit{a}) $t=0.29$, (\textit{b}) $t=0.33$ and (\textit{c}) $t=0.37$.}\label{density2}
\end{figure}

\begin{figure}
\centering
\begin{minipage}[t]{0.33\textwidth}
\centering
\includegraphics[width=\textwidth]{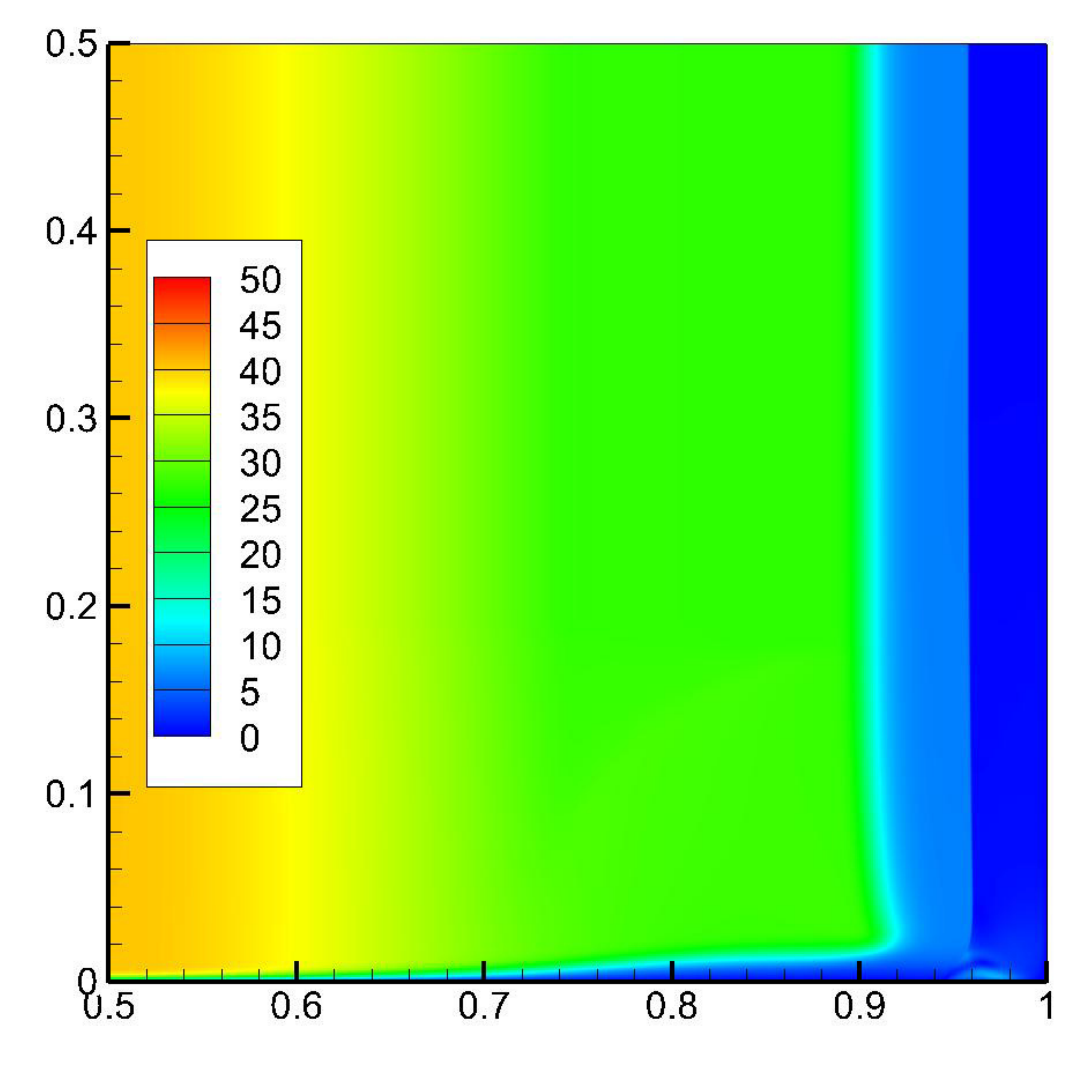}
\centerline{\footnotesize (a)}
\end{minipage}%
\begin{minipage}[t]{0.33\textwidth}
\centering
\includegraphics[width=\textwidth]{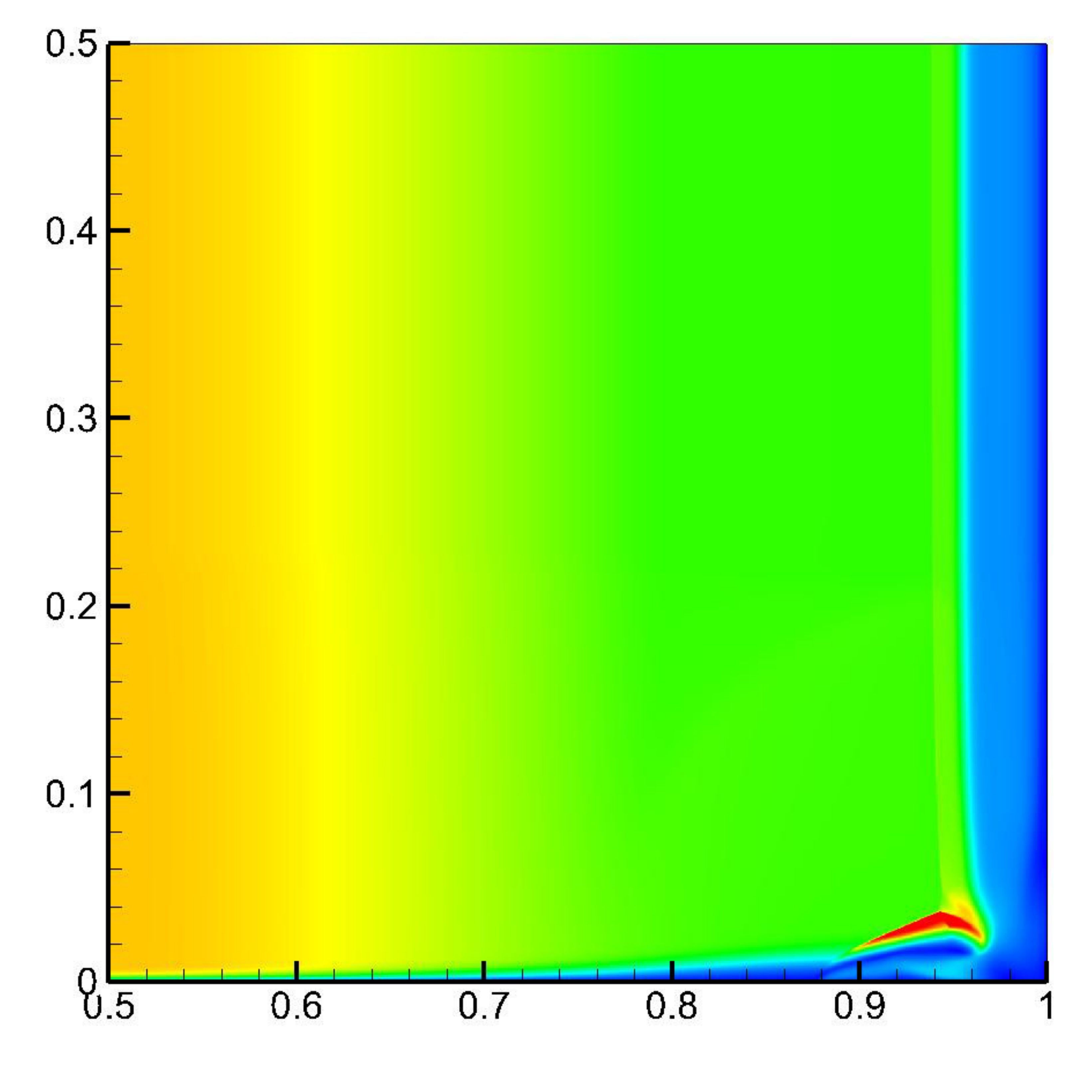}
\centerline{\footnotesize (b)}
\end{minipage}%
\begin{minipage}[t]{0.33\textwidth}
\centering
\includegraphics[width=\textwidth]{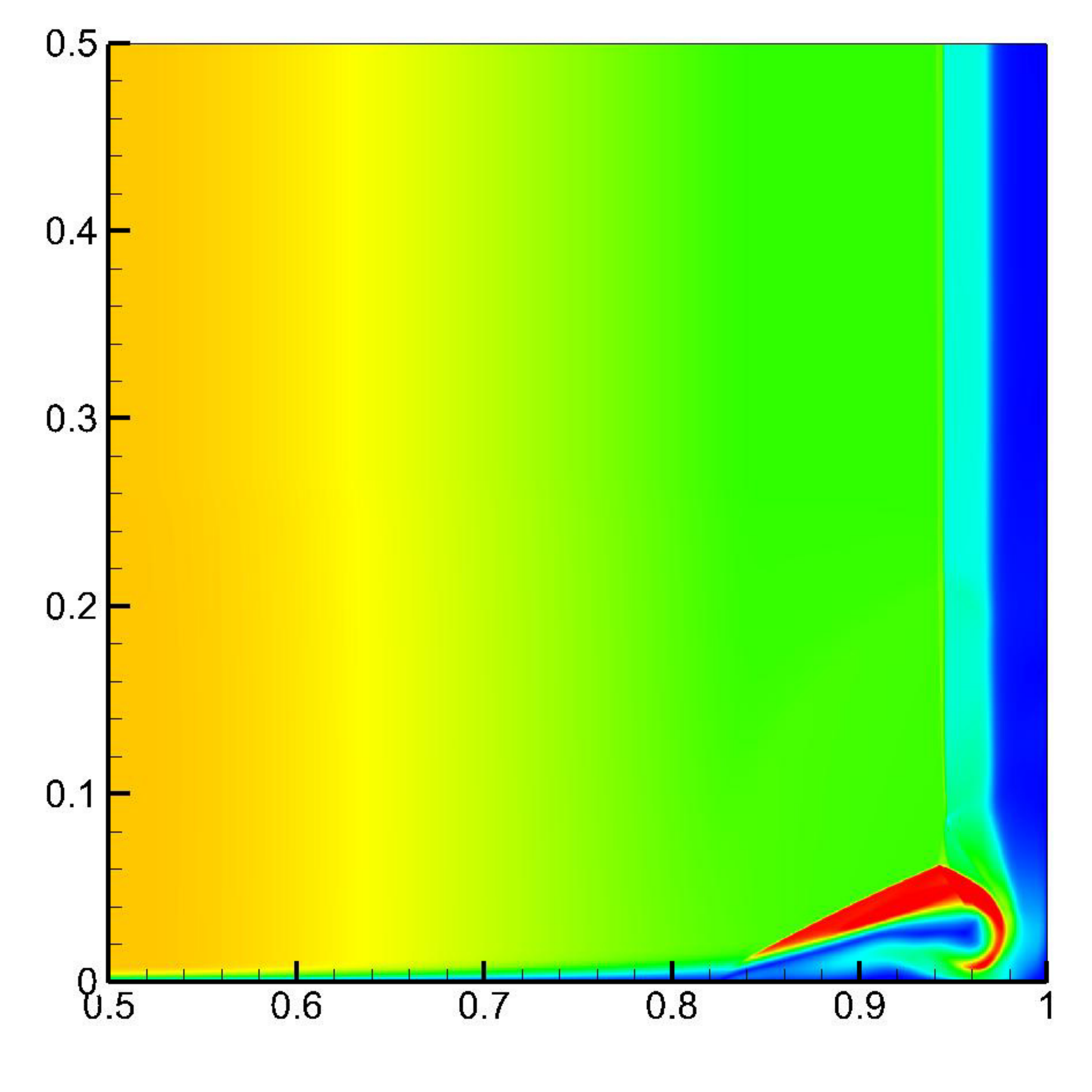}
\centerline{\footnotesize (c)}
\end{minipage}
\centering
\caption{Distribution of momentum magnitude at (\textit{a}) $t=0.25$, (\textit{b}) $t=0.30$ and \protect\\ (\textit{c}) $t=0.35$.}\label{momentum2}
\end{figure}

Later when the vortex is stronger, it dominates the local flow field. We can see from figure~\ref{density2} that the dense fluids are entrained by the vortical flow around the core of the vortex, creating a jet inserting into the bottom lighter fluids. The momentum magnitude distributions are plotted in figure~\ref{momentum2}, showing how the jet is generated at the lower right corner of the high-momentum region. 

In another view, the jet is enclosed by two contact discontinuities, one of which originates from the vertical contact discontinuity while the other originates from the oblique contact discontinuity. This mechanism is clearly shown in figure~\ref{dgm}(a), \ref{dgm}(b) and \ref{dgm}(c). In figure~\ref{dgm}(a), the two contact discontinuities with different orientations are presented in the density-gradient-magnitude contour map. Then the horizontal discontinuity encounters the oblique shock wave and the vertical contact discontinuity encounters both the normal and oblique shocks. The two contact discontinuities become stronger after getting through the shock wave, and their shape remains the same, except that the horizontal one is a little deflected up by the oblique shock. Then they are both bent and carried down by the vortex, forming the two boundaries of the jet.

\begin{figure}
\centering
\begin{minipage}[t]{0.5\textwidth}
\centering
\includegraphics[width=\textwidth]{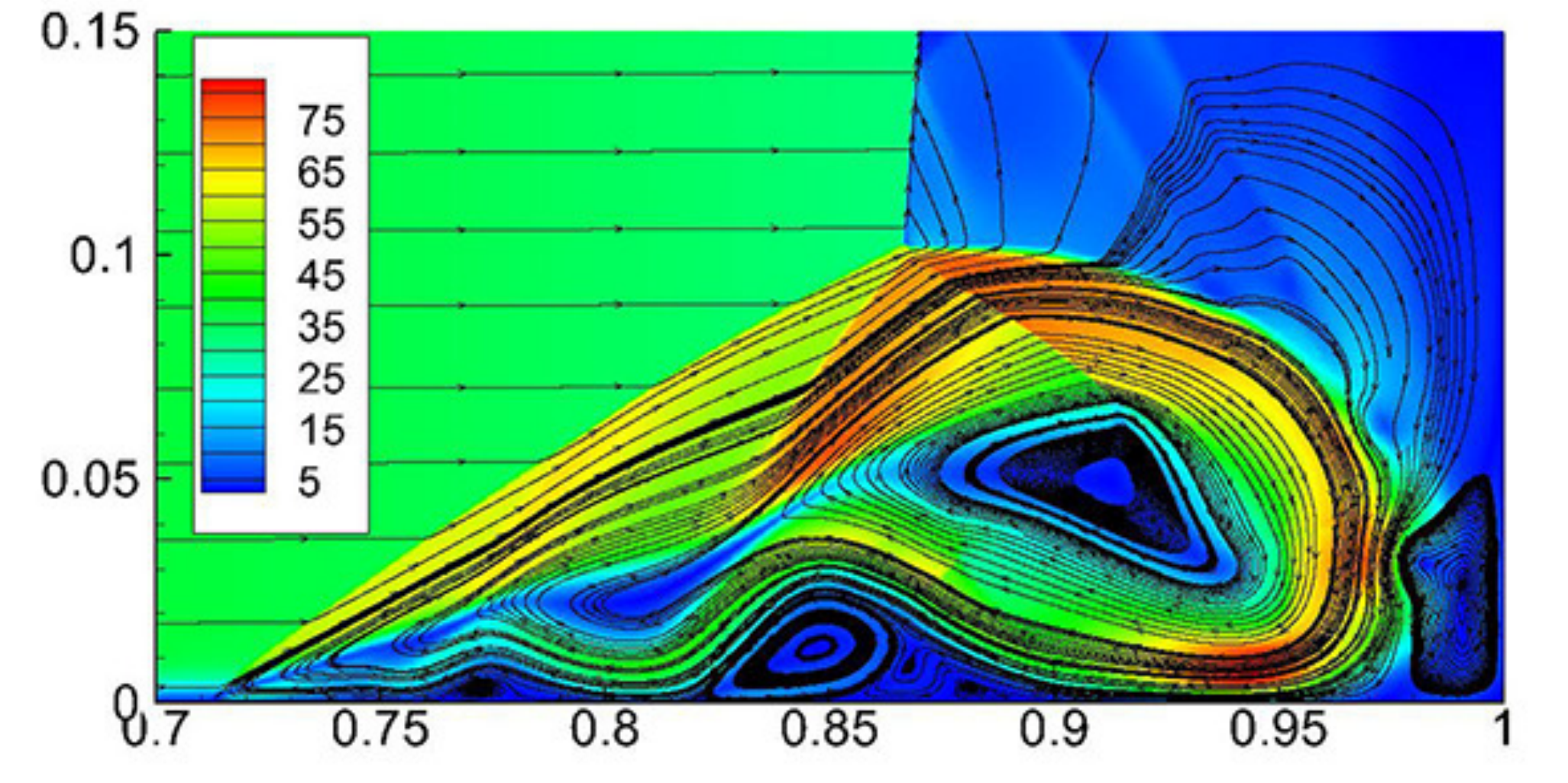}
\centerline{\footnotesize (a)}
\end{minipage}%
\begin{minipage}[t]{0.5\textwidth}
\centering
\includegraphics[width=\textwidth]{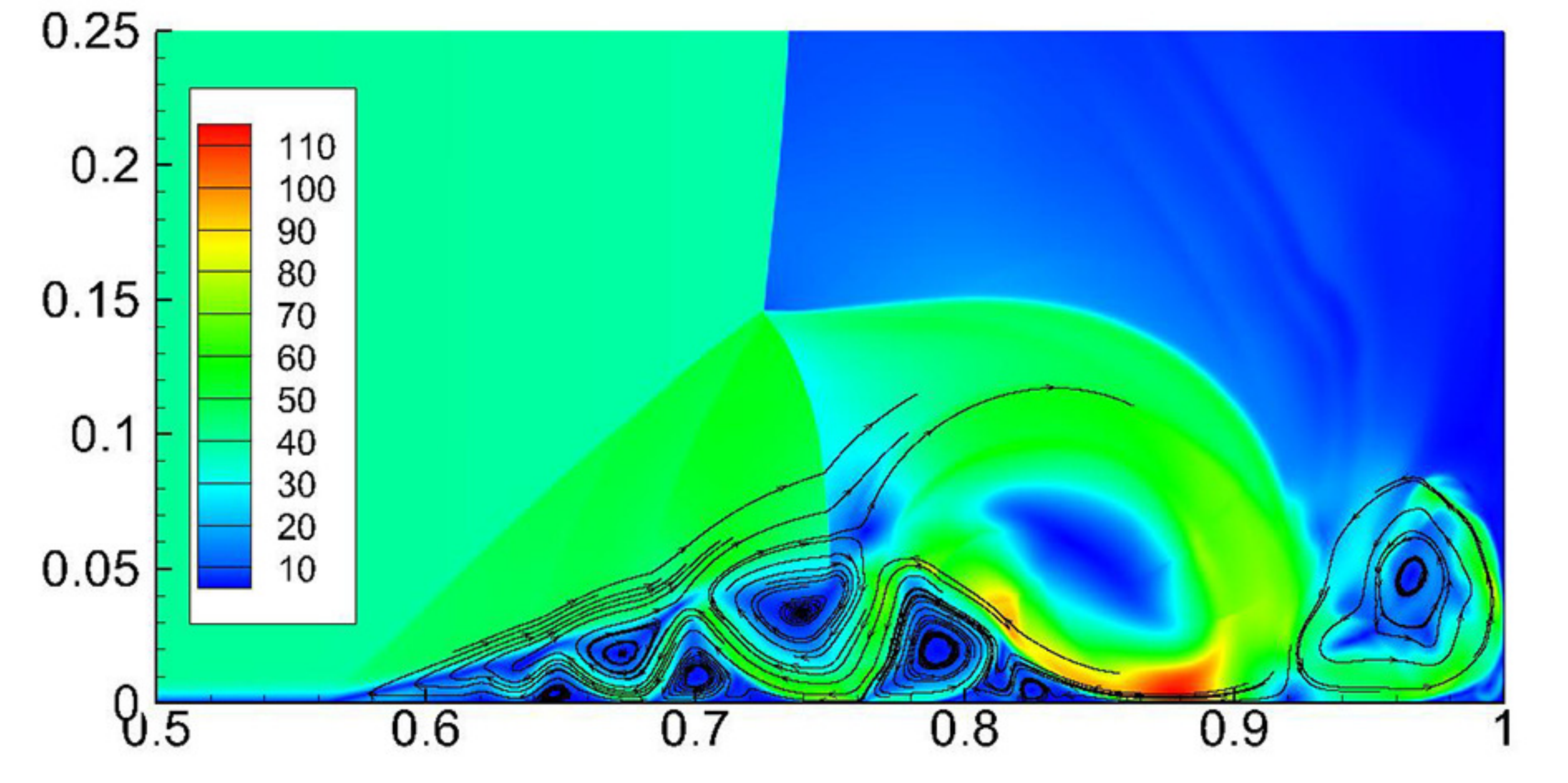}
\centerline{\footnotesize (b)}
\end{minipage}
\centering
\caption{Streamlines and momentum magnitude distribution at (\textit{a}) $t = 0.5$ and (\textit{b}) $t = 0.75$.}\label{momentum3_4}
\end{figure}

As the horizontal contact discontinuity is deflected behind the oblique shock wave, a wedge-shaped area appears between it and the bottom wall. In figure~\ref{momentum3_4}(a), we can see that the jet becomes longer and extends to the left, alternatively reflecting on the two boundaries of the wedge-shaped area. This area is then divided by the jet into several individual regions distributed on both sides of the jet. Small secondary vortices are induced by the jet in these individual regions. And these vortices may further induce smaller vortices, see the section between $x=0.85$ and $x=0.9$ in figure~\ref{momentum3_4}(a). This demonstrates the multi-scale feature of the flow field. To avoid ambiguity, the vortex formed at the beginning will be called the primary vortex hereinafter. It should be noted that a large vortex is generated by the primary vortex in the lower right corner.

From figure~\ref{momentum3_4}(a) which shows the distribution of the momentum magnitude, it is also found that the upper right edge of the primary vortex is a thin contact surface. Therefore the Kelvin-Helmholtz instability occurs around it, which is shown in figure~\ref{dgm}(f), where a sequence of vortical structures is observed near the contact surface. These structures are driven by the primary vortex down to the corner, and merged with the stationary contact discontinuity located at around $x=0.94$. After getting to the bottom wall, the vortical structures are taken over by the anticlockwise-rotating corner vortex shown in figure~\ref{momentum3_4}(a). The corner vortex carries these structures upward along its streamlines. This process is presented in figure~\ref{dgm}(g), \ref{dgm}(h) and \ref{dgm}(i). Meanwhile, the wide stationary contact discontinuity at about $x=0.97$ is rolled up. In figure~\ref{dgm}(i), we can see that the big rotating structure at the lower right corner involves at least four contact discontinuities altogether.

\begin{figure}
\centering
\begin{minipage}[t]{0.33\textwidth}
\centering
\includegraphics[width=\textwidth]{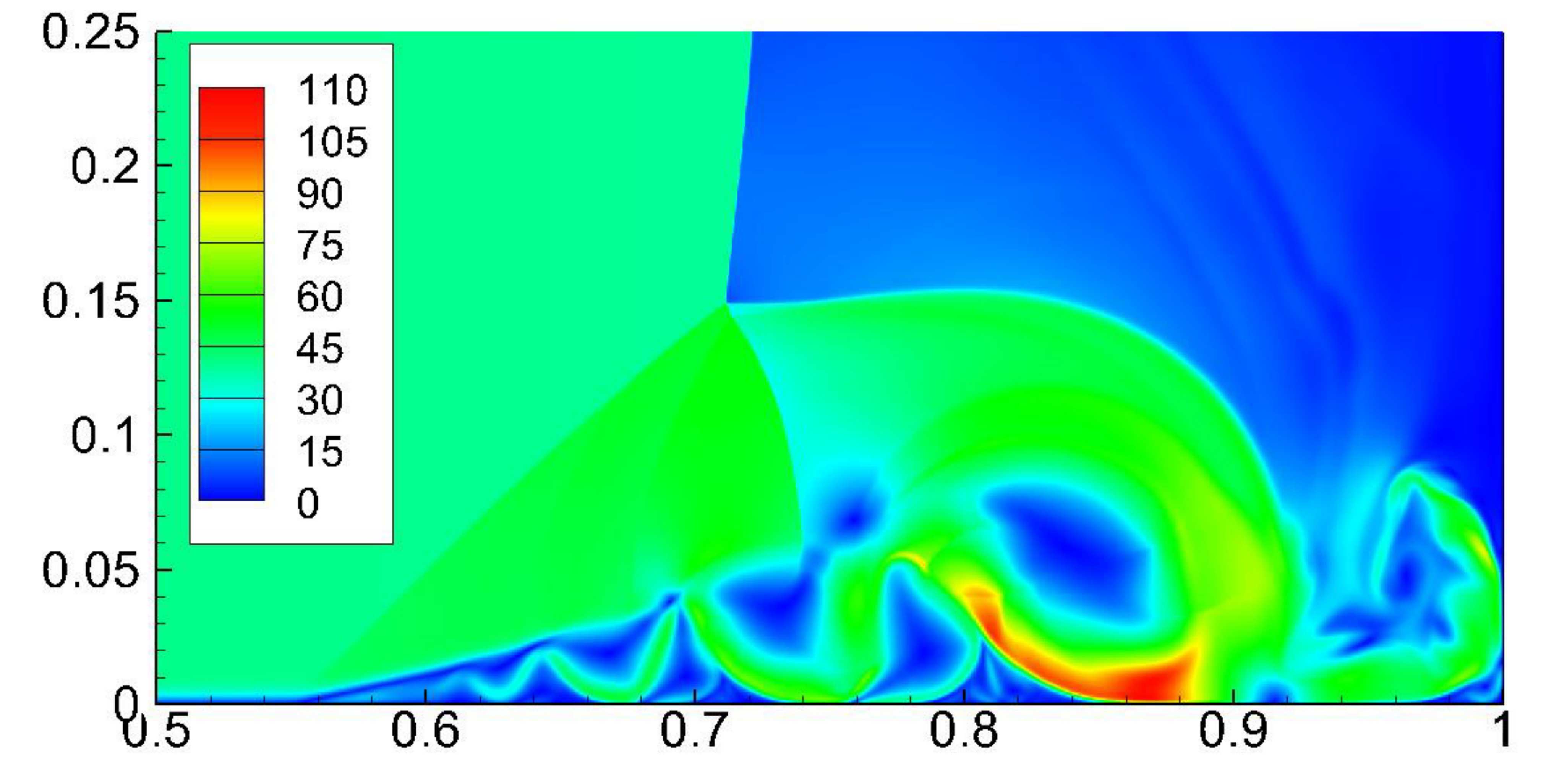}
\centerline{\footnotesize (a)}
\end{minipage}%
\begin{minipage}[t]{0.33\textwidth}
\centering
\includegraphics[width=\textwidth]{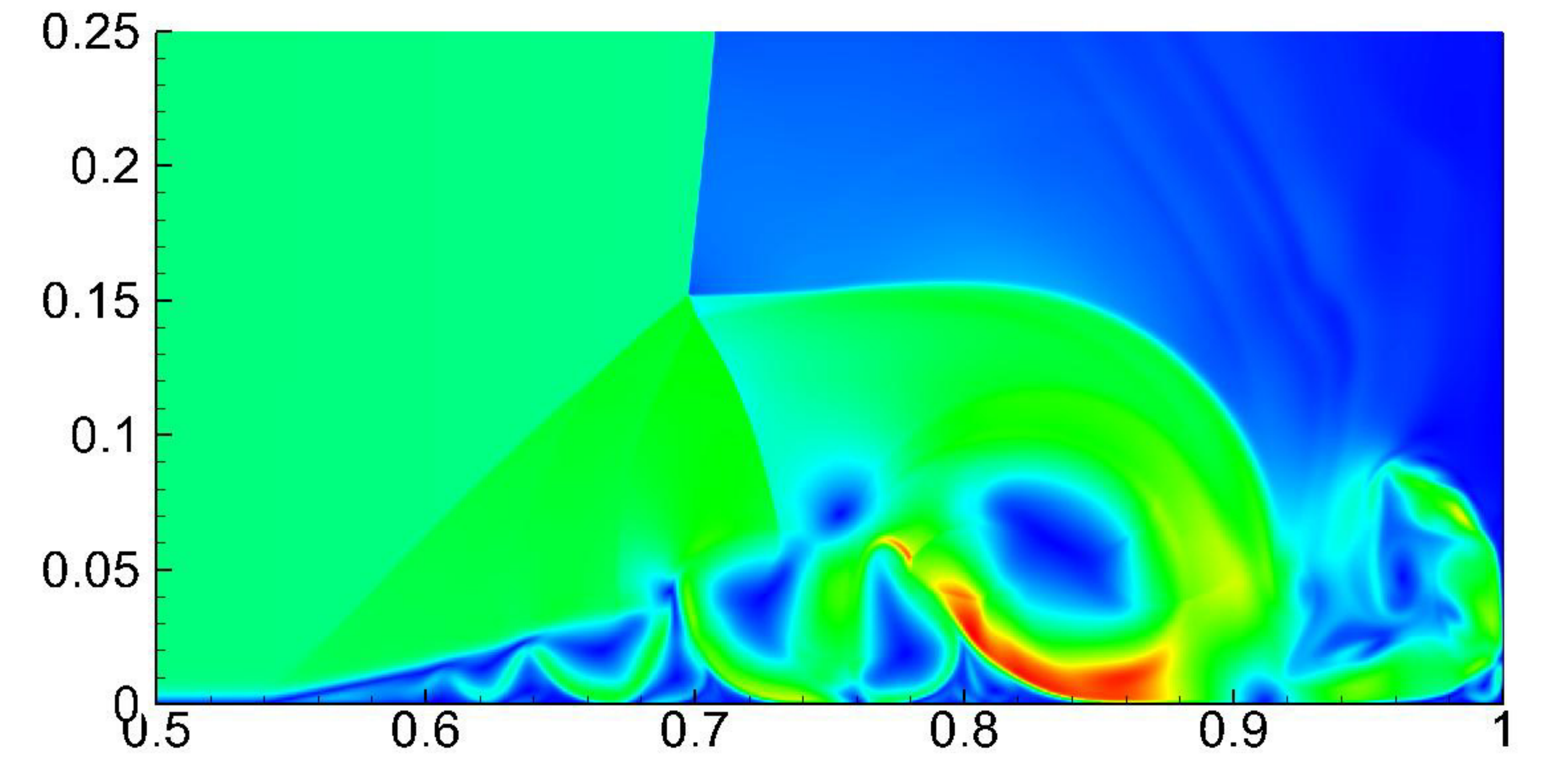}
\centerline{\footnotesize (b)}
\end{minipage}%
\begin{minipage}[t]{0.33\textwidth}
\centering
\includegraphics[width=\textwidth]{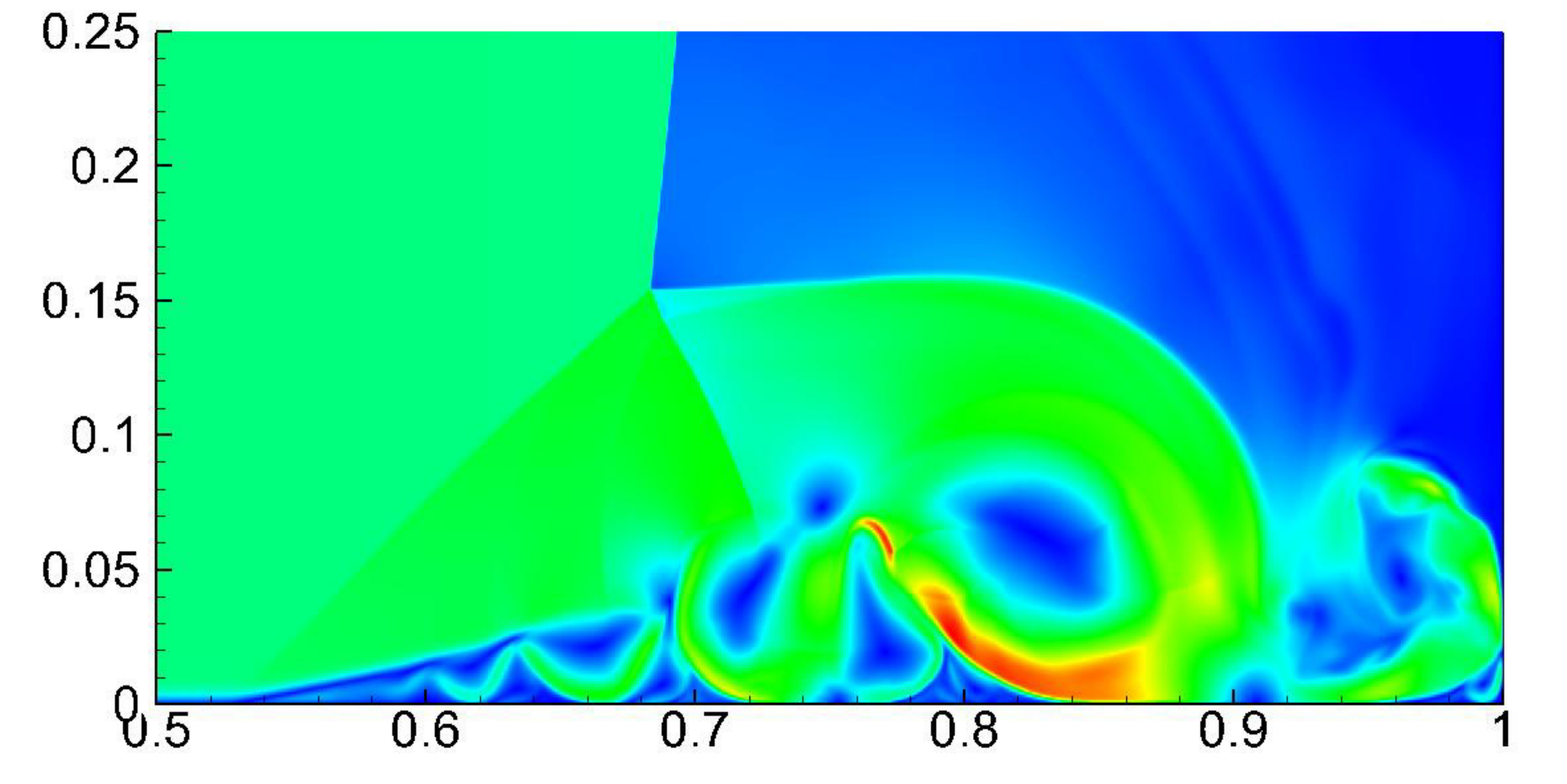}
\centerline{\footnotesize (c)}
\end{minipage}
\centering
\caption{Distribution of momentum magnitude at (\textit{a}) $t=0.775$, (\textit{b}) $t=0.800$ and \protect\\ (\textit{c}) $t=0.825$.}\label{momentum5}
\end{figure}

The distribution of the momentum magnitude at $t=0.75$ is plotted in figure~\ref{momentum3_4}(b). It is clear that besides the left-moving zigzagging jet beneath the deflected contact discontinuity, there is another jet turning right at about $x=0.92$ along the streamlines of the corner vortex. In fact, this flow pattern can be found at each point where the jet impinges on the bottom wall, which is also the lower boundary of the wedge-shaped area: The jet is split by the wall into two branches due to its high momentum, the bigger turning to the left, and the smaller to the right (see figure~\ref{momentum5}).

As for the upper boundary of the wedge-shaped area, when the jet impinges on it, a part of the jet will be ejected up leaking into the outer region above the deflected contact discontinuity. This phenomenon is demonstrated by the streamlines in figure~\ref{momentum3_4}(b). Clearly that the jet is not totally restricted in the wedge-shaped area. The ejected fluids are then taken to the right by the high-momentum flow in the outer region, producing a pulling force which makes the jet broken at the contacting points, as is shown in figure~\ref{momentum5}, where a gap is seen at about $x=0.7$.

\begin{figure}
\centering
\begin{minipage}[t]{0.33\textwidth}
\centering
\includegraphics[width=\textwidth]{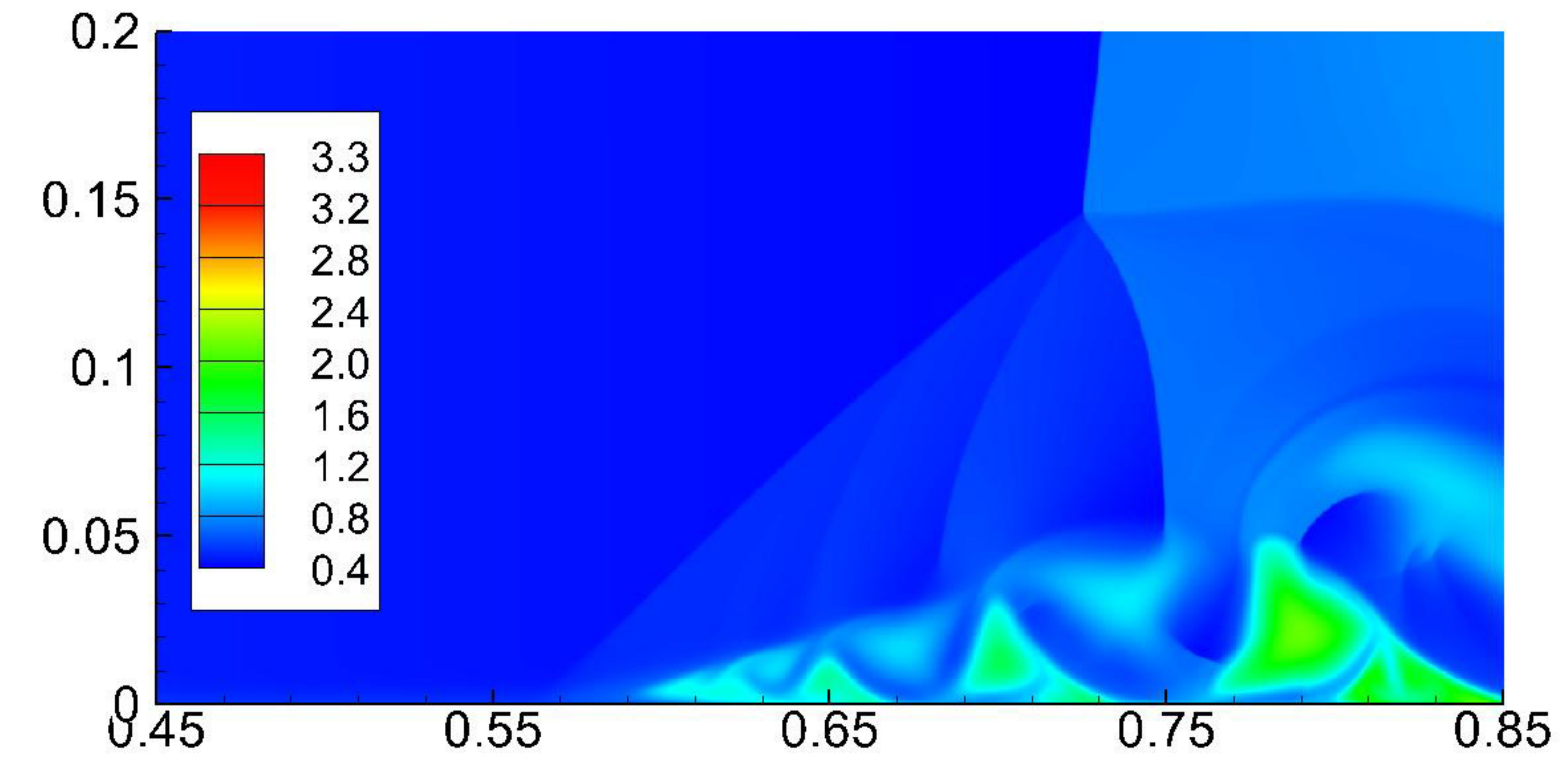}
\centerline{\footnotesize (a)}
\end{minipage}%
\begin{minipage}[t]{0.33\textwidth}
\centering
\includegraphics[width=\textwidth]{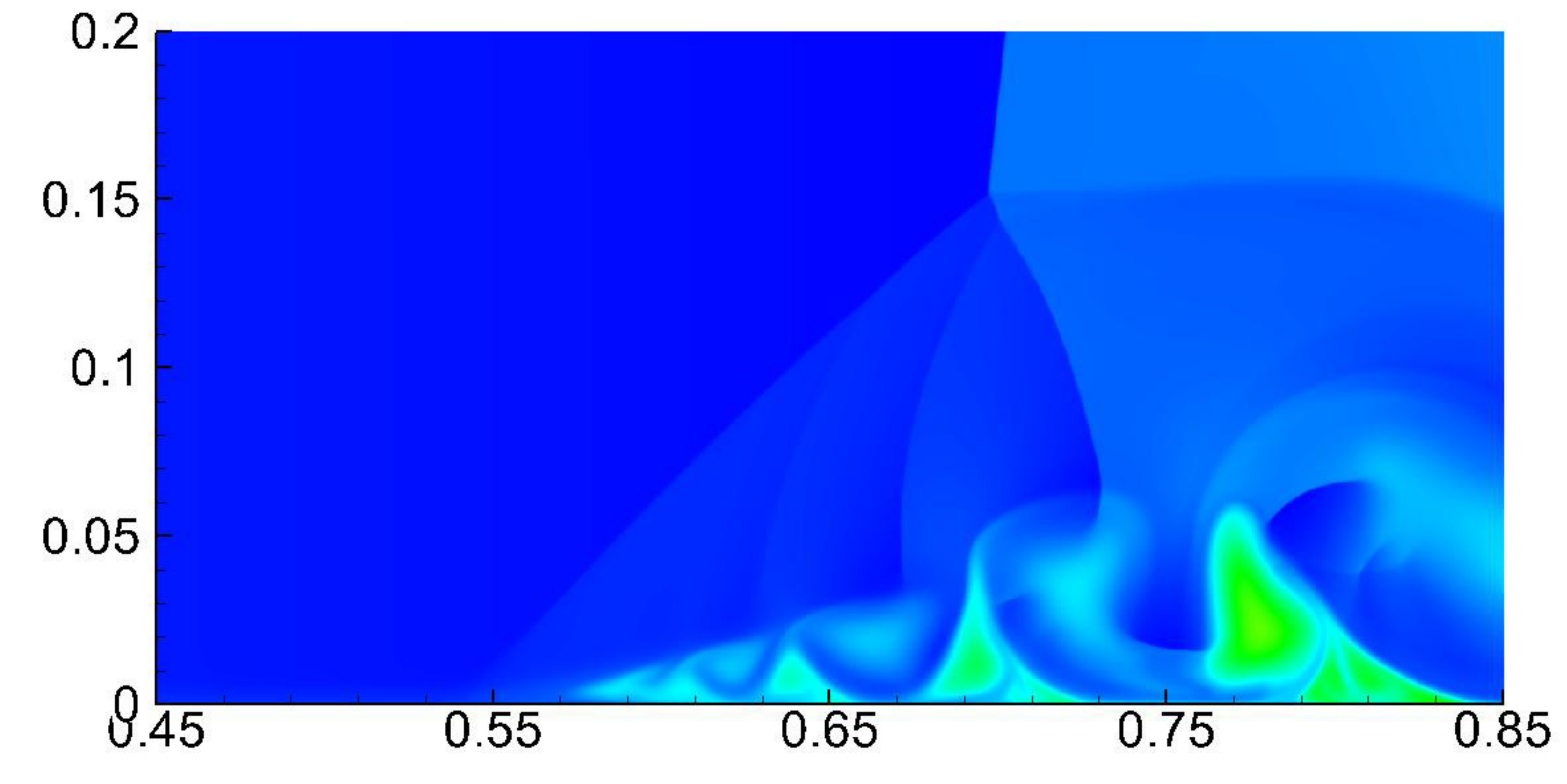}
\centerline{\footnotesize (b)}
\end{minipage}%
\begin{minipage}[t]{0.33\textwidth}
\centering
\includegraphics[width=\textwidth]{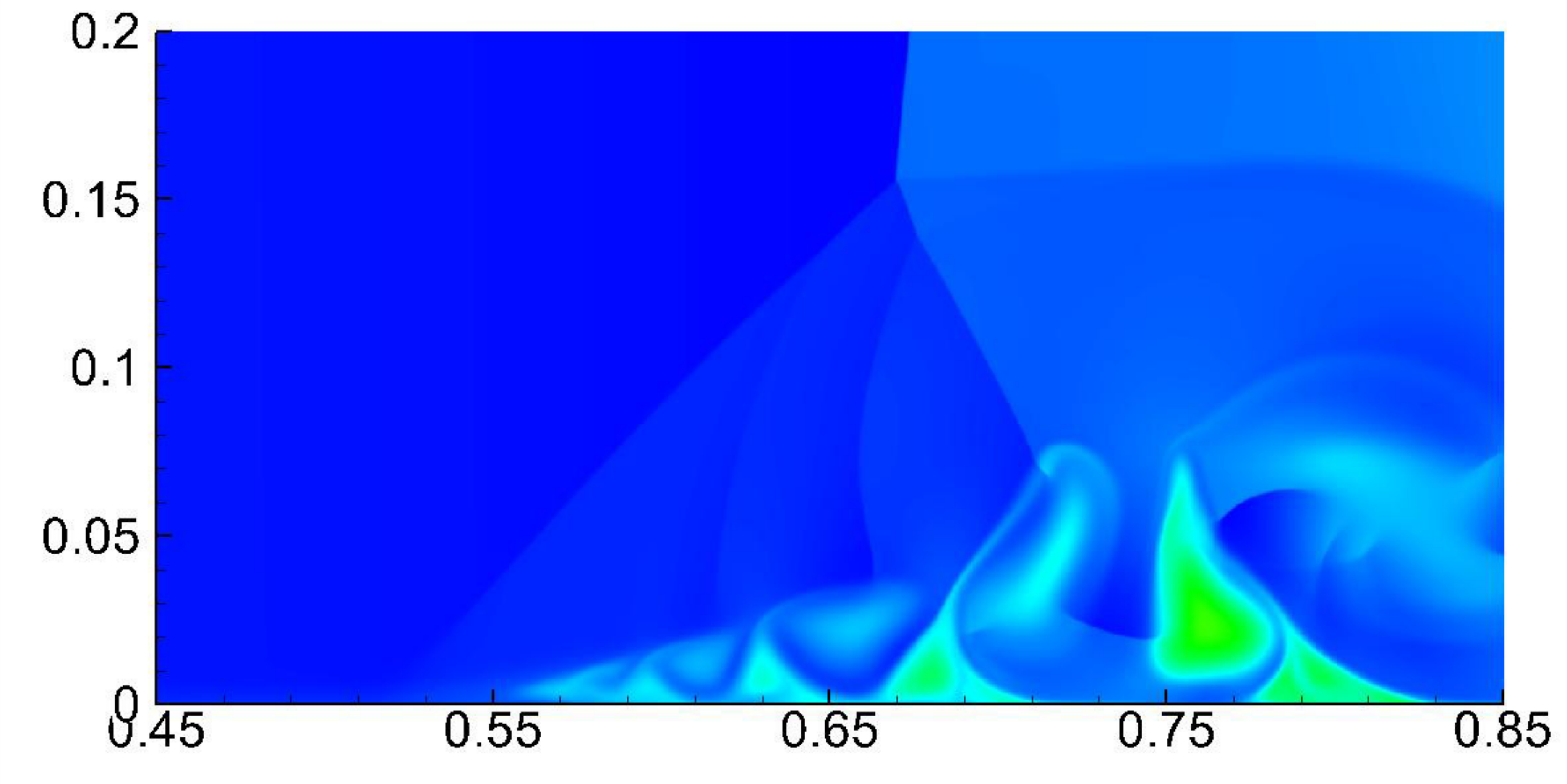}
\centerline{\footnotesize (c)}
\end{minipage}
\begin{minipage}[t]{0.33\textwidth}
\centering
\includegraphics[width=\textwidth]{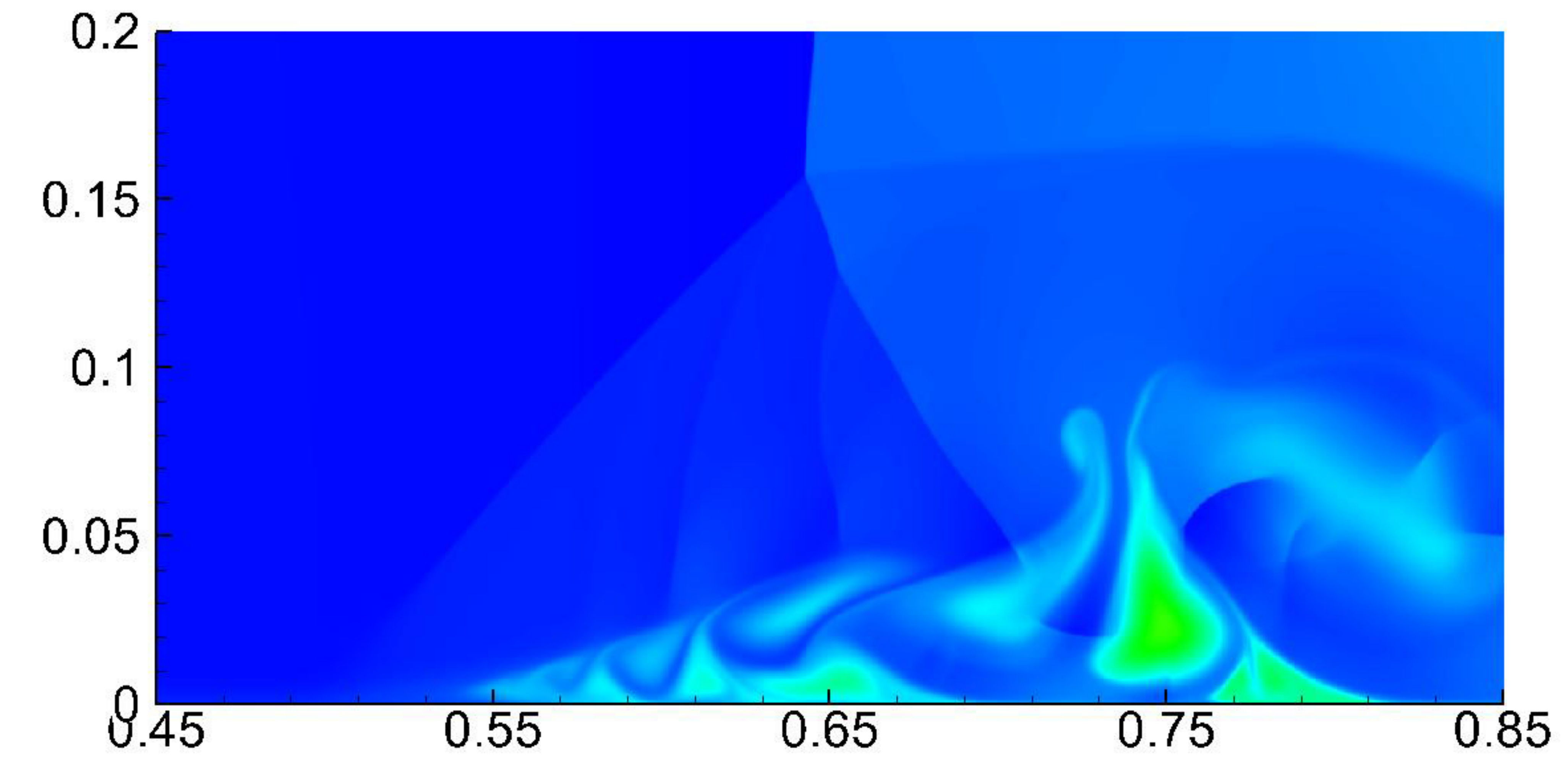}
\centerline{\footnotesize (d)}
\end{minipage}%
\begin{minipage}[t]{0.33\textwidth}
\centering
\includegraphics[width=\textwidth]{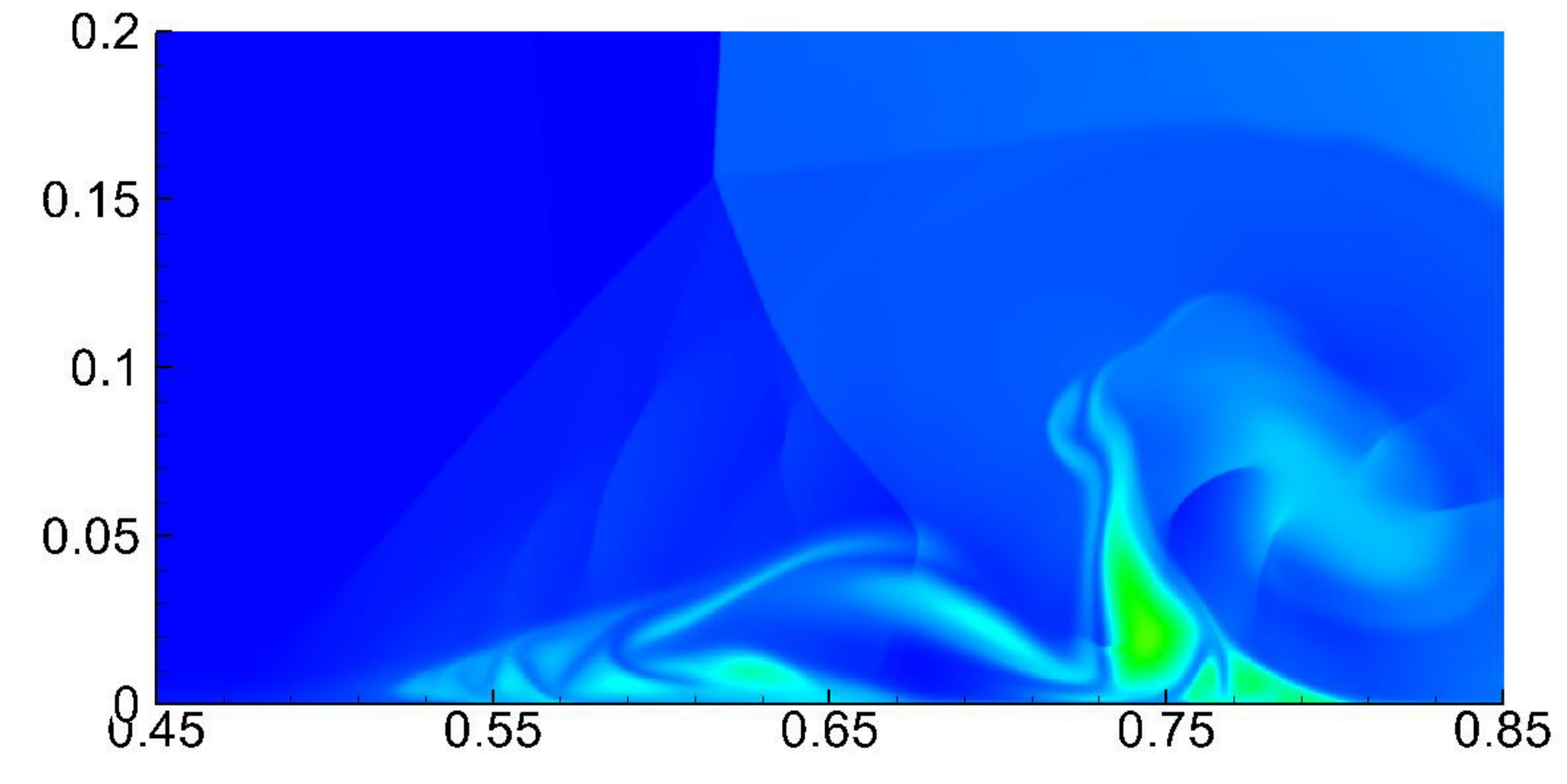}
\centerline{\footnotesize (e)}
\end{minipage}%
\begin{minipage}[t]{0.33\textwidth}
\centering
\includegraphics[width=\textwidth]{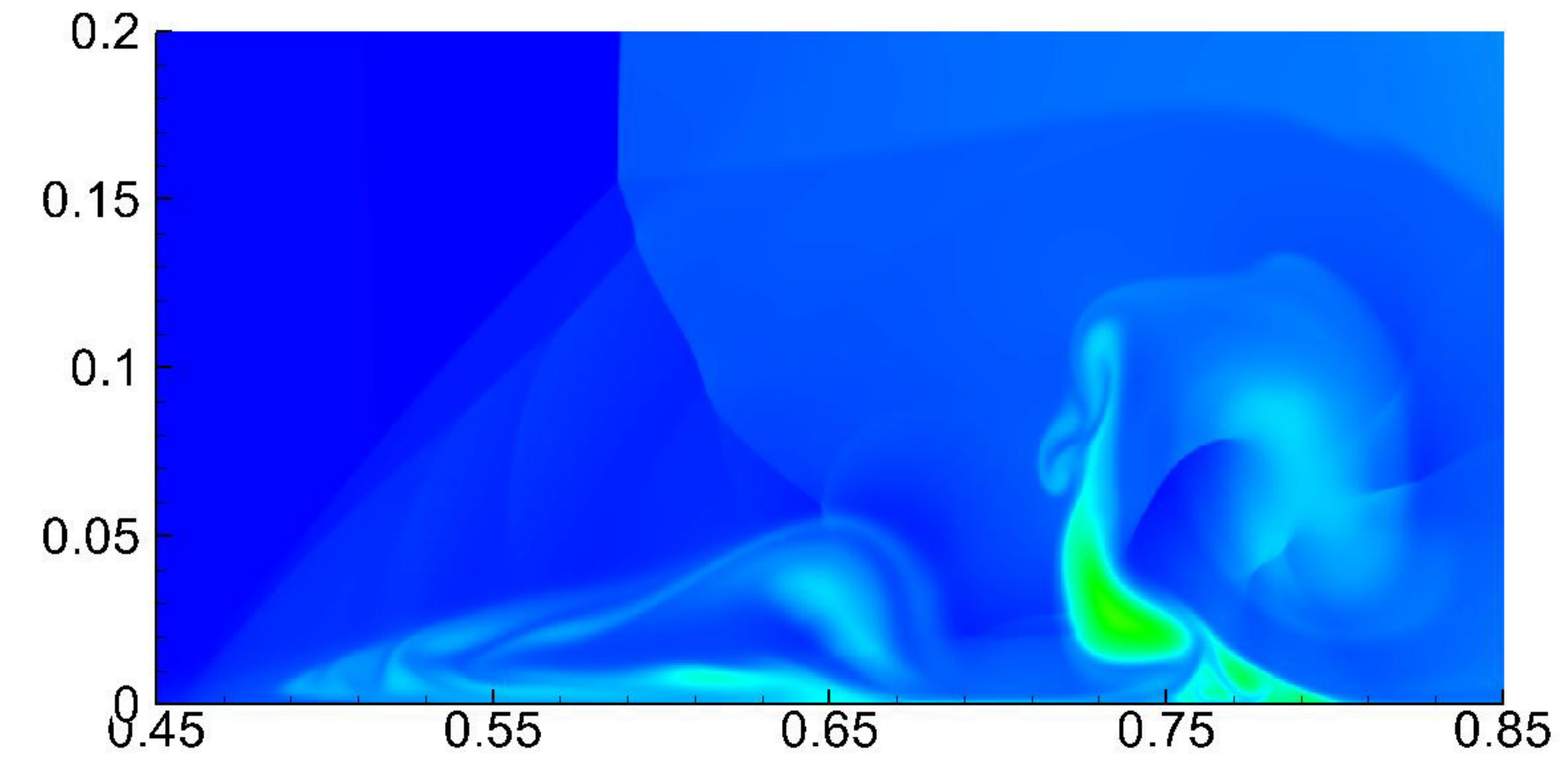}
\centerline{\footnotesize (f)}
\end{minipage}
\centering
\caption{Temperature distribution at (\textit{a}) $t=0.75$, (\textit{b}) $t=0.80$, (\textit{c}) $t=0.85$, (\textit{d}) $t=0.90$, (\textit{e}) $t=0.95$ and (\textit{f}) $t=1.00$. }\label{temperature}
\end{figure}

With presence of the gaps, the fluids under the jet are carried up by the ejected jet into the outer flow region. These hot and light fluids are also taken away by the outer high-momentum flow, forming thin filaments, the biggest among which finally bumps onto the left edge of the primary vortex. See the temperature distribution in figure~\ref{temperature}.

\begin{figure}
\centering
\begin{minipage}[t]{0.33\textwidth}
\centering
\includegraphics[width=\textwidth]{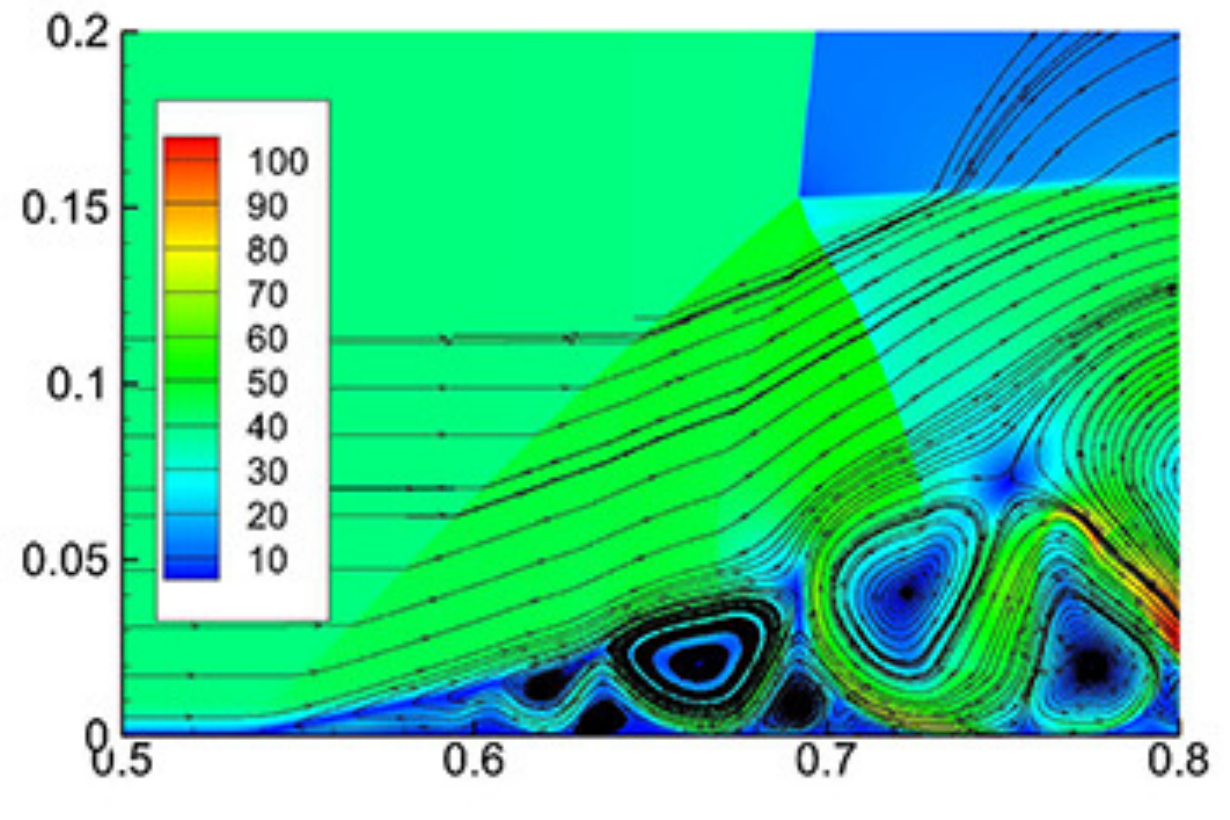}
\centerline{\footnotesize (a)}
\end{minipage}%
\begin{minipage}[t]{0.33\textwidth}
\centering
\includegraphics[width=\textwidth]{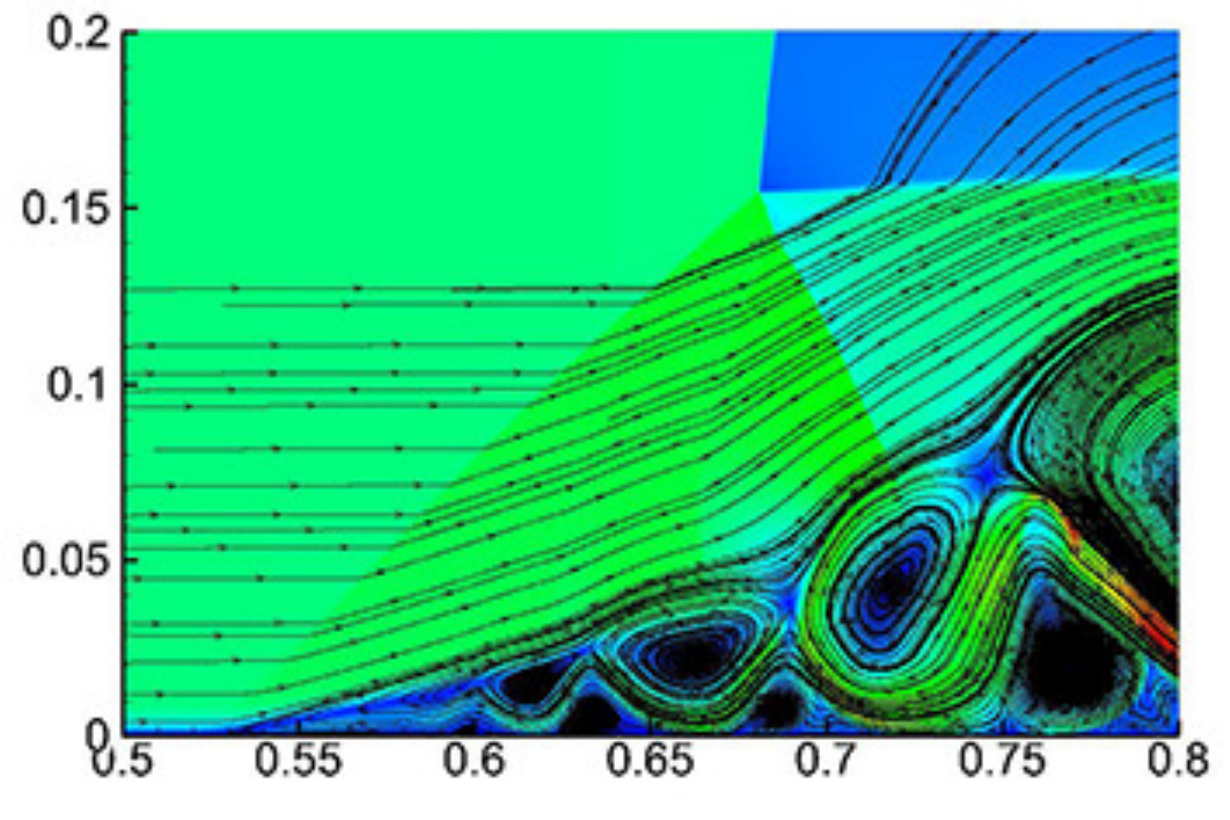}
\centerline{\footnotesize (b)}
\end{minipage}%
\begin{minipage}[t]{0.33\textwidth}
\centering
\includegraphics[width=\textwidth]{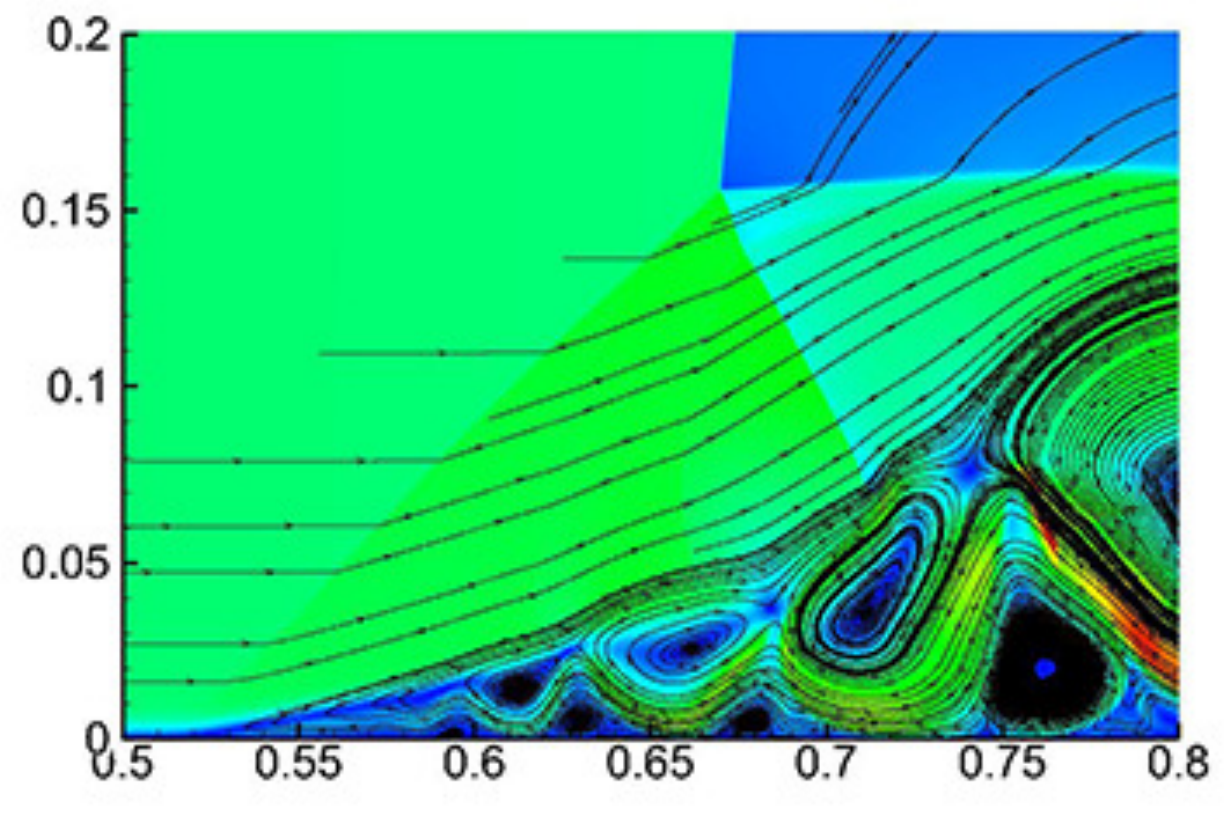}
\centerline{\footnotesize (c)}
\end{minipage}
\begin{minipage}[t]{0.33\textwidth}
\centering
\includegraphics[width=\textwidth]{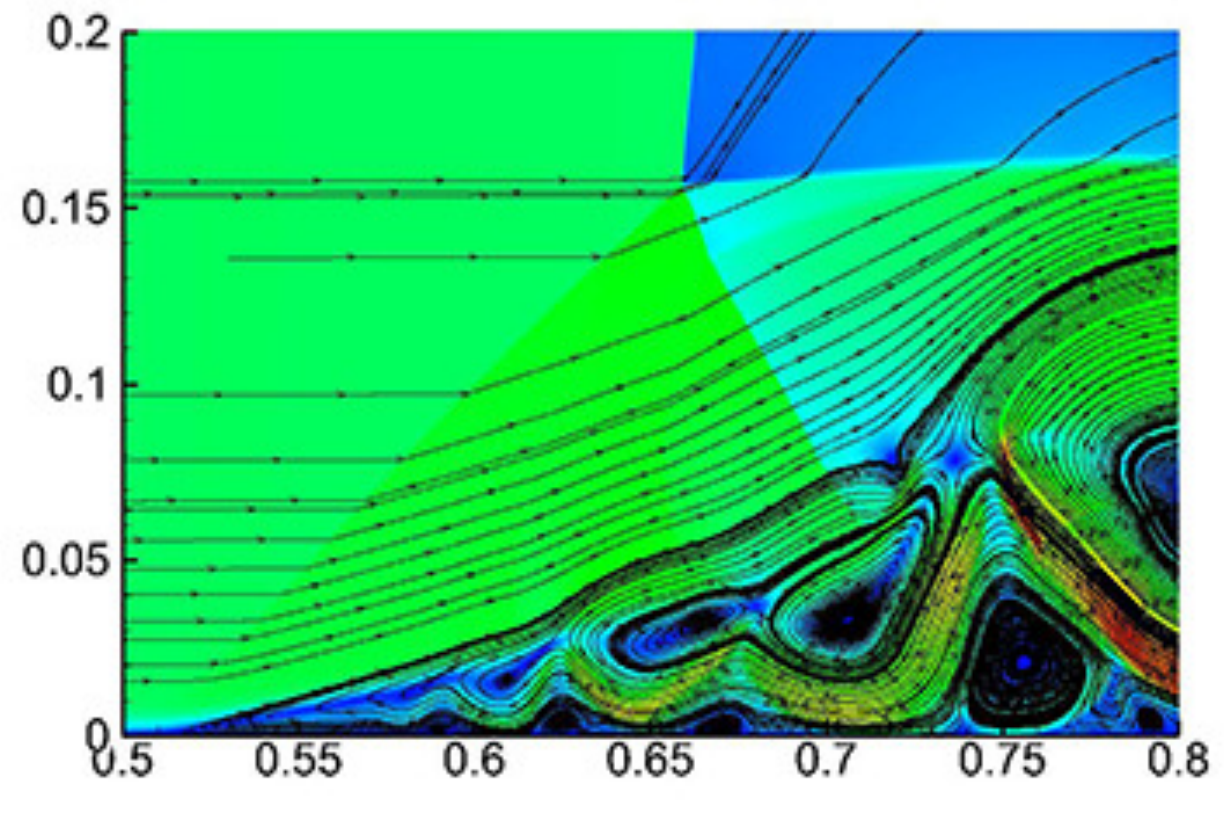}
\centerline{\footnotesize (d)}
\end{minipage}%
\begin{minipage}[t]{0.33\textwidth}
\centering
\includegraphics[width=\textwidth]{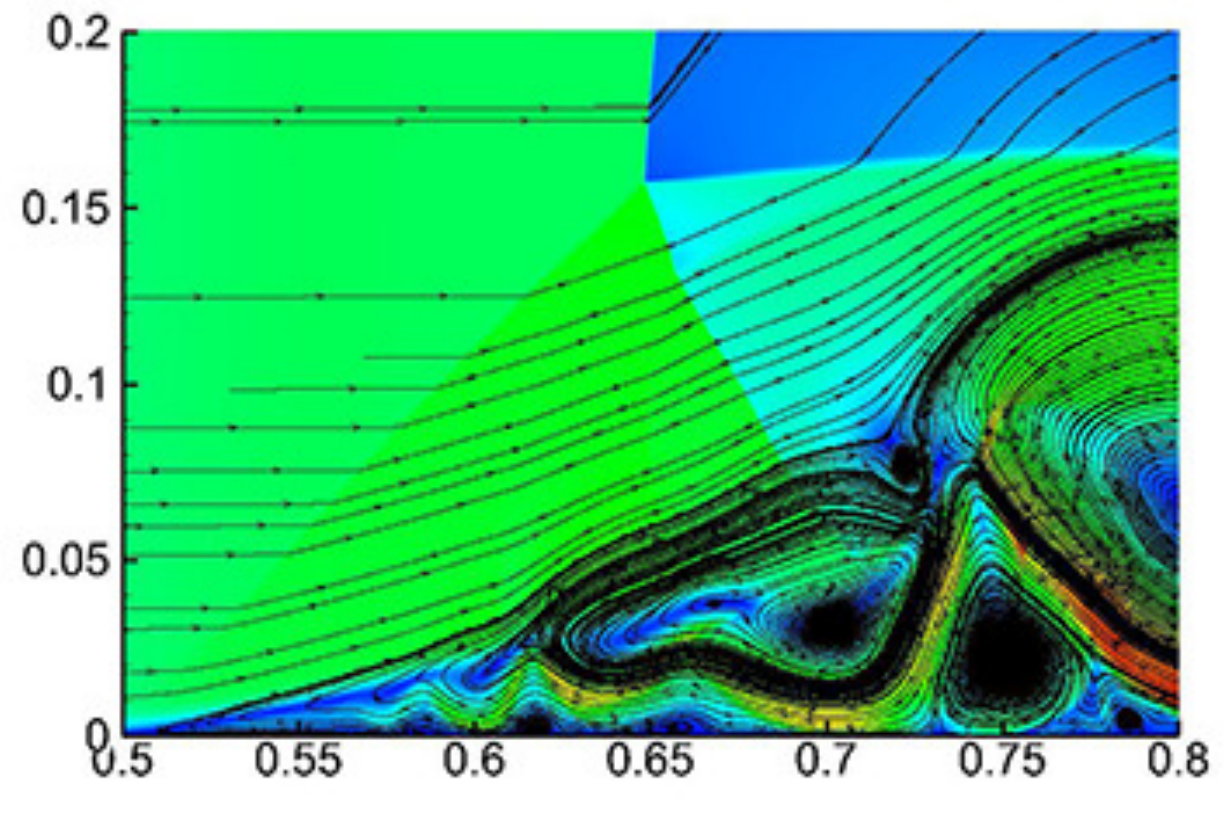}
\centerline{\footnotesize (e)}
\end{minipage}%
\begin{minipage}[t]{0.33\textwidth}
\centering
\includegraphics[width=\textwidth]{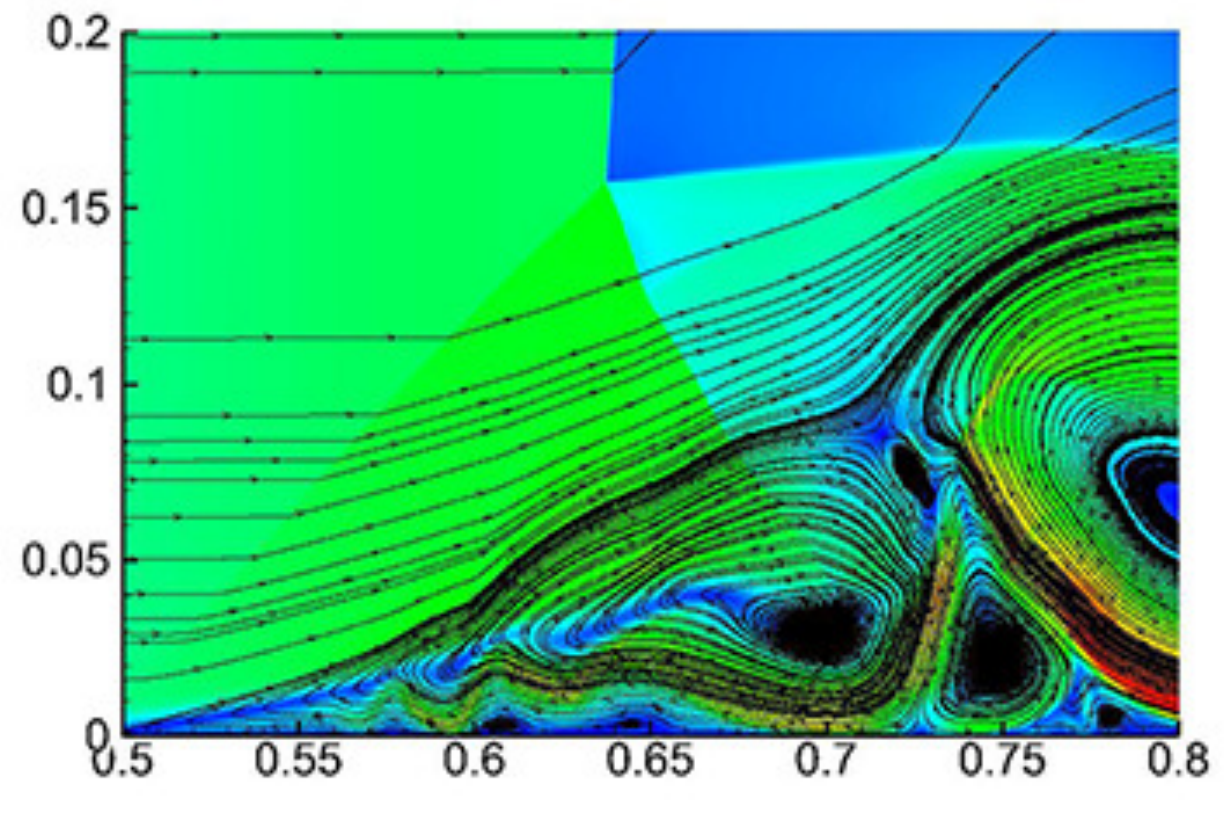}
\centerline{\footnotesize (f)}
\end{minipage}
\centering
\caption{Streamlines and distribution of momentum magnitude at (\textit{a}) $t=0.81$, \protect\\(\textit{b}) $t=0.83$, (\textit{c}) $t=0.85$, (\textit{d}) $t=0.87$, (\textit{e}) $t=0.89$ and (\textit{f}) $t=0.91$.}\label{momentum6}
\end{figure}

On the other hand, the secondary vortices above the jet are lift up as the ejected part of the jet is taken to the right by the outer fluid. Under the flushing of the high-speed outer flow, they are deformed rapidly and get closer to the neighbouring vortices, shown in figures~\ref{momentum6}(a) to \ref{momentum6}(d). Then the adjacent small vortices are merged into a big one since they share the same rotating direction, see figures~\ref{momentum6}(e) and \ref{momentum6}(f). The rotation of the new big vortex tends to make the jet become straight. Also notice that the amount of the fluid under the jet has decreased due to the ejection from the gaps. The final result is that the small vortices in the wedge-shaped area all vanish, and the jet becomes very flattened.

\section{Conclusion} \label{section conclusion}
The viscous shock tube problem is simulated by an efficient high-order gas-kinetic scheme. Grid-convergence studies by using the GCI approach are presented for the two cases at $\Rey = 200$ and $\Rey = 1000$. Grid-converged solutions are achieved on the $500 \times 250$ grid for the $\Rey = 200$ case and on the $3000 \times 1500$ grid for the $\Rey = 1000$ case. Nevertheless, critical points on the curve of  the density distribution along the bottom wall are extracted from the result obtained on the finest grid ($1500 \times 750$ for $\Rey = 200$ and $5000 \times 2500$ for $\Rey = 1000$) as benchmark data. The viscous shock tube problem is a good test case for accuracy, resolution and efficiency of high-order high-resolution schemes. We hope the present results can serve as a benchmark solution.

The dynamic process of the $\Rey = 1000$ case is analysed. Important evolution stages, flow structures and physical phenomena are interpreted in detail, including the downward flow due to the shock curvature, the origin of the primary vortex, the shock bifurcation (the formation of the lambda-shaped shock), the Kelvin-Helmholtz instability, the components of the corner vortex, the secondary vortices and their breaking up. These processes demonstrate the complexity of the interactions between shock waves, contact discontinuities, boundary layers, and multi-scale vortices.

\bibliographystyle{jfm}
\bibliography{jfm-instructions}

\end{document}